\documentclass[12pt,pre,showpacs,preprint]{revtex4}
\usepackage{amsmath,graphicx,color,setspace,subfigure}
\usepackage{amsthm,amssymb}
\usepackage{psfrag}
\renewcommand{\eqref}[1]{(\ref{#1})}
\newcommand{\figref}[1]{Fig.~\ref{fig:#1}}
\newcommand{\secref}[1]{Sec.~\ref{sec:#1}}
\newcommand{\appref}[1]{Appendix~\ref{app:#1}}
\newcommand{\tabref}[1]{Table~\ref{#1}}

\begin{document}

\title{Physical and Mathematical Properties of a Quasi-Geostrophic Model of Intermediate Complexity of the Mid-Latitudes Atmospheric Circulation}

\author{Valerio Lucarini}
\email{valerio.lucarini@unicam.it}
\author{Antonio Speranza}
\author{Renato Vitolo}
\affiliation{
  PASEF -- Physics and Applied Statistics of Earth Fluids,\\
  Dipartimento di Matematica ed Informatica,\\
  Universit\`{a} di Camerino,\\
  Via Madonna delle Carceri, 62032 Camerino (MC), Italy}
\homepage{www.unicam.it/matinf/pasef}

%05.45.-a Nonlinear dynamics and nonlinear dynamical systems
%47.10.+g General theory
%47.20.-k Hydrodynamic stability
%47.52.+j Chaos
%92.60.-e Meteorology
%47.11.+j Computational methods in fluid dynamics

%General circulation, Climate dynamics, Global change
%\subjclass{Primary 47A15; Secondary 46A32, 47D20}
\clearpage

\begin{abstract}
A quasi-geostrophic intermediate complexity model is considered,
providing a schematic re\-pre\-sen\-tation of the baroclinic
conversion processes which characterize the physics of the
mid-latitudes atmospheric circulation. The model is relaxed
towards a given latitudinal temperature profile, which acts as
baroclinic forcing, controlled by a parameter $T_E$ determining
the forced equator-to-pole temperature gradient. As $T_E$
increases, a transition takes place from a stationary regime to a
periodic regime, and eventually to an earth-like chaotic regime
where evolution takes place on a strange attractor. The dependence
of the attractor dimension, metric entropy, and bounding box
volume in phase space is studied by varying both $T_E$ and model
resolution. The statistical properties of observables having
physical relevance, namely the total energy of the system and the
latitudinally averaged zonal wind, are also examined. It is
emphasized that while the attractor's properties are quite
sensitive to model resolution, the global physical observables
depend less critically on it. For more detailed physical
observables, such as the latitudinal profiles of the zonal wind,
model resolution again may be critical: the effectiveness of the
zonal wind convergence, acting as barotropic stabilization of the
baroclinic waves, heavily relies on the details of the latitudinal
structure of the fields. The necessity and complementarity of both
the dynamical systems and physical approach is underlined.
\end{abstract}

\maketitle

\newpage

\tableofcontents
\newpage

\section{Introduction: Atmospheric Circulation as a problem in Physics and in Mathematics}
In the scientific context, Climate is defined by the statistical
properties of the Climatic System. In its most complete
definition, the Climatic system is composed of four intimately
interconnected sub-systems, Atmosphere, Hydrosphere, Cryosphere,
and Biosphere. These subsystems interact nonlinearly with each
other on various time-space scales~\cite{Peix,Luc02}.

The Atmosphere is the most rapid component of the Climatic System.
The Atmosphere is very rich in microphysical structure and
composition and evolves under the action of macroscopic driving
and modulating agents - solar heating and Earth's rotation and
gravitation, respectively. The Atmospheric Circulation is the
basic engine which transforms solar heating into the energy of the
atmospheric motions determining weather and climate as we commonly
perceive them. The Atmosphere features both many degrees of
freedom, which makes it complicated, and nonlinear interactions of
several different components coupling a vast range of time-space
scales, which makes it complex. In many cases, the dynamics of
such a system is strongly chaotic  - in the sense that the
autocorrelation function of any variable vanishes on finite time
scales - and is characterized by a large natural variability on
different time scales~\cite{Lor69,Lor76}

The understanding of the physical mechanisms operating in the
Atmosphere critically influences important human activities like
weather forecast, territorial planning, etc. This is one reason
why, more than half a century ago, von Neumann posed the
Atmospheric Circulation in the core of the ongoing development of
numerical modelling~\cite{Char50}. However, the General
Atmospheric Circulation (GAC) also poses problems of general
physical nature as a realization - in fact the one we can best
observe - of planetary scale thermodynamic transformations in a
rotating, stratified fluid.

Historically - see the classical monograph and paper by
Lorenz~\cite{Lor67,Lor83} - the problem of GAC has been
essentially approached in terms of time-mean circulation and the
processes which generate and maintain it. Almost one century ago
Jeffrey~\cite{Jef24,Jef25} realized that in order to maintain the
observed time-mean circulation at middle latitudes, it is
necessary to take into account the momentum and heat transfer
properties of the eddies, \textit{i.e.} the fluctuating component
of the atmospheric flows~\cite{Palmen}.

Among all the physical processes involved in the GAC, the
so-called \textit{baroclinic} conversion (baroclinic comes from
ancient Greek: constant pressure surfaces not parallel to constant
density surfaces) plays a central role because it is through this
mechanism that rotating, stratified fluids  convert the available
potential energy~\cite{Mar,Lor55,Lor60}, stored in the form of
thermal fluctuations, into the vorticity and kinetic energy of the
air flows as we observe them. At mid-latitudes of both
hemispheres, the baroclinic conversion process can be taken as
responsible for the destabilization of the fixed point given by
the zonally (longitudinally) symmetric Atmospheric Circulation
characterized by a purely zonal wind (jet)~\cite{HH}. Baroclinic
unstable waves can be actually observed (see for
instance~\cite{Black,SP,DellAquila}). The definition of the basic
ingredients in the mechanism of baroclinic instability has been
one the main successes of the dynamical Meteorology of this
century~\cite{Char47,Eady49}.

Within the, virtually innumerable, papers devoted to the subject
of GAC, a few happened to suggest new methodologies and concepts
of general interest for fundamental disciplines, such as Physics
and Mathematics, as well as more empirical natural and social
sciences, such as Biology, Medicine and Economics. A leading
example is that of Lorenz' attractor~\cite{Lor63}. But, apart from
such exceptions, the problem of GAC has remained confined within
the boundaries of Geophysical (mostly Meteorological) literature,
with all the ensuing language barriers with respect to Physics and
Mathematics. Also the relatively recent (last fifteen years)
public attention on Climate issues has been attracted essentially
on phenomenological and/or numerical modelling issues rather than
on fundamental mechanisms~\cite{IPCC}. One consequence of this
cultural separation has been that, for example, the knowledge that
in dynamical systems the stability properties of the time mean
state do not even provide a zeroth-order approximation of the
dynamical properties of the full nonlinear system has been, and
still is, quite systematically ignored in specialized literature -
see~\cite{HAL} and~\cite{Kuo} for enlightening examples - despite
both theoretical arguments~\cite{Far} and simple counter-examples
of physical significance~\cite{SM,MTS} indicated throughout the
years. Note that this, somewhat methodological, issue bears
relevance also in practical problems like the provision of the
so-called extended range weather forecasts, which extend beyond
the deterministic predictability horizon of about 10-15 days (see
\textit{e.g.} Lorenz~\cite{Lor67,Lor83}). Suppose, in fact, that
the forecaster was given the next month average atmospheric
fields: what practical information would he derive from that? Of
course, if dynamical information is stored in the average fields -
for example in the form of dominant regimes of instability
derivable from the stability analysis of time-mean
flow~\cite{Fred} - we could obtain useful information from the
prediction of such time mean fields. Unfortunately, as remarked
above, this picture is far from being true, and the problem of
extended range is still open even in terms of formulating clearly
what we should forecast!

In order to address some of the above mentioned issues, in this
work we consider a quasi-geostrophic model of
\textit{intermediate} complexity for the atmospheric circulation.
By intermediate we mean that the number of variables ($48$ to
$384$) lies between the few degrees of freedom of, say, the Lorenz
models~\cite{Lor63,Lor80}, and the state-of-the-art Global
Circulation Models~\cite{IPCC}, which feature over $10^6$ degrees
of freedom. The model used here has no seasonal cycle and provides
an earth-like representation of the turbulent baroclinic
jet~\cite{SM,MTS}. It is vertically discretized into two layers,
which is the minimum for baroclinic conversion to take
place~\cite{Ped,Phil54}, and latitudinally discretized by a
Fourier half-sine pseudo-spectral expansion up to order $JT$. We
have used $JT=8$, $16$, $32$, $64$, yielding a hierarchy of
quasi-geostrophic models having increasing resolution. A
fundamental property of these models is semi-linearity: the eddy
field is truncated to one wavenumber in the longitudinal (zonal)
direction, so that the evolution equation is linear in terms of
the time-varying zonal flow. This provides a dynamical meaning for
the separation between zonal and eddy flow that is only
geometrical - and originally just geographical - in the
traditional approach: in our case the zonal flow is an integrator
of the nonlinear self-interactions of the wave-field which
propagates and grows linearly on the zonal flow self.

In \secref{ab-initio} we present a detailed general derivation of
the evolution equations for the two-level quasi-geostrophic model
starting from the \textit{ab-initio} equations and explaining the
approximation involved in the derivation of the 3D
quasi-geostrophic equations. This derivation allows a clear
understanding of the physics involved in the considered hierarchy
of quasi-geostrophic equations and is alternative to the
non-dimensional formulations which are common in the
meteorological literature~\cite{Ped}. We further obtain the
equations of the one-wavenumber model examined in this study, in
the form adopted for the numerical integration.

The main results of this work are presented in
\secref{mainresults} and \secref{energy}. We study the sensitivity
of the model behavior with respect to the parameter $T_E$
determining the forced equator-to-pole temperature gradient, which
acts as baroclinic forcing. The influence of the order of
(spectral) discretization in the latitudinal direction is also
analyzed. For low values of $T_E$ there occurs a transition from a
stationary to a earth-like chaotic regime. Here chaotic means that
an attractor is detected having a positive maximal Lyapunov
exponent, \textit{i.e.}, a \emph{strange} attractor (see~\cite{ER}
for terminology). In \secref{mainresults} we characterize the
transition from stationary to chaotic dynamics in terms of
bifurcation theory and study the dependence on $T_E$ and on model
resolution $JT$ of the dimension of the strange attractor, of the
metric entropy, and of the volume of its bounding box in the phase
space. In \secref{energy} we analyze the statistical properties of
two physically meaningful observables, namely the total energy of
the system and the latitudinally averaged zonal wind. An
inspection of the latitudinal wind profiles is also presented. In
\secref{conclusions} we give our conclusive remarks and
perspectives for future works.

\newpage
\section{The \textit{ab-initio} formulation of the model equations of motion}
\label{sec:ab-initio}
\subsection{Initial Remarks}
The dynamics and thermodynamics of the dry atmosphere  for an
observer in the Earth's uniformly rotating frame of reference is
described by the following equations~\cite{Peix}:

\begin{align}
  \label{conti}
  &\frac{D}{Dt}\rho+\rho\vec{\nabla}\cdot\vec{u}=0\\
  \label{Navier}
  &\frac{D}{Dt}\vec{u}+2\vec{\Omega}\times\vec{u}=
  -\frac{\vec{\nabla}p}{\rho}-\vec{\nabla}\Phi+\vec{F}\\
  \label{entropy}
  &\frac{D}{Dt}h-\frac{1}{\rho}\frac{D}{Dt}p={Q}+D\\
  \label{state}
  &\rho=\rho\left(p,T\right).
\end{align}
Here $\rho$ is the density, $\vec{u}$ is the velocity vector,
$\Omega$ is the Earth's rotation angular velocity, $p$ is the
pressure, $\Phi$ is the geopotential, $\vec{F}$ is the resultant
of the frictional forces per unit mass, $h$ is the specific
enthalpy, $Q$ is the diabatic heating, $D$ represents the effect
of heat diffusion processes, and $T$ is the temperature of the
fluid. The material derivative $D/Dt$ is defined as follows:
\begin{equation}%%noref
  \frac{D}{Dt}\bullet=\frac{\partial }{\partial t}
  \bullet+\left(\vec{u}\cdot \vec{\nabla}\right)\bullet.
\end{equation}%%noref
Equations~\hbox{\eqref{conti}-\eqref{entropy}} are commonly
referred to as mass continuity, Navier-Stokes, and thermodynamics,
respectively. They express the dynamic balances of mass, forces,
and specific enthalpy of the system, while~\eqref{state} is the
equation of state of the fluid under consideration, which, in the
case of dry air, can be well-represented as a perfect gas.

The description of the macroscopic behavior of the atmosphere is
based on the systematic use of dominant balances derived on a
phenomenological basis. Suitable approximations to
equations~\hbox{\eqref{conti}-\eqref{entropy}} are obtained by
assuming that the actual evolution departs only slightly from the
balances. In fact, different balances have to be applied depending
on the time and space scales we are focusing on. In this way, it
is possible to filter out (exclude) all solutions corresponding to
physical processes that are heuristically assumed to contribute
only negligibly to the dynamics of the system, at the time and
space scale under examination. The magnitudes of various terms the
governing equations for a particular type of motion are estimated
using the so-called scale analysis technique~\cite{SPELUC}. The
resulting models usually give good approximation to the observed
fields when sufficiently large spatial or temporal averages are
selected~\cite{Ped,Holton,Peix}.

\subsection{The hydrostatic and quasi-geostrophic approximations}
For the dynamics of the atmosphere at mid-latitudes, on spatial
and temporal scales comparable with or larger than those of the
synoptic weather (about 1000 $Km$ and 1 day, respectively), it is
phenomenologically well-established that the hydrostatic balance
is obeyed with excellent approximation~\cite{Ped,Holton,Peix}:
\begin{equation}
  \label{hydro}
  \hat{k}\cdot\vec{\nabla}p=-\rho g,
\end{equation}
where:
\begin{equation}%%noref
  \hat{k}=-\frac{\vec{\nabla}\Phi}{\left\vert\vec{\nabla}\Phi\right\vert}
\end{equation}%%noref
This expresses the balance between the gravitational force and the
vertical pressure gradient, the vertical direction $\hat{k}$ being
defined by the gradient of $\Phi$. Since the atmosphere is shallow
with respect to the radius of the Earth, one can use the
approximation $\Phi\sim gz$, where $z$ is the local geometric
vertical coordinate. The hydrostatic balance also allows the usage
of $p$ as vertical coordinate.

Moreover, in the just mentioned synoptic scales, the atmosphere is
close to the geostrophic equilibrium, which is realized when the
local horizontal pressure gradient exactly balances the Coriolis
acceleration. In the geostrophically balanced flows, when pressure
is taken as vertical coordinate, the wind can be expressed as
\begin{equation}
  \label{geostro}
  \vec{u}_g=\left(u_g,v_g,0\right)=
  \frac{1}{f_0}\hat{k}\times\vec{\nabla}\Phi=\hat{k}\times\vec{\nabla}\psi_g,
\end{equation}
where $f_0=2\Omega\sin\varphi$ is the the orthogonal projection of
the Coriolis parameter on the surface of the planet at latitude
$\varphi$ and $\psi_g=\Phi/f_0$ is defined as the streamfunction
of the flow. The geostrophic wind~\eqref{geostro} is horizontal
and non-divergent. This implies that the geostrophic vorticity
vector is parallel to the vertical direction and its non-vanishing
component can be expressed as:
\begin{equation}%%noref
  \xi_g=\hat{k}\cdot\left(\nabla\times\vec{u}_g\right)=
  \frac{1}{f_0}\Delta_H\Phi=\Delta_H\psi_g,
\end{equation}%%noref
where $\Delta_H$ is the horizontal Laplacian operator, see
\emph{e.g.}~\cite{Ped,Holton,Peix}.

Equations~\eqref{hydro} and~\eqref{geostro} are only diagnostic,
so that no information on the evolution of the system can be
obtained. From the set~\hbox{\eqref{conti}-\eqref{state}} of
\textit{ab-initio} dynamic and thermodynamic equations of the
atmosphere it is possible to obtain a set of simplified prognostic
equations for the synoptic weather atmospheric fields by assuming
that the fluid obeys the hydrostatic balance and undergoes small
departures from the geostrophic balance. Moreover, we assume that
the domain is centered at mid-latitudes and it is such that $f$
can be well-approximated by the linear expansion
$f\left(\varphi\right)\sim f\left(\varphi_0\right)
+2\Omega\cos\left(\varphi_0\right)\left(\varphi-\varphi_0\right)$.

\begin{figure}[p]
  {\includegraphics[angle=270,width=0.9\textwidth]{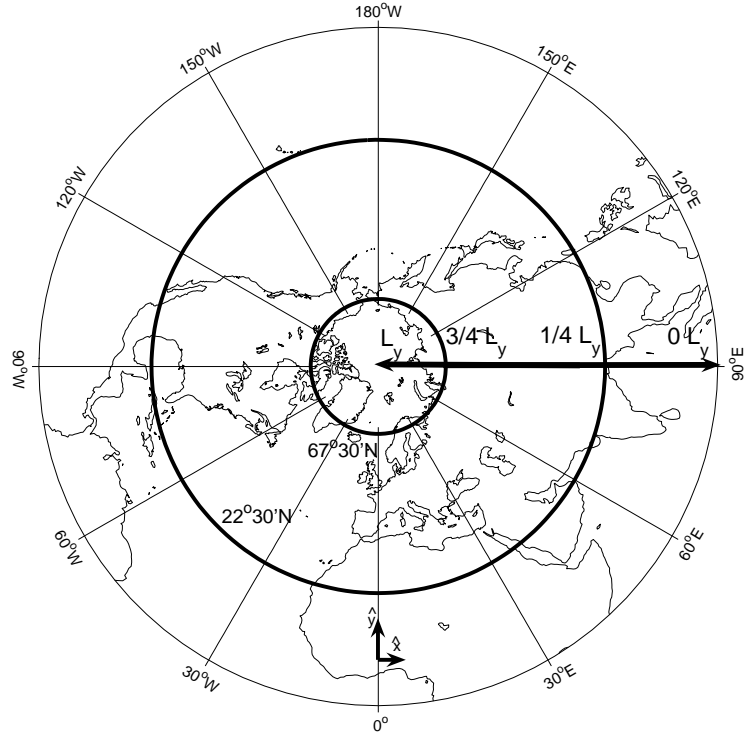}}
  \caption{
    Sketch of the actual geographical area
    corresponding to the simplified $\beta$ channel.
    The  local $x$  and $y$ directions and the $\beta$-channel width $L_y$ are indicated.
    The mid-latitudes range from $1/4 L_y$ to $3/4 L_y$,
    corresponding to a $45^o$ latitudinal belt centered at $45^o N$.
  }
  \label{fig:geogr}
\end{figure}

Local Cartesian coordinates $(x,y)$ and pressure coordinate $p$
are introduced for the horizontal and vertical directions,
respectively, with $x$ denoting the zonal and $y$ the latitudinal
coordinate. The resulting domain is periodic in $x$, with
wavelength $L_x$, and bounded in $y$ and $p$, yielding
\begin{equation}%%noref
  x\in \mathbf{R}/2\pi L_x,\quad y\in\left[0,L_y\right],\quad
  p\in\left[0,p_0\right],
\end{equation}%%noref
and $f$ is approximated as $f\sim f_0+\beta \left(y-L_y/2\right)$.
In the meteorological \textit{jargon} this is usually referred to
as the $\beta$-\textit{channel}. A sketch of the actual
geographical area corresponding to the $\beta$-channel is
presented in \figref{geogr}. We remark that in this work, in order
to avoid problems in the definition of the boundary conditions of
the system, due to the prescription of the interaction with the
polar and the equatorial circulations at the northern and southern
boundary, respectively~\cite{SM}, we consider a domain extending
from the pole to the equator. We remark that the quasi-geostrophic
approximation is not appropriate for the equatorial region, so
that we do not expect to capture any realistic feature of the
tropical circulation, and that the mid-latitude channel is
determined by $y$ ranging from $1/4$ $L_y$ to $3/4$ $L_y$,
corresponding to a latitudinal belt centered about $45^o N$ with
an extension of $45^o$.

Proceeding further with simplifying assumptions, the equation of
state $\rho=p/R T$ is adopted for~\eqref{state}, where $R$ is the
gas constant for dry air, so that the following relation holds:
\begin{equation}
  \label{tempe}
  \frac{\partial \psi_g}{\partial p}=-\frac{R}{f_0p}T,
\end{equation}
and the specific enthalpy for the dry air is expressed as
$h=C_pT$. We introduce the quasi-geostrophic material derivative:
\begin{equation}%%noref
  \frac{D_g}{Dt}\bullet=\frac{\partial }{\partial
    t}\bullet+\left(\vec{u}_g\cdot
    \vec{\nabla}\right)\bullet=\frac{\partial }{\partial
    t}\bullet+J\left(\psi_g,\bullet\right),
\end{equation}%%noref
where $J$ is the conventional Jacobian operator defined as
$J\left(A,B\right)=\partial_xA\partial_yB-\partial_yA\partial_xB$.
Physically, this means that advection occurs along constant
pressure levels and is performed by the geostrophic wind. This
yields the so-called quasi-geostrophic equations for the
streamfunction $\psi_g$:
\begin{align}
  \label{dyna}
  &\frac{D_g}{Dt}\left(\Delta\psi_g+f_0+\beta y\right)-f_0\frac{\partial\omega}{\partial p}=\hat{k}\cdot\vec{\nabla}\times F+\nu\left(\Delta_H\right)^2\psi_g\\
  \label{thermo}
  &\frac{D_g}{Dt}\left(-\frac{\partial \psi_g}{\partial
      p}\right)+\frac{Rp}{f_0^2}\frac{T}{\Theta}\frac{\partial
    \Theta}{\partial
    p}\frac{f_0}{p^2}\omega=\kappa\Delta_H\left(-\frac{\partial
      \psi_g}{\partial p}\right)+\frac{R}{p f_0}\frac{Q}{C_p}
\end{align}
where $\omega$ is the velocity in the direction of $p$, the
frictional forces are represented as viscous processes with
diffusion constant $\nu$, the heat diffusion is parameterized by
the coefficient $\kappa$, and $\Theta$ is the potential
temperature:
\begin{equation}%%noref
  \Theta=T\left(\frac{p_0}{p}\right)^{\frac{R}{C_p}},
\end{equation}%%noref
which is related to the specific entropy $s$ of the air by
\begin{equation}%%noref
  s=C_p\ln\Theta.
\end{equation}%%noref
The quasi-geostrophic approximation is very useful because the
resulting evolution equations~\hbox{\eqref{dyna}-\eqref{thermo}}
focus on the process of slanted convection which is responsible
both for the baroclinic conversion of potential energy into eddy
energy and for the generation of vorticity. These are the
essential ingredients underlying the generation of atmospheric
disturbances at mid-latitudes~\cite{Lor67,Ped,Holton,Peix}

The non-geostrophic velocity component $\omega$ does not have an
evolution equation and can be diagnosed from the thermodynamics
equation~\eqref{thermo}. The following boundary conditions apply
for the ageostrophic velocity $\omega$:
\begin{align}
  \label{bcomega1}
  &\omega\left(x,y,p=0\right)=0\\
  \label{bcomega2}
  &\omega\left(x,y,p=p_0\right)=-E_0\xi_g\left(x,y,p=p_0\right)
\end{align}
where the condition at $p=p_0$ is due to the Ekman description of
the coupling of the free atmosphere with the planetary boundary
layer~\cite{Holton}. We adopt the phenomenologically-based
approximation:
\begin{equation}%%noref
  \frac{Rp}{f_0^2}\frac{T}{\Theta}\frac{\partial \Theta}{\partial
    p}\sim-H(p)^2,
\end{equation}%%noref
where $H\left(p\right)$ is the vertical scale related to the
stratification of the atmosphere which depends only on $p$. By
substituting the vertical velocity $\omega$ obtained in
equation~\eqref{thermo} into equation~\eqref{dyna}, we obtain that
the quasi-geostrophic potential vorticity $q_g$, defined as:
\begin{equation}
  \label{qg2}
  q_g=\Delta_H\psi_g+f_0+\beta y+\frac{\partial}{\partial
    p}\left(\frac{p^2}{H^2}\frac{\partial \psi_g}{\partial
    p}\right),
\end{equation}
satisfies the following canonical equation~\cite{Hosk,Ped,Holton}:
\begin{equation}
  \label{qg1}
  \frac{D_g}{Dt}q_g=\kappa\frac{\partial}{\partial
    p}\left[\frac{p^2}{H^2} \left(-\frac{\partial \psi_g}{\partial
        p}\right)\right]+R \frac{\partial}{\partial
    p}\left(\frac{p}{H^2}\frac{Q}{C_p}\right)+\nu\left(\Delta_H\right)^2\psi_g.
\end{equation}
The quantity $q_g$ (as well as all of its powers) is conserved
along motion if no diabatic forcing is applied ($Q=0$) and if the
diffusion and viscous effects are discarded, \textit{e.g.} by
setting in our case $\nu=\kappa=0$.

We remind that, formally, the quasi-geostrophic
equations~\hbox{\eqref{dyna}-\eqref{thermo}} can be derived from
the \textit{ab-initio}
equations~\hbox{\eqref{conti}-\eqref{state}} by retaining the
zeroth and first order term in the expansion performed on the
Rossby number:
\begin{equation}%%noref
  Ro=\frac{U}{f_0 L}\ll1,
\end{equation}%%noref
with $\beta L\ll f_0$, where $U$ and $L$ are typical values of the
horizontal velocity and horizontal space scale~\cite{Ped}.

\subsection{The two-level model}
A simplified version of system~\hbox{\eqref{dyna} and
\eqref{thermo}} is produced by discretizing the vertical direction
into a finite number of pressure levels. This vertical
discretization approach has been first introduced by
Phillips~\cite{Phil54} and retains the baroclinic conversion
process, which is the basic physical feature of the
quasi-geostrophic approximation.

\begin{figure}[p]
  {\includegraphics[angle=270,width=0.9\textwidth]{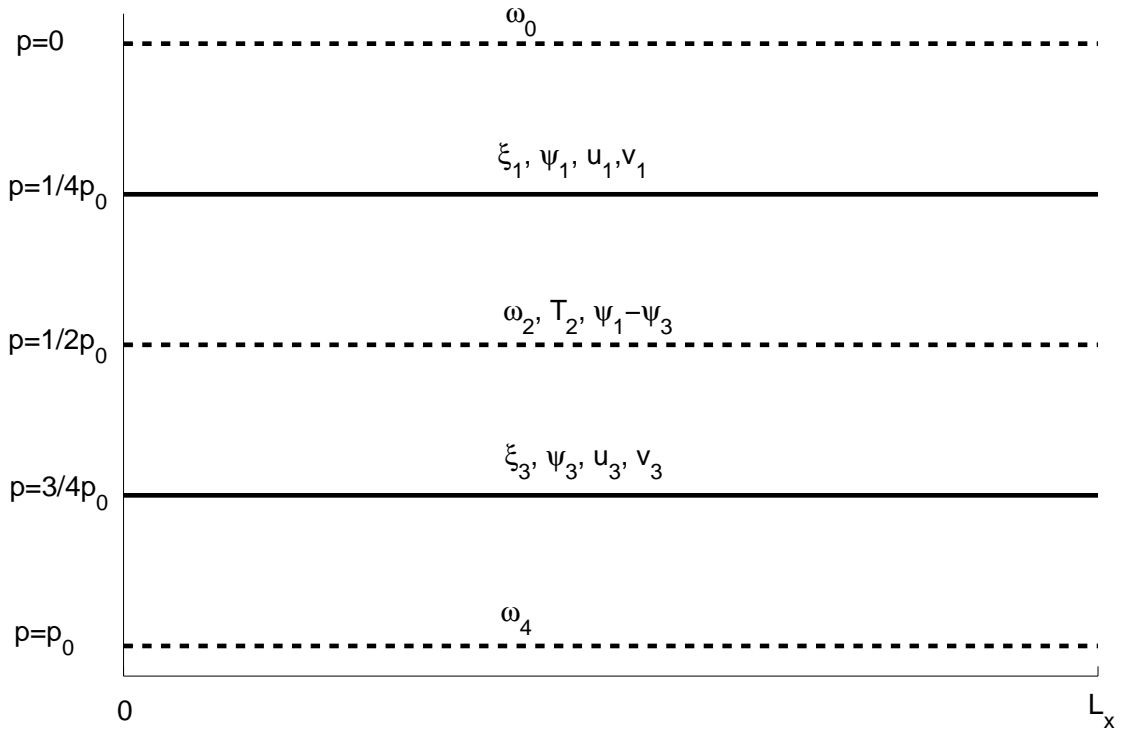}}
  \caption{
    Sketch of the vertical-longitudinal section of the system domain.
    The domain is periodic in the zonal direction $x$ with wavelength $L_x$.
    At each pressure level, the relevant variables are indicated.
  }
  \label{fig:system}
\end{figure}

We refer to \figref{system} for a sketch of the vertical geometry
of the two-layer system. The streamfunction $\psi_g$ is thus
defined at pressure levels $p=p_1=p_0/4$ and $p=p_3=3/4p_0$, while
$\omega$ is defined at the pressure levels $p=0$ (top boundary),
$p=p_2=p_0/2$, and $p=p_0$ (surface boundary). The pressure level
pertaining to the vertical derivative of the streamfunction
$\partial \psi_g/\partial p$ as well as the stratification height
$H$ is $p=p_2$. We note that $\delta p=p_3-p_1=p_2=p_0/2$. This
system is described by the following equations of motion:
\begin{align}
  \label{level1}
  &\frac{D_{1}}{Dt}\left(\Delta_H\psi_1+f_0+\beta
    y\right)-f_0\frac{\omega_2-\omega_0}{\delta p}=0,\\
  \label{level3}
  &\frac{D_{3}}{Dt}\left(\Delta_H\psi_3+f_0+\beta
    y\right)-f_0\frac{\omega_4-\omega_2}{\delta p}=0,\\
  \label{level2}
  &\frac{D_{2}}{Dt}\left(\frac{\psi_1-\psi_3}{\delta
      p}\right)-H_2^2\frac{f_0}{p_2^2 }\omega_2=\kappa
  \Delta_H\left(\frac{\psi_1-\psi_3}{\delta p}\right)+\frac{R}{p_2
    f_0}\frac{Q_2}{C_p},
\end{align}
where we have neglected the viscous dissipation by setting
$\nu=0$, dropped the $g$ subscript for simplicity, and have
adopted the notation
\begin{align}%%noref
  &\psi_j=\psi\left(p_j\right), \hspace{5pt}j=1,3,
  \hspace{25pt}Q_2=Q(p_2),\\
  &\omega_j=\omega\left(p_j\right), \hspace{5pt}j=0,2,4,
  \hspace{15pt}H_2=H(p_2),\\
  &\frac{D_{j}}{Dt} \bullet=\frac{\partial }{\partial t} \bullet
  +J\left(\psi_j,\bullet\right)\hspace{5pt}j=1,3.&
\end{align}%%noref
The boundary conditions~\hbox{\eqref{bcomega1}-\eqref{bcomega2}}
on the vertical velocity are implemented as
\begin{align}
  \label{bcon20}&\omega_0=0,\\
  \label{bcon24}&\omega_4=-E_0\Delta_H\psi_3,
\end{align}
where the streamfunction at the top of the boundary layer has been
approximated by the streamfunction $\psi_3$~\cite{Ped}. The
streamfunction at the intermediate level $p_2$ is computed as
average between the streamfunctions of the levels $p_1$ and $p_3$,
so that the material derivative at the level $p_2$ can be
expressed as:
\begin{equation}%%noref
  \frac{D_{2}}{Dt}=
  \frac{1}{2}\left(\frac{D_{1}}{Dt}+\frac{D_{3}}{Dt}\right)=
  \frac{\partial}{\partial
    t}\bullet+\frac{1}{2}J\left(\psi_1+\psi_3,\bullet\right).
\end{equation}%%noref
Along the lines of the derivation
of~\hbox{\eqref{qg1}-\eqref{qg2}}, by substituting $\omega_2$,
$\omega_0$, and $\omega_4$ (as defined in
~\eqref{level2},~\eqref{bcon20}, and~\eqref{bcon24}, respectively)
into~\eqref{level1} and~\eqref{level3}, we obtain the evolution
equations for the quasi-geostrophic potential vorticity at the two
levels:
\begin{align}
  \label{level1q}
  &\frac{D_{1}}{Dt}q_1=-\frac{\kappa}{H_2^2}
  \Delta_H\left(\psi_1-\psi_3\right)-\frac{R}{f_0H_2^2}\frac{Q_2}{C_p},\\
  \label{level3q}
  &\frac{D_{3}}{Dt}q_3=-\frac{f_0 E_0}{\delta
    p}\Delta_H\left(\psi_1-\psi_3\right)+\frac{\kappa}{H_2^2}
  \Delta_H\left(\psi_1-\psi_3\right)+\frac{R}{f_0H_2^2}\frac{Q_2}{C_p}.
\end{align}
Here the $q_i$'s are defined as:
\begin{equation}%%noref
  q_i=\Delta_H\psi_1+f_0+\beta
  y+\left(1-2\delta_{i,1}\right)\frac{1}{H_2^2}\left(\psi_1-\psi_3\right),
  \hspace{5mm}i=1,3,
\end{equation}%%noref
where $\delta_{i,1}$ is the Kronecker's \textit{delta}, which is
equal to $1$ when the two indexes are mutually equal and $0$
otherwise.

It is possible to derive the following expression for the
horizontal energy density of the system:
\begin{equation}
  \label{energydensity}
  e\left(x,y\right)=\frac{\delta
    p}{g}\left[\frac{1}{2}\left(\vec{\nabla}\psi_1\right)^2+
    \frac{1}{2}\left(\vec{\nabla}\psi_3\right)^2+
    \frac{1}{2H_2^2}\left(\psi_1-\psi_3\right)^2\right].
\end{equation}
Here the factor $\delta p/g$ is the mass per unit surface in each
level, the last term and the first two terms inside the brackets
represent the potential and kinetic energy, respectively, thus
featuring a clear similarity with the functional form of the
energy of a harmonic oscillator. We emphasize that in
~\eqref{energydensity} the potential energy term is half of what
reported in~\cite{Ped}, which contains a trivial algebraic mistake
in the derivation of the energy density, as discussed with the
author of the book.

We choose the following simple functional form for the diabatic
heating:
\begin{equation}
  \label{heating}
  Q_2=\nu_NC_p\left(T^\star-T_2\right)=\nu_NC_p\frac{f_0p_2}{R}\left(\frac{2\tau^\star}{\delta
      p}-\frac{\psi_1-\psi_3}{\delta p}\right),
\end{equation}
where $\tau^\star$ has been introduced for later convenience and,
consistently with equation~\eqref{tempe}, $T_2$ is evaluated at
the pressure level $2$ and is defined by
\begin{equation}
  \label{tempe2}
  \frac{\psi_3-\psi_1}{\delta p}=-\frac{R}{f_0p_2}T_2.
\end{equation}
The functional form of equation~\eqref{heating} implies that the
system is relaxed towards a prescribed temperature profile
$T^\star$ with a characteristic time scale of $1/\nu_N$. $T^\star$
and $\tau^\star$ are respectively defined as follows:
\begin{equation}
  \label{deftau}
  T^\star=\frac{T_E}{2}\cos\left(\frac{\pi y}{L_y}\right),
  \qquad
  \tau^\star=\frac{R}{f_0}\frac{T_E}{4}\cos\left(\frac{\pi y}{L_y}\right),
\end{equation}
so that $T_E$ is the forced temperature difference between the low
and the high latitude border of the domain. Since we assume no
time dependence for the forcing parameter $T_E$, we discard the
seasonal effects. Considering that the thermal wind relation:
\begin{equation}%%noref
\frac{\partial  \vec{u}_g }{\partial
p}=\hat{k}\times\vec{\nabla}\frac{\partial \psi_g}{\partial p}
\end{equation}%%noref
can be discretized as follows for the two level system:
\begin{equation}%%noref
  \frac{\vec{u}_1-\vec{u}_3}{\delta
    p}=\hat{k}\times\vec{\nabla}\frac{ \psi_1-\psi_3}{\delta p},
\end{equation}%%noref
we have that the diabatic forcing $Q_2$ in~\eqref{heating} causes
a relaxation of the vertical gradient of the zonal wind
$u_{1}-u_{3}$ towards the following prescribed profile $2m^\star$:
\begin{equation}
  \label{defmstar}
  2m^\star=\frac{R}{f_0}\frac{\pi}{L_y}\frac{T_E}{2}\sin\left(\frac{\pi
      y}{L_y}\right),
\end{equation}
where the constant $2$ has been introduced for later convenience.
We introduce the baroclinic and barotropic components
$(\tau,\phi)$ as
\begin{align}
\label{tau1}
  \tau&=\frac{1}{2}\left(\psi_1-\psi_3\right),\\
\label{phi1}
  \phi&=\frac12\left(\psi_1+\psi_3\right).
\end{align}
From equations~\hbox{\eqref{level1q}-\eqref{level3q}} one obtains
the equations of motion for $(\tau,\phi)$:
\begin{align}
  \label{tau}
  \frac{\partial}{\partial t}\Delta_H\tau
  -\frac{2}{H_2^2}\frac{\partial}{\partial t}\tau
  +J\left(\tau,\Delta_H\phi+\beta y + \frac{2}{H_2^2}\phi\right)
  +J\left(\phi,\Delta_H\tau\right)=\nonumber\\
  \frac{2\nu_E}{H_2^2}\Delta_H\left(\phi-\tau\right)
  -\frac{2\kappa}{H_2^2}\Delta_H\tau
  +\frac{2\nu_N}{H_2^2}\left(\tau-\tau^\star\right),\\
  \label{phi}
  \frac{\partial}{\partial t}\Delta_H\phi
  +J\left(\phi,\Delta_H\phi+\beta y\right)
  +J\left(\tau,\Delta_H\tau\right)=
  -\frac{2\nu_E}{H_2^2}\Delta_H\left(\phi-\tau\right).
\end{align}
where $\nu_E=f_0E_0H_2^2/\left(2\delta\right)$ and the meaning of
$\tau^\star$ is made clear. Notice that this system only features
quadratic nonlinearities. The two-level quasi-geostrophic
system~\hbox{\eqref{tau}-\eqref{phi}} can be brought to the
non-dimensional form, which is more usual in the meteorological
literature and is easily implementable in computer codes. This is
achieved by introducing length and velocity scales $l$ and $u$ and
performing a non-dimensionalization of both the system variables
$(x,y,t,\phi,\tau,T)$ (as described in \tabref{tab1}) and of the
system constants (\tabref{tab2}). When assessing, as in our case,
atmospheric phenomena from synoptic to planetary scales, suitable
choices for the length and velocity scales are $l=10m^6$ and
$u=10ms^{-1}$. With the choices of the constants described in
table~\ref{tab2}, our system is equivalent to that of Malguzzi and
Speranza~\cite{SM}, where the following correspondences hold:
\begin{equation}%%noref
  \frac{1}{H_2^2}\leftrightarrow F,\quad
  \frac{2\nu_E}{H_2^2}\leftrightarrow \frac{\nu_E}{2},\quad
  \frac{2\kappa}{H_2^2}\leftrightarrow\nu_S,\quad
  \text{and}\quad
  \nu_N\leftrightarrow \nu_H.
\end{equation}%%noref

\subsection{The single zonal wave two-level model}
\label{sec:singlewave} In this section we derive the evolution
equations used in the present study. We Fourier-expand the $\phi$
and $\tau$ fields in the zonal direction $x$ as follows:
\begin{align}
  \label{phifourier}
  &\phi\left(x,y,t\right)=
  \sum_{n=0}^\infty{A_n\left(y,t\right)
    \exp{\left(\textrm{i}2n\pi x/L_x\right)}}+\textrm{c.c.}\\
  \label{taufourier}
  &\tau\left(x,y,t\right)=
  \sum_{n=0}^\infty{B_n\left(y,t\right)
    \exp{\left(\textrm{i}2n\pi x/L_x\right)}}+\textrm{c.c.},
\end{align}
where c.c. stands for complex conjugate. By definition we have
\begin{equation}
  \label{defUm}
  U\left(y,t\right)=-\frac{\partial A_0\left(y,t\right)}{\partial y},
  \qquad
  m\left(y,t\right)=-\frac{\partial B_0\left(y,t\right)}{\partial y},
\end{equation}
so that $U$ represents the zonal average of the mean of the zonal
wind at the two pressure levels $1$ and $3$ (see previous
Section), while $m$ represents the zonal average of the halved
difference between the the zonal wind at the two pressure levels
$1$ and $3$. In this work we focus on the interaction between the
average zonal wind and waves, thus neglecting the wave-wave
nonlinear interactions. We therefore only retain the zonally
symmetric component (\textit{i.e.}, that of order $n=0$) and one
of the non-zonal components (\textit{i.e.}, for a fixed $n\ge1$)
in the Fourier
expansions~\hbox{\eqref{phifourier}-\eqref{taufourier}} and in the
equations of motion. Since quadratic nonlinearities like those
described in equations~\hbox{\eqref{tau}-\eqref{phi}} generate
terms with Fourier components corresponding to the sum and
difference of the Fourier components of the two factors, no
wave-wave interactions can take place. Note that if cubic
nonlinearities were present, direct wave-wave interaction would
have been possible~\cite{Luc05}. In the present case, the wave can
self-interact only indirectly through the changes in the values of
the zonally symmetric fields $U$ and $m$. This amounts to building
up a semi-linear equation for the wave on top of a nonlinear
dynamics for the zonally symmetric parts of the fields.

As the only retained non-zonal component we select that of order
$n=6$, since we intend to represent the baroclinic conversion
processes, that in the real atmosphere take place on scales of
$L_x/6$ or smaller \cite{DellAquila}. With this choice, setting
$\chi=6\times 2\pi/L_x$ in order to simplify the notation,
equations~\hbox{\eqref{phifourier}-\eqref{taufourier}} reduce to
\begin{align}
  \label{phifourier2}
  &\phi\left(x,y,t\right)=
  -\int_{\pi/2}^y{U\left(z,t\right)\rm{d}z}
  +A\exp{\left(\textrm{i} \chi x\right)}+\textrm{c.c.},\\
  \label{taufourier2}
  &\tau\left(x,y,t\right)=
  -\int_{\pi/2}^y{m\left(z,t\right)\rm{d}z}
  +B\exp{\left(\textrm{i} \chi x\right)}+\textrm{c.c.},
\end{align}
where the choice of the lower integration limit will be explained
later. By
substituting~\hbox{\eqref{phifourier2}-\eqref{taufourier2}} into
equations~\hbox{\eqref{tau}-\eqref{phi}} and projecting onto the
Fourier modes of order $n=0$ and $n=6$, we obtain the equations:
\begin{align}
  \label{dotA}
  &\begin{aligned}
    \dot{A}_{yy}-\chi^2\dot{A}
&+\left(\textrm{i} \chi U+\frac{2\nu_E}{H_2^2}\right)A_{yy}
    -\left(\textrm{i}  \chi^3U+\textrm{i}  \chi U_{yy}+\frac{2\nu_E}{H_2^2}\chi^2-\textrm{i}  \chi\beta\right)A
    \\
    &+\left(\textrm{i}  \chi m-\frac{2\nu_E}{H_2^2}\right)B_{yy}
    -\left(\textrm{i}  \chi^3m+\textrm{i}  \chi m_{yy}-\frac{2\nu_E}{H_2^2}\chi^2\right)B
    =0,
  \end{aligned}\\
  \label{dotB}
  &\begin{aligned}
    \dot{B}_{yy}-\chi^2\dot{B}-&\frac{2}{H_2^2}\dot{B}
    +\left(\textrm{i}   \chi U+\frac{2\nu_E}{H_2^2}+\frac{2\kappa}{H_2^2}\right)B_{yy}\\
   &-\left(\textrm{i}  \chi^3U+\textrm{i}  \chi U_{yy}+\frac{2\nu_E}{H_2^2}\chi^2-\textrm{i}  \chi\beta
    +\frac{2\kappa}{H_2^2}\chi^2+\frac{2\nu_N}{H_2^2}+\frac{2}{H_2^2}\textrm{i}  \chi U\right)B\\
    &+\left(\textrm{i}  \chi m-\frac{2\nu_E}{H_2^2}\right)A_{yy}
    -\left(\textrm{i}  \chi^3m+\textrm{i}  \chi m_{yy}-\frac{2\nu_E}{H_2^2}\chi^2
     -\frac{2}{H_2^2}\textrm{i}  \chi m\right)A
    =0,
  \end{aligned}\\
  \label{dotU}
  &\dot{U}+\frac{2\nu_E}{H_2^2}(U-m)+2\chi\,\mathrm{Im}(AA^*_{yy}+BB^*_{yy})=0,\\
  \label{dotm}
  &\begin{aligned}
    \dot{m}_{yy} - \frac{2}{H_2^2}\dot{m} &+\frac{2\kappa}{H_2^2} m_{yy}
    -\frac{2\nu_E}{H_2^2}(U-m)_{yy}
    -\frac{2\nu_N}{H_2^2}(m-m^*)\\
    &+\frac{4}{H_2^2}\chi\,\mathrm{Im}(A^*B)_{yy}
    +2\chi\,\mathrm{Im}(AB^*_y+BA^*_y)_{yyy}
    =0,
  \end{aligned}
\end{align}
where \hbox{\eqref{dotA}-\eqref{dotB}} and
\hbox{\eqref{dotU}-\eqref{dotm}} refer respectively to the
non-zonal and zonal components, the dot indicates time
differentiation, and $X^*$ denotes the complex conjugate of $X$.
This is a set of $6$ equations for the real fields $A^1$, $A^2$,
$B^1$, $B^2$, $U$, $m$, where $A^1$ and $A^2$ are the real and
imaginary parts of $A$ and similarly for $B$. Rigid walls are
taken as boundaries at $y=0,L_y$, so that all fields have
vanishing boundary conditions. We emphasize that, by construction,
no wave-wave interactions occur
in~\hbox{\eqref{dotA}-\eqref{dotm}}. Moreover, only quadratic
nonlinear terms are present, due to the fact that the same holds
for~\hbox{\eqref{tau}-\eqref{phi}}. A Fourier half-sine expansion
of the fields is carried out, with time-varying coefficients:
\begin{align}
  \label{weight1}
  &A^i=\sum_{j=1}^{JT}A^i_j\sin\left(\frac{\pi j y}{L_y}\right), \hspace{5pt}i=1,2,\\
  \label{weight2}
  &B^i=\sum_{j=1}^{JT}B^i_j\sin\left(\frac{\pi j y}{L_y}\right), \hspace{5pt}i=1,2,\\
  \label{weight3}
  &U=\sum_{j=1}^{JT}U_j\sin\left(\frac{\pi j y}{L_y}\right),\\
  \label{weight4}
  &m=\sum_{j=1}^{JT}m_j\sin\left(\frac{\pi j y}{L_y}\right),
\end{align}
truncating at order $JT$. Therefore, the lower integration limit
in~\hbox{\eqref{phifourier2}-\eqref{taufourier2}} is such that the
two fields $\tau$ and $\phi$ as reconstructed
from~\hbox{\eqref{weight1}-\eqref{weight4}} have automatically
zero mean when latitudinally integrated. Such a choice allows for
the fact that the energy density~\eqref{energydensity}, and
consequently the total energy of the system, does not depend on
the latitudinally averaged value of $\tau$, which has no physical
relevance. We denote by $\Pi_j\left(\cdot\right)$ the projection
operator onto the basis function $\sin(\frac{\pi j y}{L_y})$.
Because of computational speed, we chose a collocation (also known
as pseudospectral) projection, see \appref{numerical} for a
details. By linearity of $\Pi_j\left(\cdot\right)$, its action on
linear terms in~\hbox{\eqref{dotA}-\eqref{dotm}} is obvious. For
example, terms like $A^1_{yy}$ are represented as
\begin{equation}%%noref
  A^1_{yy}=-\sum_{j=1}^{JT}w_j^2A^1_j\sin\left(\frac{\pi j y}{L_y}\right),
  \qquad\text{where}\ \
  w_j=\frac{\pi j}{L_y}.
\end{equation}%%noref
So by plugging expansion~\hbox{\eqref{weight1}-\eqref{weight4}}
into the equations~\hbox{\eqref{dotA}-\eqref{dotm}}, and by
applying $\Pi_j\left(\cdot\right)$, we eventually obtain a set of
$6\times JT$ ordinary differential equations in the coefficients
$A^1_j$, $A^2_j$, $B^1_j$, $B^2_j$, $U_j$, $m_j$, with
$j=1,\dots,JT$:

\begin{align}
  \label{dotA1j}
  &\begin{aligned}
    \dot{A}^1_j= \frac{1}{\chi^2+w_j^2}
    &\left[
      -\frac{2\nu_E}{H_2^2}(\chi^2+w_j^2)A^1_j-\chi\beta A^2_j
      +\frac{2\nu_E}{H_2^2}(\chi^2+w_j^2)B^1_j
      +\right.\\
    &\left.
      \Pi_j\left(
        -\chi U A^2_{yy} + \chi^3UA^2 + \chi U_{yy}A^2
        -\chi m B^2_{yy} + \chi^3mB^2 + \chi m_{yy}B^2
      \right)
    \right],
  \end{aligned}\\
  \label{dotA2j}
  &\begin{aligned}
    \dot{A}^2_j= \frac{1}{\chi^2+w_j^2}
    &\left[
      -\frac{2\nu_E}{H_2^2}(\chi^2+w_j^2)A^2_j+\chi\beta A^1_j
      +\frac{2\nu_E}{H_2^2}(\chi^2+w_j^2)B^2_j
      +\right.\\
    &\left.
      \Pi_j\left(
        \chi U A^1_{yy} - \chi^3UA^1 - \chi U_{yy}A^1 +
         \chi m B^1_{yy} - \chi^3mB^1 - \chi m_{yy}B^1
      \right)
    \right],
  \end{aligned}\\
  \label{dotB1j}
  &\begin{aligned}
    \dot{B}^1_j= &\frac{1}{\chi^2+w_j^2+\frac{2}{H_2^2}}
    \left[
      -\left(\frac{2\nu_E}{H_2^2}+\frac{2\kappa}{H_2^2}\right)(\chi^2+w_j^2)B^1_j-\chi\beta B^2_j
      +\frac{2\nu_E}{H_2^2}(\chi^2+w_j^2)A^1_j
      +\right.\\
    &\left.\hspace*{-60pt}
      \Pi_j\left(
        -\chi U B^2_{yy} + \chi^3UB^2 + \chi U_{yy}B^2 + \frac{2}{H_2^2}\chi UB^2
        -\chi m A^2_{yy} + \chi^3mA^2 + \chi m_{yy}A^2 - \frac{2}{H_2^2}\chi mA^2
      \right)
    \right],
  \end{aligned}\\
  \label{dotB2j}
  &\begin{aligned}
    \dot{B}^2_j=& \frac{1}{\chi^2+w_j^2+\frac{2}{H_2^2}}
    \left[
      -\left(\frac{2\nu_E}{H_2^2}+\frac{2\kappa}{H_2^2}\right)(\chi^2+w_j^2)B^2_j+\chi\beta B^1_j
      +\frac{2\nu_E}{H_2^2}(\chi^2+w_j^2)A^2_j
      +\right.\\
    &\left.\hspace*{-60pt}
      \Pi_j\left(
        \chi U B^1_{yy} - \chi^3UB^1 - \chi U_{yy}B^1 - \frac{2}{H_2^2}\chi UB^1 +
        \chi m A^1_{yy} - \chi^3mA^1 - \chi m_{yy}A^1 + \frac{2}{H_2^2}\chi mA^1
      \right)
    \right],
  \end{aligned}\\
  \label{dotUj}
  &\begin{aligned}
    \dot{U}_j=
    -\frac{2\nu_E}{H_2^2}(U-m)-2\chi\Pi_j\left(
      -A^1A^2_{yy} +A^2A^1_{yy} -B^1B^2_{yy} +B^2B^1_{yy}\right),
  \end{aligned}\\
  \label{dotmj}
  &\begin{aligned}
    \dot{m}_j= &\frac{1}{\frac{2}{H_2^2}+w_j^2}
    \left[-w_j^2\frac{2\kappa}{H_2^2} m_j + w_j^2\frac{2\nu_E}{H_2^2}(U_j-m_j)
      -\frac{2\nu_N}{H_2^2}(m_j-m^*_j)\right.+\\
    &\left.\Pi_j\left(
        4\chi \frac{1}{H_2^2}(A^1B^2 - A^2B^1)_{yy}
        +2\chi(-A^1B^2_{yy} +A^2B^1_{yy} -B^1A^2_{yy} -B^2A^1_{yy})_{yy}
      \right)
    \right].
  \end{aligned}
\end{align}
System~\hbox{\eqref{dotA1j}-\eqref{dotmj}} constitutes the base
model of our study. For the truncation order $JT$ we have used the
values: $JT=8, 16, 32, 64$.

\newpage
\section{Dynamical and statistical characterization of the model's attractor}
\label{sec:mainresults}
\subsection{Hadley Equilibrium}

The system of equations~\hbox{\eqref{tau}-\eqref{phi}} has the
following stationary solution for zonally symmetric flows:
\begin{align}
  &\label{phiHAD}
  \phi\left(y\right)=\tau\left(y\right),\\
  &\label{hadTAU}
  \frac{2\kappa}{H_2^2}\frac{d^2\tau\left(y\right)}{dt^2}+\frac{2\nu_N}{H_2^2}\left(\tau\left(y\right)-\tau^\star\left(y\right)\right)=0.
\end{align}
Considering the functional form~\eqref{deftau} for
$\tau^\star\left(y\right)$, the following expression for
$\tau\left(y\right)$ holds:
\begin{equation}
  \label{elim1}
  \tau\left(y\right)=\phi\left(y\right)=\frac{R}{f_0}\frac{T_E}{4}\frac{\cos\left(\frac{\pi
        y}{L_y}\right)}{1+\frac{\kappa}{\nu_N}\left(\frac{\pi}{L_y}\right)^2}=\frac{\tau^\star\left(y\right)}{1+\frac{\kappa}{\nu_N}\left(\frac{\pi}{L_y}\right)^2}.
\end{equation}
When expressing this solution in terms of the average zonal wind
$U\left(y\right)$ and of half of the wind shear $m\left(y\right)$,
both defined in~\eqref{defUm}, we have:
\begin{equation}
  \label{mTAU}
  m\left(y\right)=U\left(y\right)=\frac{R}{f_0}\frac{\pi}{L_y}\frac{T_E}{4}\frac{\sin\left(\frac{\pi
        y}{L_y}\right)}{1+\frac{\kappa}{\nu_N}\left(\frac{\pi}{L_y}\right)^2}=\frac{m^\star\left(y\right)}{1+\frac{\kappa}{\nu_N}\left(\frac{\pi}{L_y}\right)^2},
\end{equation}
where we have used the definition~\eqref{defmstar} for
$m^\star\left(y\right)$. Moreover, since the temperature $T_2$ is
proportional to $\tau$ (compare~\eqref{tempe2} and~\eqref{tau1}),
the following temperature profile $T_2\left(y\right)$ is realized:
\begin{equation}
  \label{mT2}
  T_2\left(y\right)=\frac{T_E}{2}\frac{\cos\left(\frac{\pi
        y}{L_y}\right)}{1+\frac{\kappa}{\nu_N}\left(\frac{\pi}{L_y}\right)^2}=\frac{T^\star\left(y\right)}{1+\frac{\kappa}{\nu_N}\left(\frac{\pi}{L_y}\right)^2}.
\end{equation}

This solution describes a zonally symmetric circulation
characterized by the instantaneous balance between the horizontal
temperature gradient and the vertical wind shear, which
corresponds on the Earth system to the idealized pattern of the
Hadley equilibrium~\cite{HH,Ped,Holton}. In particular, since
$m\left(y\right)=U\left(y\right)$, we have that
$u_1\left(y\right)=2U\left(y\right)$ and $u_3\left(y\right)=0$,
\textit{i.e.} all the dynamics takes place in the upper pressure
level. Since the the lower pressure level experiences no motion,
the Ekman sucking process is switched off and, consistently, the
solution does not depend on the corresponding coupling constant
$\nu_E$.

There is a value of the equator-to-pole temperature gradient
$T_E^{H}$ such that if $T_E<T_E^{H}$ the Hadley
equilibrium~\hbox{\eqref{phiHAD}-\eqref{mTAU}} is stable and has
an infinite basin of attraction, whereas if $T_E>T_E^{H}$ it is
unstable.

In the stable regime with $T_E<T_E^{H}$, after the decay of
transients, the fields $\phi$, $\tau$, $m$, $U$, and $T$ are
time-independent and feature zonal symmetry - they only depend on
the variable $y$. Moreover, they are proportional by the same
near-to-unity factor to the corresponding relaxation profiles,
compare~\hbox{\eqref{phiHAD}-\eqref{mT2}}. In particular, this
implies that all the equilibrium fields are proportional to the
parameter $T_E$.

In our model, since the forcing $m^\star\left(y\right)$ only
projects onto the first latitudinal Fourier mode (see
\eqref{defmstar}), the Hadley equilibrium is fully described as
follows:
\begin{align}%%noref
  &A^i_j\left(t\right)=B^i_j\left(t\right)=0,\quad i=1,2,
  \quad j=1,\ldots,JT,\\
  &m_j\left(t\right)=U_j\left(t\right)=0,\quad j=2,\ldots,JT,\\
  &m_1\left(y\right)=U_1\left(y\right)=\frac{m^\star\left(y\right)}{1+\frac{\kappa}{\nu_N}\left(\frac{\pi}{L_y}\right)^2}.
\end{align}%%noref

When increasing the values of the control parameter $T_E$ beyond
$T_E^{H}$, the equilibrium described
by~\hbox{\eqref{phiHAD}-\eqref{hadTAU}} becomes unstable. The
physical reason for this, as first pointed out by Charney and Eady
on vertically continuous models~\cite{Char47,Eady49} and by
Phillips on the two two-level model~\cite{Phil54}, is that for
high values of the meridional temperature gradient the Hadley
equilibrium is unstable with respect to the process of baroclinic
conversion, which allows the transfer of available potential
energy of the zonal flow stored into the meridional temperature
gradient into energy of the eddies, essentially transferring
energy from the latter term to the first two terms of the energy
density expression~\eqref{energydensity}. The two-level model, as
first pointed out by Phillips~\cite{Phil54}, is the minimal model
allowing for the representation of this process.

At mathematical level, in our model we have that for $T_E=T_E^H$ a
complex conjugate pair of eigenvalues of the linearization
of~\hbox{\eqref{dotA1j}-\eqref{dotmj}} cross the imaginary axis so
that their real part turns positive, which suggests the occurrence
of a Hopf bifurcation~\cite{Kuz}.

The observed value of $T_E^{H}$ changes with the considered
truncation order $JT$. Results are reported in \tabref{hopf1} for
the choice of constants reported in \tabref{tab2} and for $JT=$
$8$, $16$, $32$, $64$. We have that $T_E^{H}$ increases with the
value of $JT$. The reason for this is that the finer is the
resolution, the more efficient are the stabilizing mechanisms
which counteract the baroclinic instability. Such mechanisms are
the barotropic stabilization of the jet, increasing the horizontal
shear through the convergence of zonal momentum
\cite{Kuo73,SIMM,RAND,JAMES}, and the viscous dissipation, which
both act preferentially on the small scales since they involve the
spatial derivatives of the fields $\phi$ and $\tau$. This is a
clarifying example that in principle it is necessary to include
suitable renormalizations in the parameters of a model when
changing the resolution $JT$, in order to keep correspondence with
the resulting dynamics~\cite{Lor80}. Nevertheless, in our case the
values of $T_E^{H}$ obtained for the adopted resolutions are
rather similar.

Moreover, in our model the number of linearly unstable modes of
the Hadley equilibrium~\hbox{\eqref{phiHAD}-\eqref{mTAU}}
increases with the value of $T_E$. As shown in \figref{linearly},
at each \textit{jump} in the graphs an additional pair of complex
conjugate eigenvalues crosses the imaginary axis. The increase of
the number of linearly unstable modes of the Hadley equilibrium
can be framed at physical level in the fact that for larger values
of $T_E$ a larger pool of available potential energy is available
for conversion and faster latitudinally varying modes can become
unstable, similarly to case of the Phillips model~\cite{Phil54}.
We remind that the system under investigation obeys a Squires
condition~\cite{DraReid81}, so that the fastest growing among the
unstable modes is the latitudinally gravest one. The signature of
the relevance of the stabilizing mechanisms and of the geometrical
properties of the linearly unstable modes developing for higher
values of $JT$ can be confirmed by observing respectively that
while for low values of $T_E$ ($T_E\lesssim12$) the number of
linearly unstable modes decreases with $JT$, the converse is true
for high values of $T_E$ ($T_E\gtrsim25$).

\subsection{Transition to Chaos}
\label{sec:transition} In this section we analyze the route to the
formation of a strange attractor of
model~\hbox{\eqref{dotA1j}-\eqref{dotmj}} as the parameter $T_E$
is increased. Throughout the section, $JT$ is fixed at $32$, but
the results are similar for the other considered values of $JT$.

\begin{figure}[p]
  \includegraphics[angle=270,width=0.49\textwidth]{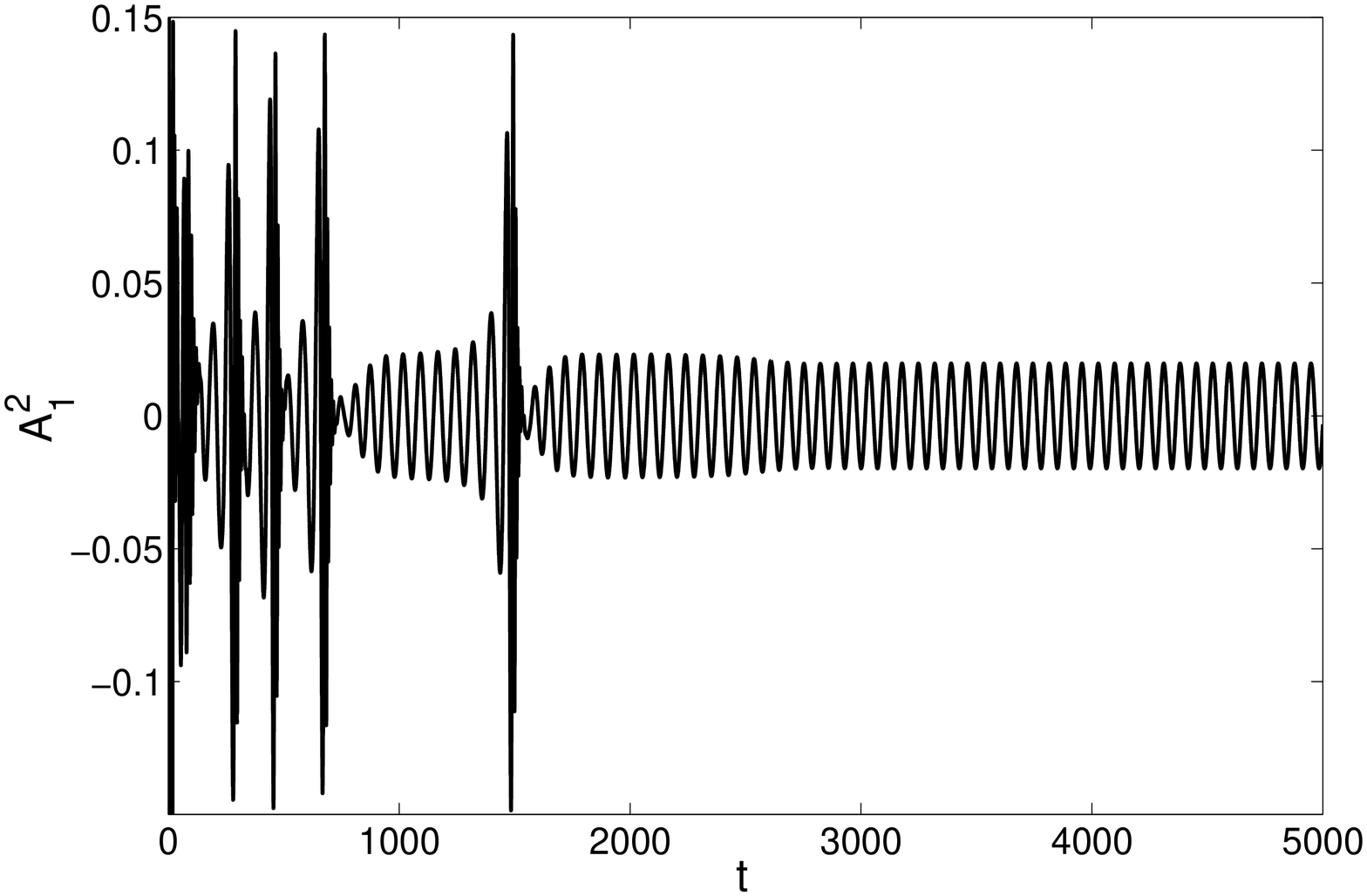}
  \includegraphics[angle=270,width=0.49\textwidth]{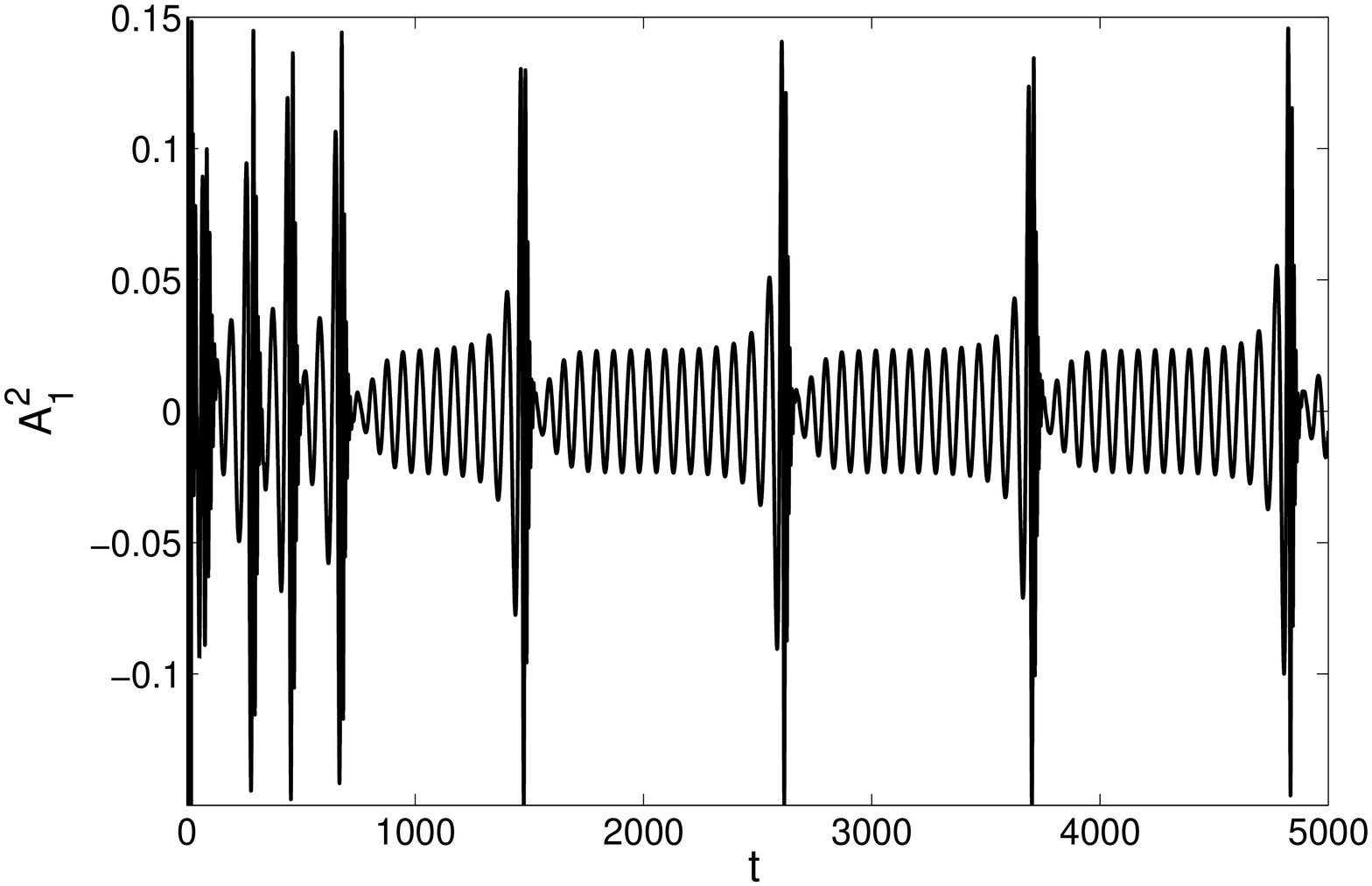}
 \caption{
   Time-evolution of the component $A^2_1$ of
   system~\hbox{\eqref{dotA1j}-\eqref{dotmj}},
   starting from the initial condition mentioned
   in \appref{numerical}.
   Left: $T_E=8.484631$. Right: $T_E=8.484632$. No transient has been discarded.
  }
  \label{fig:interm}
\end{figure}

\begin{figure}[p]
 \includegraphics[angle=270,width=0.49\textwidth]{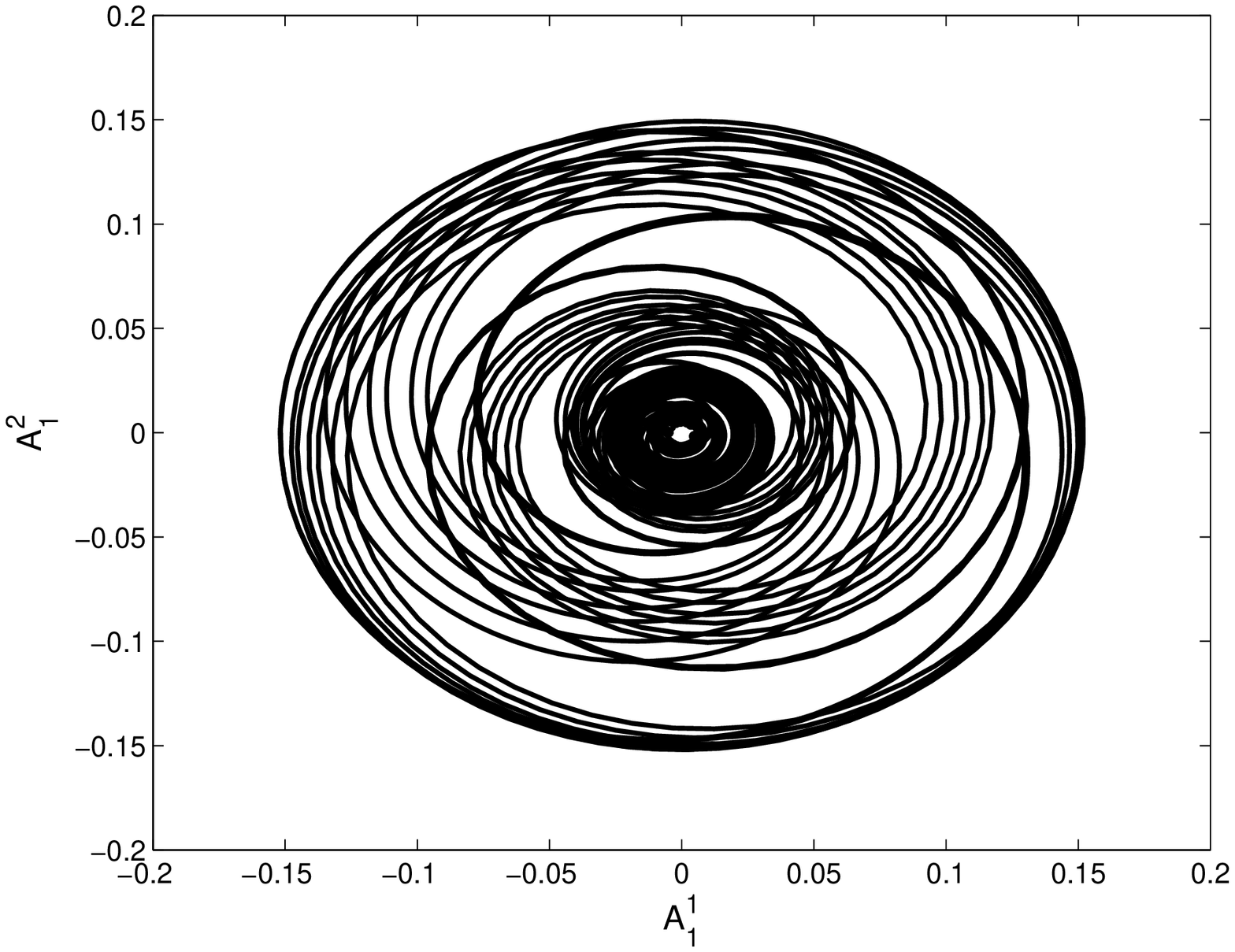}
 \includegraphics[angle=270,width=0.49\textwidth]{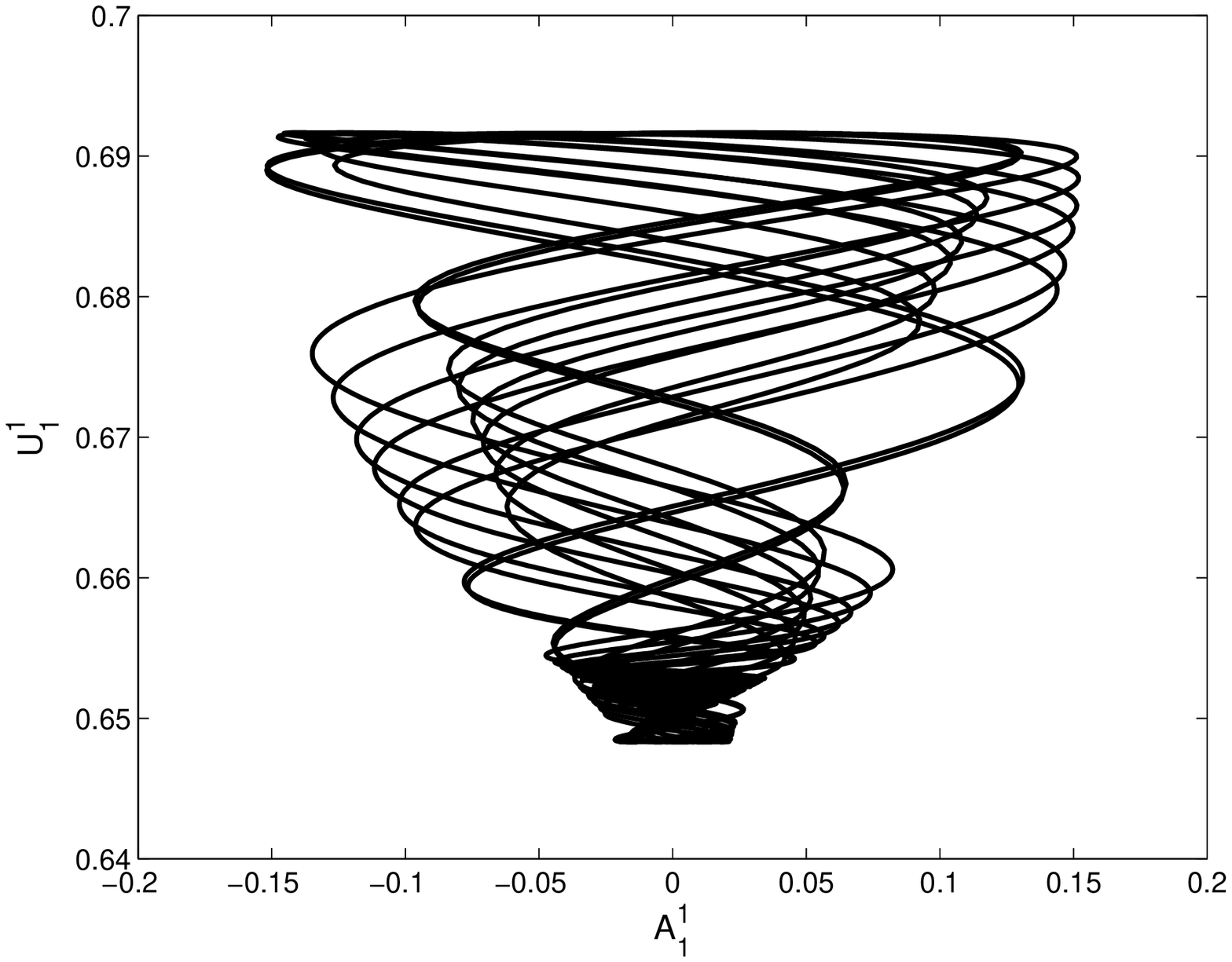}
  \caption{
    Left: projection on $(A^1_1,A^2_1)$ of an orbit on the
    attracting two-torus of~\hbox{\eqref{dotA1j}-\eqref{dotmj}}
    for $T_E=8.484632$.
    Right: same as Left, projection on $(A^1_1,U^1_1)$.
    Units as indicated in \tabref{tab1}.
    A five-year transient has been discarded.
    The phase-space region where the orbit accumulates
    more densely is due to intermittency of saddle-node type
    near the location of the periodic orbit occurring
    for $T_E=8.484631$, compare \figref{interm}
    and see text for details.
  }
  \label{fig:2torus}
\end{figure}

A stable periodic orbit (\figref{interm}~(A)) branches off from
the Hadley equilibrium~\hbox{\eqref{phiHAD}-\eqref{mTAU}} as $T_E$
increases above $T_E^{H}\sim 8.275$. We recall that the Hadley
equilibrium loses stability at $T_E^H$ as a pair of complex
conjugate eigenvalues of the linearization
of~\hbox{\eqref{dotA1j}-\eqref{dotmj}} crosses the imaginary axis,
(see previous section). This strongly suggests the occurrence of a
supercritical Hopf bifurcation, which might be checked by center
manifold reduction and normal form analysis (see
\emph{e.g.}~\cite{Kuz}), but it is beyond the scope of the present
work. The attracting periodic orbit persists for $T_E$ in a narrow
interval, up to approximately $T_E=8.485$, where it disappears
through a saddle-node bifurcation taking place on an attracting
invariant two-torus, see \figref{2torus}. Intermittency of
saddle-node type~\cite{PM} on the two-torus is illustrated in
\figref{interm}~(B): after an initial transient the orbit is
attracted to a quasi-periodic evolution characterized by long
time-spans, resembling the periodic evolution of
\figref{interm}~(A), alternated by relatively short bursts in
which the orbit \textit{explores} the rest of the two-torus. In
other words, for $T_E$ right after the saddle-node bifurcation,
the orbit on the two-torus \textit{slows down} in the phase space
region where the saddle-node has taken place. This yields a higher
density of points in that region, see \figref{2torus}.

\begin{figure}[p]
  \includegraphics[angle=270,width=0.32\textwidth]{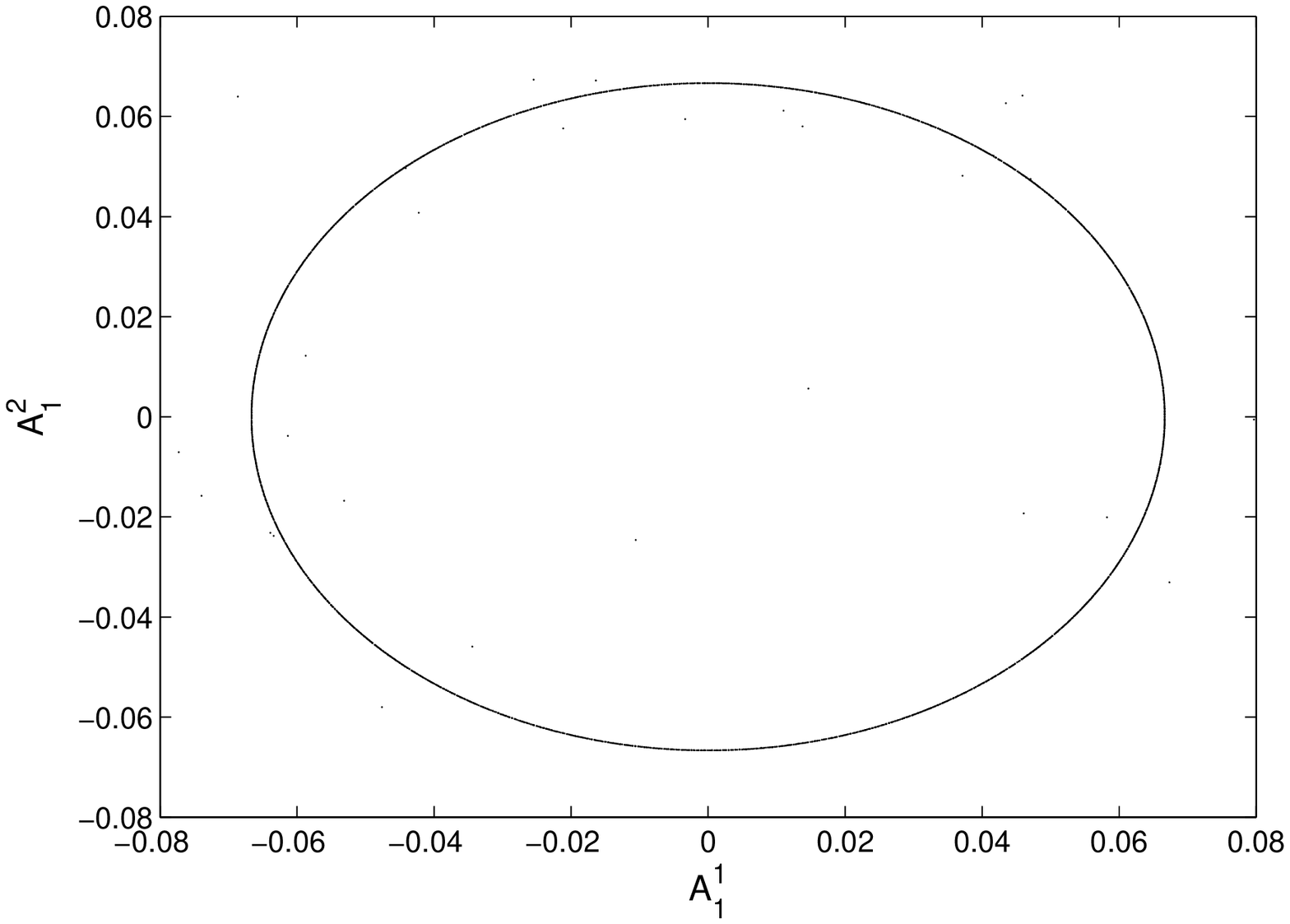}
  \includegraphics[angle=270,width=0.32\textwidth]{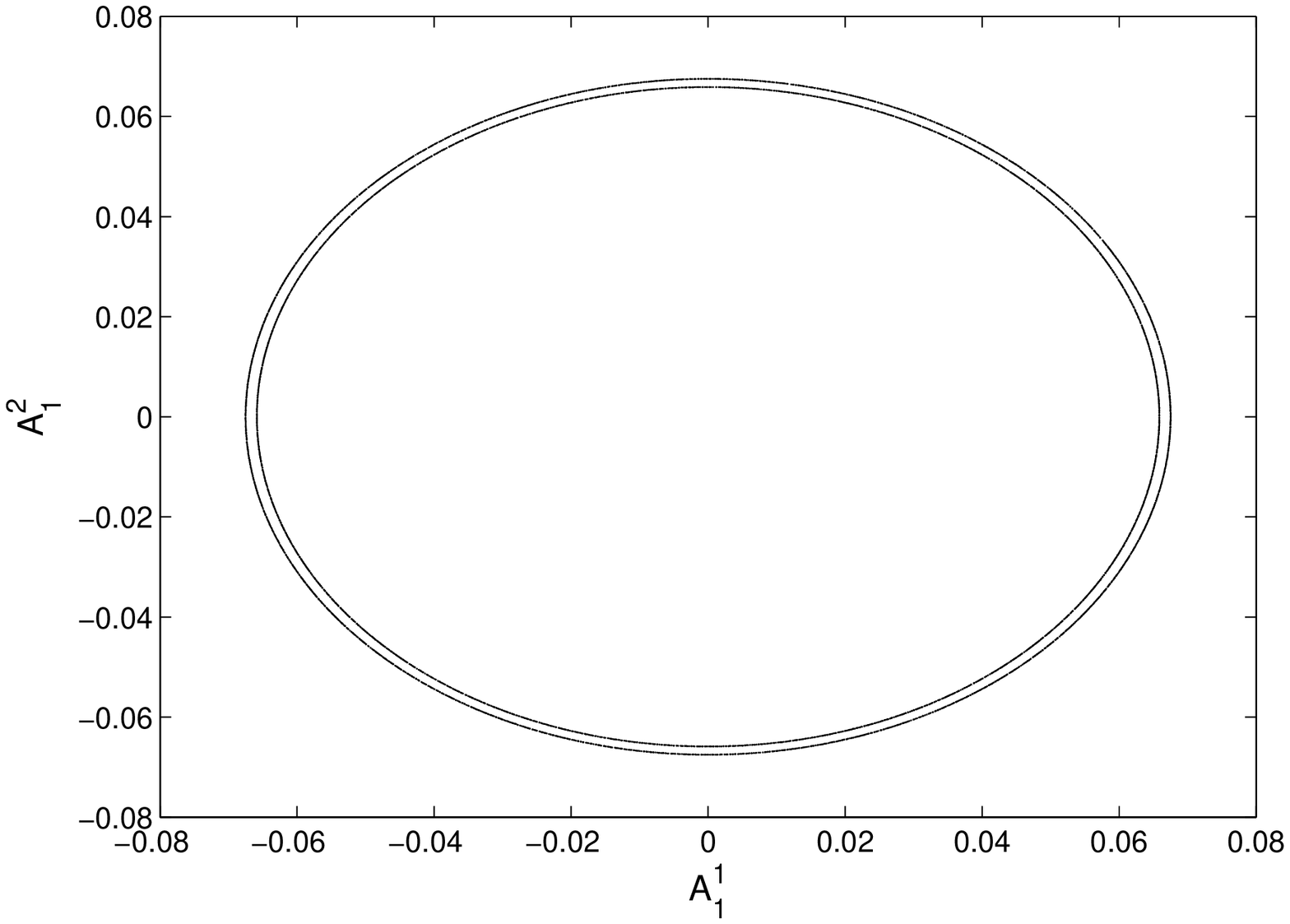}
  \includegraphics[angle=270,width=0.32\textwidth]{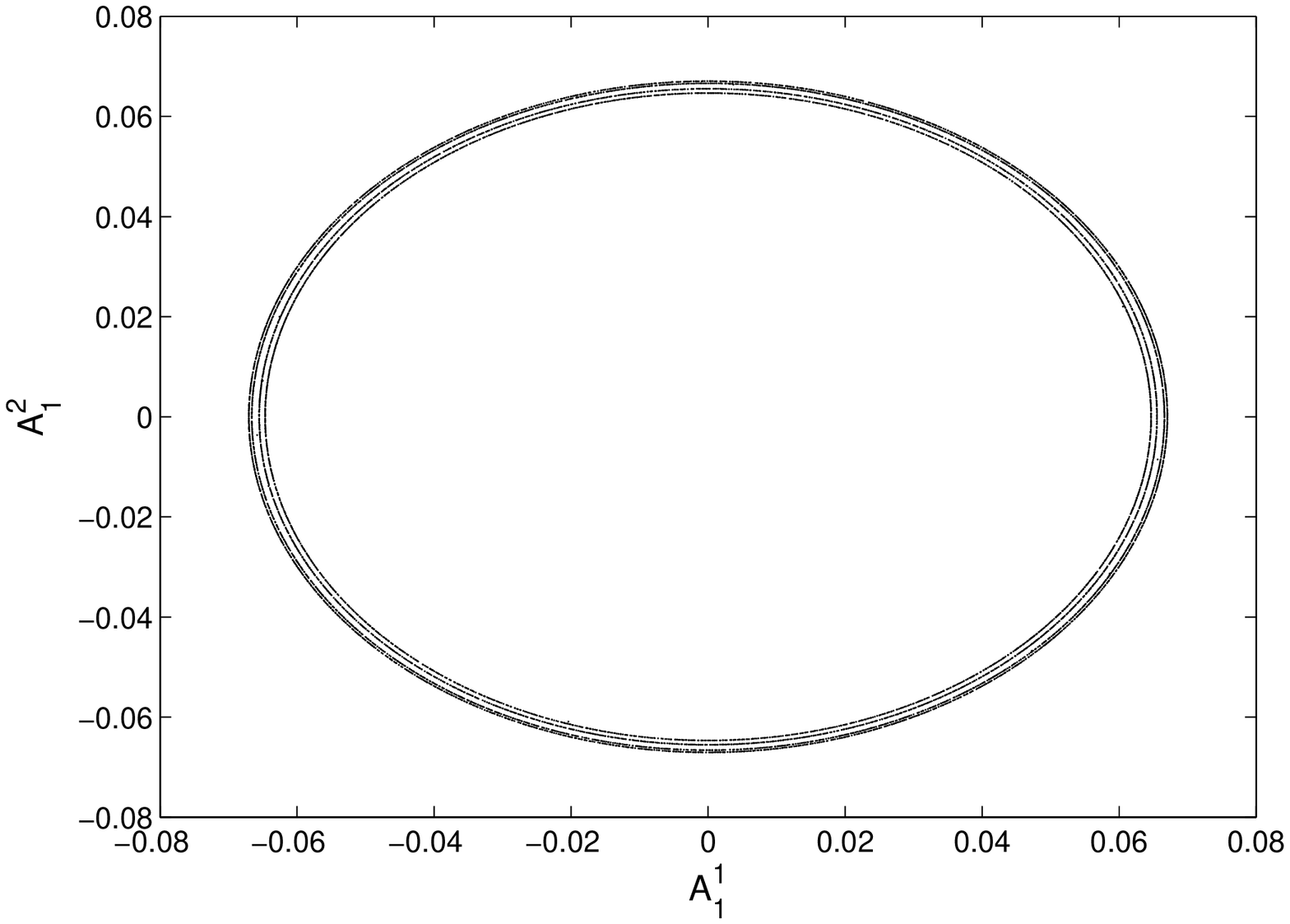}\\
  \includegraphics[angle=270,width=0.32\textwidth]{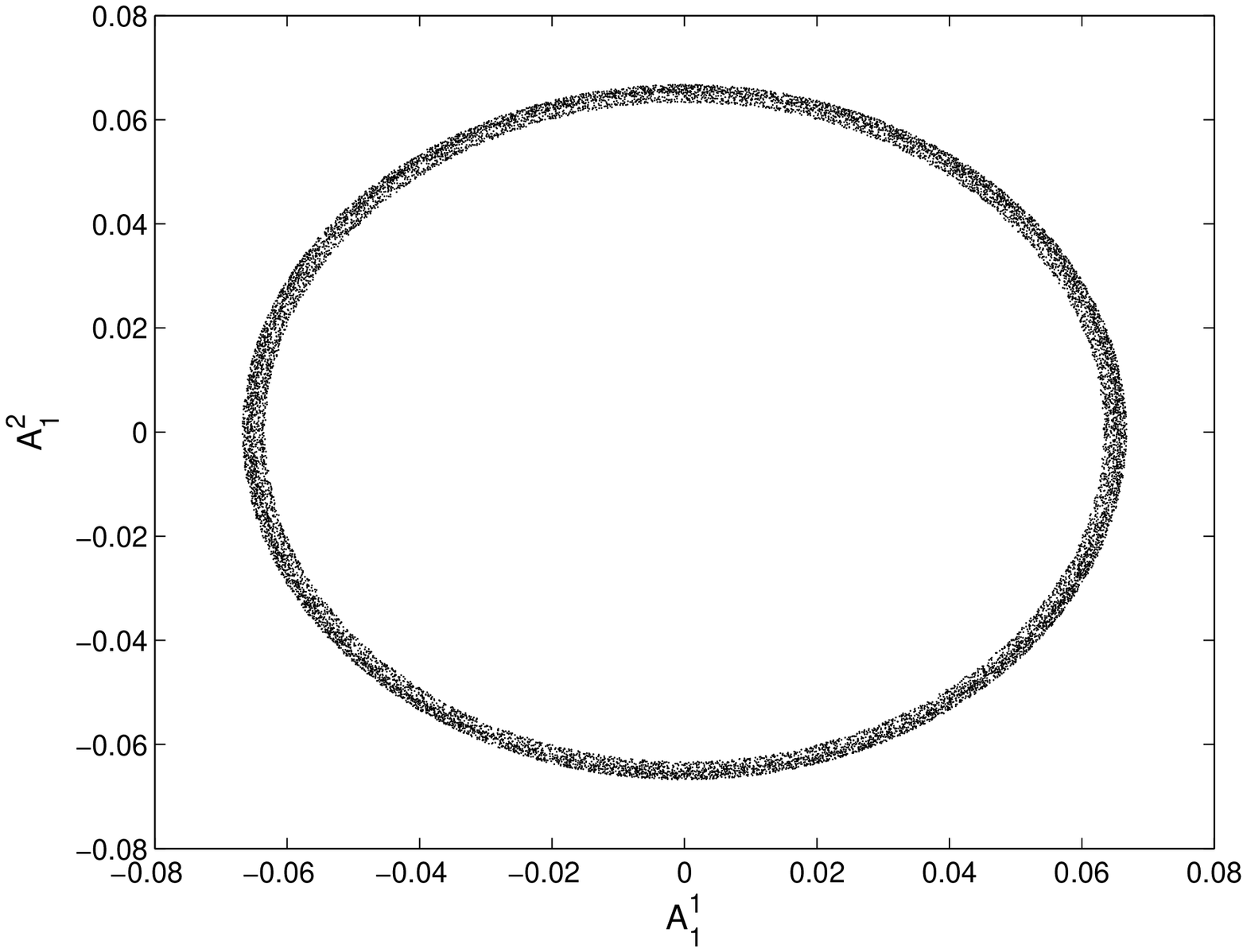}
  \includegraphics[angle=270,width=0.32\textwidth]{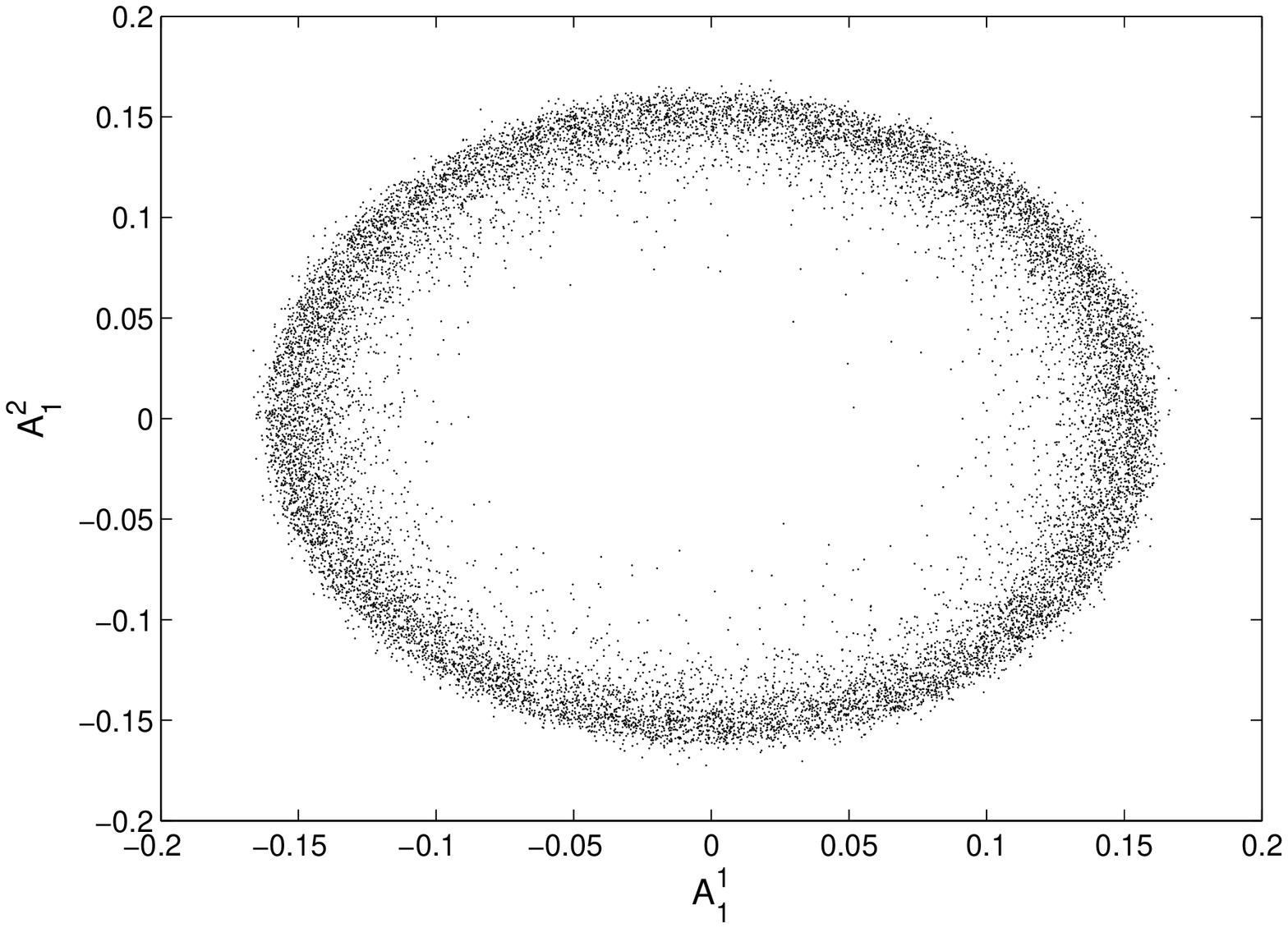}
  \includegraphics[angle=270,width=0.32\textwidth]{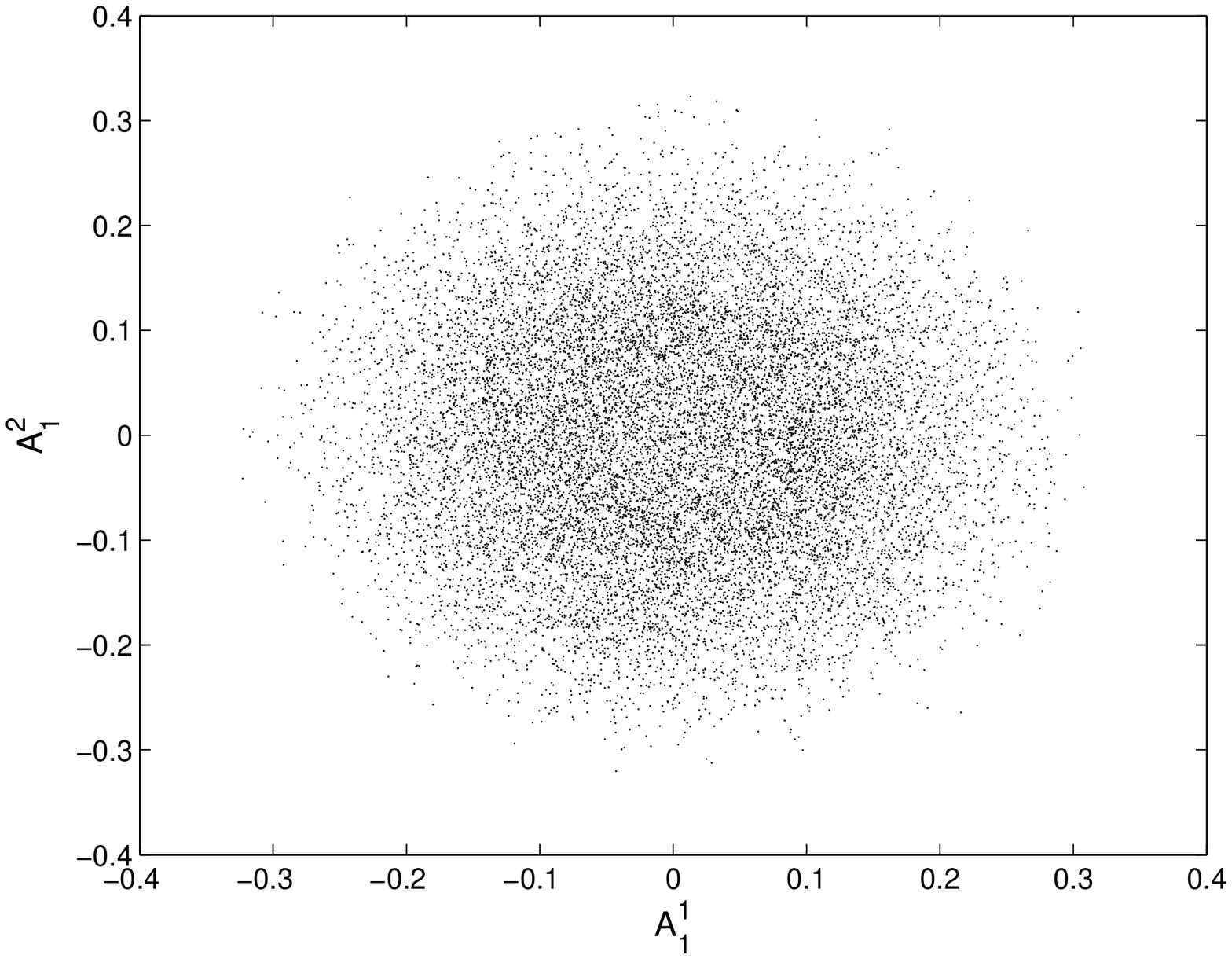}
  \caption{
    Projections on $(A^1_1,A^2_1)$ of a Poincar\'e section
    of the attractor of~\hbox{\eqref{dotA1j}-\eqref{dotmj}},
    obtained by intersecting it with a hyperplane $U_1=c_0$
    for several values of $T_E$.
    From (A)~to~(F) $T_E$ is, respectively, $8.516$, $8.52$,
    $8.521$, $8.522$, $8.58$, $10$. The value $c_0$ of the section
    is fixed at $0.66$ (A)~to~(D) and is 0.7 and 0.8 for
    (E)~and~(F) respectively. Also notice the different axis scale
    for the last two plots.
  }
  \label{fig:poincare}
\end{figure}

For slightly larger values of $T_E$, a strange attractor develops
by so-called quasi-periodic breakdown of a \textit{doubled} torus.
This is one of the most typical routes for onset of chaos (weak
turbulence) in fluid dynamics experiments and low-dimensional
models, compare~\cite{BS,BSV1,FHW,RFMR05} and references therein.
We describe this route by means of a Poincar\'e section of the
attractor of~\hbox{\eqref{dotA1j}-\eqref{dotmj}}, obtained by
intersecting an orbit with a hyperplane $U_1=c_0$ for a suitable
constant $c_0$. In this Poincar\'e section, the two-torus yields a
circle ($T_E=8.516$) which is invariant and attracting under the
Poincar\'e (return) map, see \figref{poincare}~(A). At first, at
$T_E=8.52$ the two-torus loses stability through a
\emph{quasi-periodic period doubling} (see~\cite[Sec. 4.3]{BHS}
and references therein for the theory of quasi-periodic
bifurcations). Thereby a period two circle attractor is created,
meaning a pair of disjoint circles mapped onto each other by the
Poincar\'e map (\figref{poincare}~(B)). By further increasing
$T_E$ up to $T_E=8.521$, a second doubling occurs, where a period
four circle attractor is born (\figref{poincare}~(C)). Then for
$T_E=8.522$ approximately a transition to chaotic motion occurs:
the period four circle turns into a strange attractor having a
narrow band-like structure (\figref{poincare}~(D)), which is
likely to be a \emph{quasi-periodic} H\'enon-like strange
attractor, see~\cite{BSV1,BSV2}. We remark that:
\begin{itemize}
\item
  For smaller values of $T_E$, a sort of \textit{doubling bubble} occurs,
  \textit{i.e.} two consecutive doublings
  (at approximately $T_E=8.504$ and $T_E=8.509$)
  resulting in an attractor like in \figref{poincare}~(C),
  followed by two \textit{undoublings} (at $T_E=8.513$ and $T_E=8.516$)
  where the circle attractor in \figref{poincare}~(A) reappears.
  This sort of direct-inverse finite sequence
  is not uncommon in dynamical systems, see \emph{e.g.}~\cite{BST98}.
\item
  Both the breakdown of a quasi-periodic circle attractor
  and the resulting quasi-periodic H\'enon-like strange attractor
  are dynamical phenomena occurring rather frequently
  but which are not completely understood from the theoretical
  viewpoint (see~\cite{BSV1,BSV2}).
\end{itemize}
As $T_E$ further increases, the band widens and blurs
(\figref{poincare}~(E), for $T_E=8.58$) until no significant
structure can be visually detected (\figref{poincare}~(F)) for
$T_E=10$.

\begin{figure}[p]
  \includegraphics[angle=270,width=0.49\textwidth]{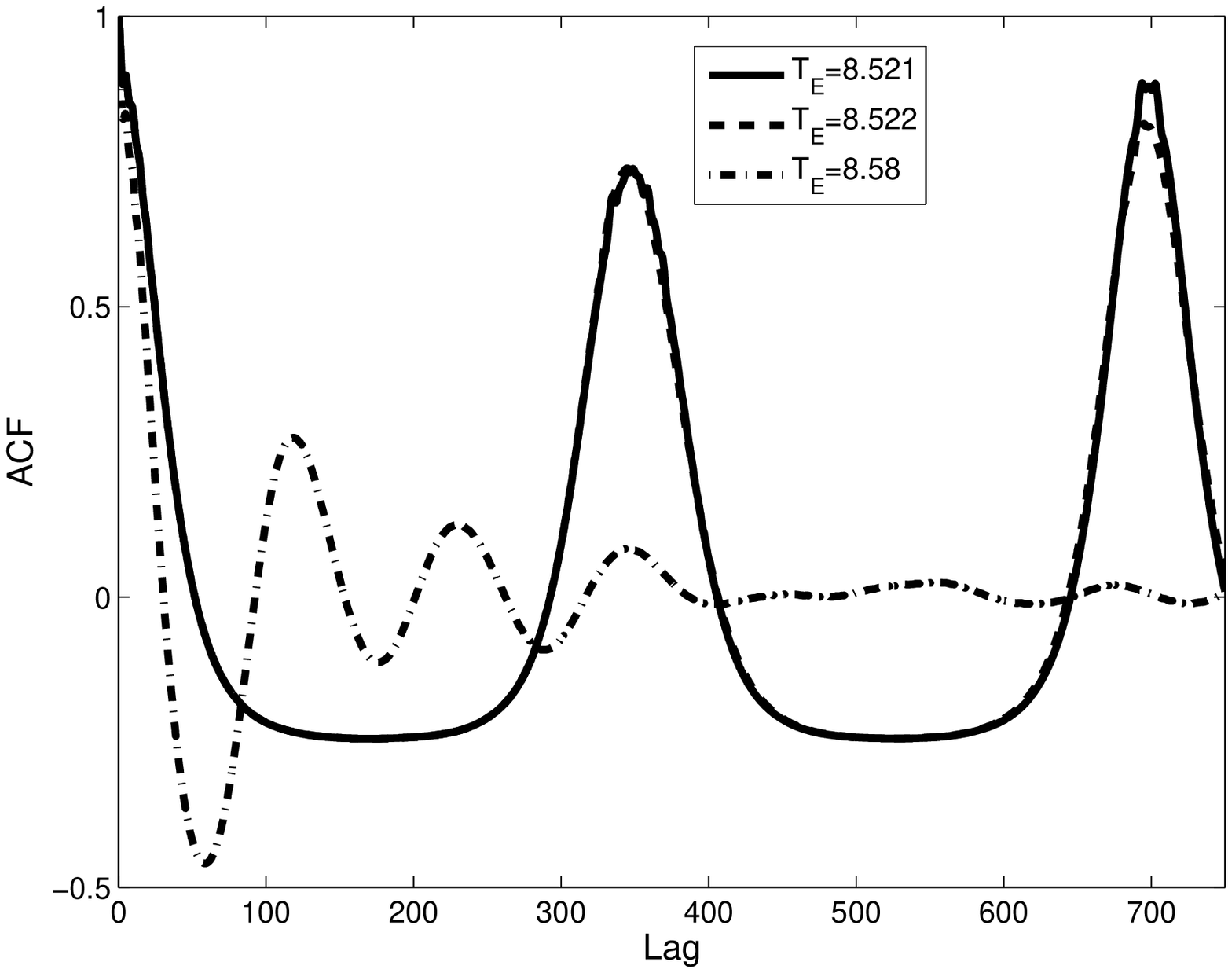}
  \includegraphics[angle=270,width=0.49\textwidth]{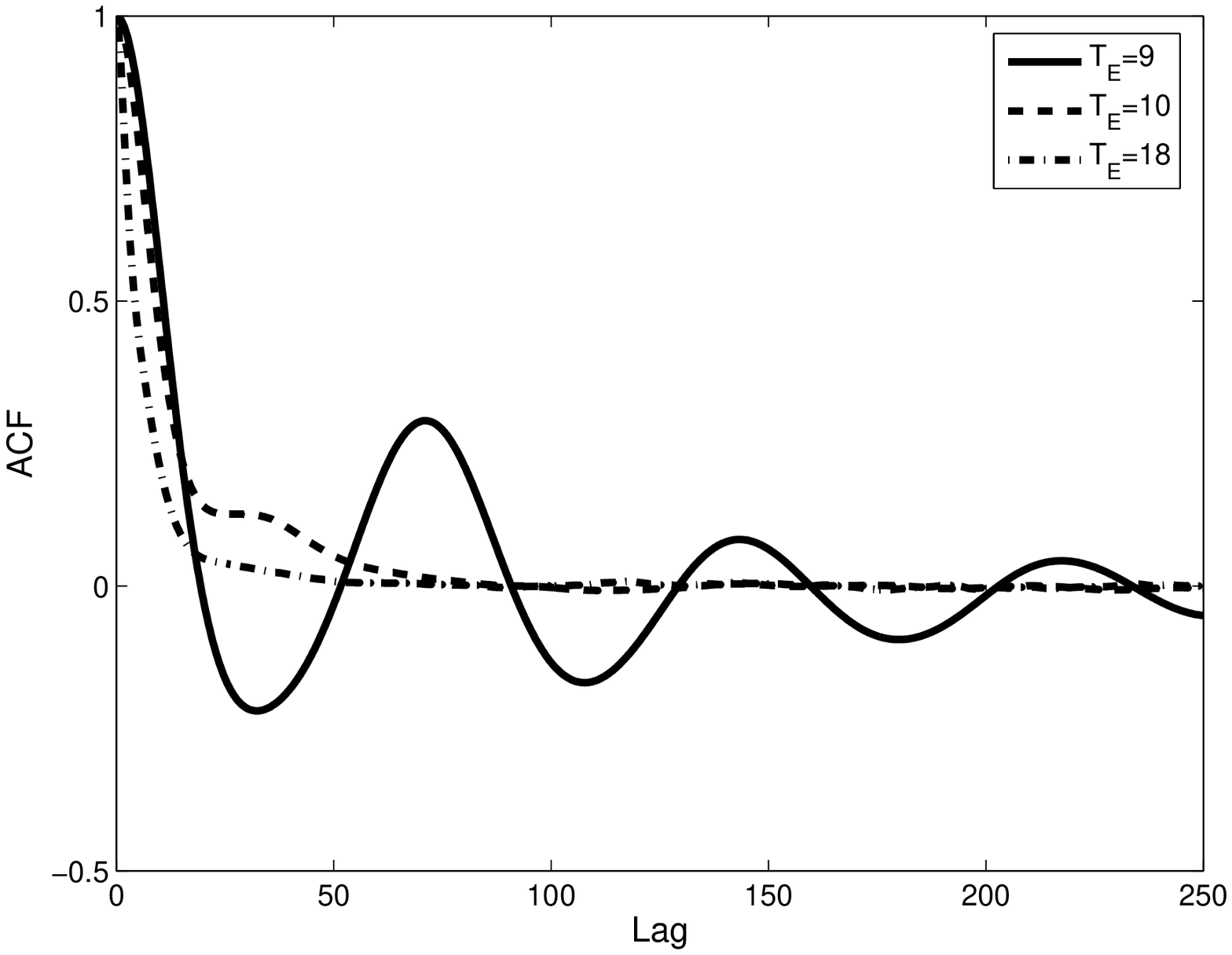}
  \caption{
    Autocorrelations of the total energy time series on the
    attractor of~\hbox{\eqref{dotA1j}-\eqref{dotmj}},
    for various values of $T_E$.
    Left: $T_E=$ $8.521$, $8.522$, $8.58$;
    Right: $T_E=$ $9$, $10$, $18$.
  }
  \label{fig:autocorr}
\end{figure}

The statistical properties of the attractor
of~\hbox{\eqref{dotA1j}-\eqref{dotmj}} also display a typical
evolution. For values of $T_E$ nearby the two-torus breakdown,
quasi-periodic intermittency is observed, \textit{i.e.} the
autocorrelations of an \emph{observable} (a function of state
space variables) typically decay very slowly. We consider the
total energy $E(t)$ of~\hbox{\eqref{dotA1j}-\eqref{dotmj}},
defined as:
\begin{equation}
  \label{eq:energy}
  E(t)=\int_0^{L_y}\int_0^{L_x}e(x,y,t)dxdy=
  6\int_0^L\int_0^{\frac{2\pi}{\chi}}e(x,y,t)dxdy,
\end{equation}
where $e(x,y,t)$ is the energy density in~\eqref{energydensity}
(details on the algorithm used for the computation of time series
of $E(t)$ are given in \appref{numerical}). In
\figref{autocorr}~(A), we display the lagged autocorrelation
$ACF[E(t),Lag]$ of the time series of the total energy for $T_E=$
$8.521$, $8.522$ and $T_E=8.58$. These cases are representative of
the qualitatively distinct observed behaviors. For $T_E=8.522$ the
autocorrelation is very similar to what obtained for $T_E=8.521$,
in spite of the fact that the former value corresponds to chaotic
behavior whereas the latter to regular (quasi-periodic) dynamics.
This occurs because for $T_E=8.522$ the chaoticity is very weak
and the quasi-periodic intermittency rather strong. A much faster
decay of the autocorrelation, albeit still with the signature of
intermittency, is observed for $T_E=8.58$. The quasi-periodic
intermittency for $T_E$ near $T_E^{crit}$ is also illustrated by
the geometrical structure of the attractor, which still bears
resemblance with that of the formerly existing torus. See the
Poincar\'e sections in \figref{poincare}~(D) and~(E).

For larger values of $T_E$ (\figref{autocorr}), we have that at
$T_E=9$ the autocorrelation decays quite similarly to the case
$T_E=8.58$ (the Poincar\'e section, not shown, is also quite
similar), but already at $T_E=10$ the quasi-periodic intermittency
is no longer present (compare \figref{poincare}~(F)).
Correspondingly, the autocorrelation decays rather quickly for
$T_E=10$ and, \textit{a fortiori}, for $T_E=18$. Again compare
with~\cite{BS,BSV1,FHW,RFMR05}.

\begin{figure}[t]
  \includegraphics[angle=270,width=0.9\textwidth]{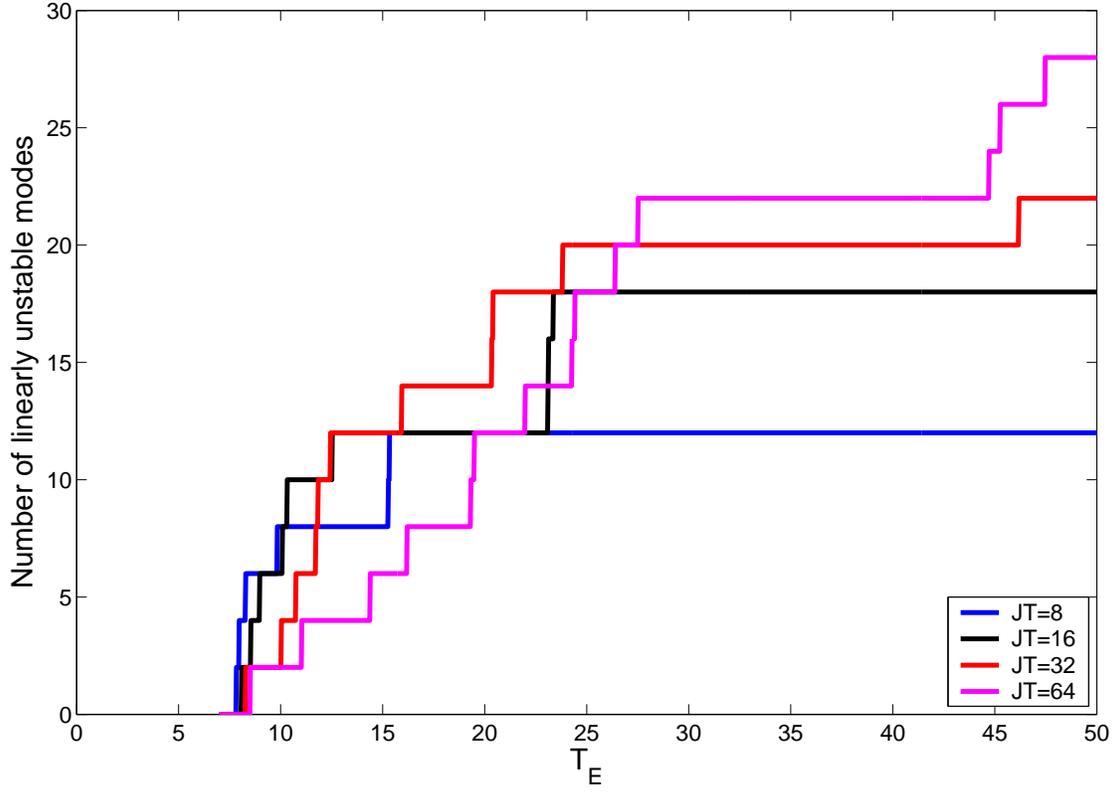}
  \caption{
    Number of linearly unstable modes at the Hadley equilibrium
    as a function of the parameter $T_E$ for $JT=8$, $16$, $32$, $64$.
  }
  \label{fig:linearly}
\end{figure}

Completely analogous routes to chaos occur for
model~\hbox{\eqref{dotA1j}-\eqref{dotmj}} with $JT=16$ and $64$.
However, the locations on the $T_E$-axis of the various
bifurcations are slightly shifted with respect to the case
$JT=32$, compare \figref{linearly} and \tabref{hopf1}. Moreover,
for $JT=8$, a different route takes place, involving a
quasi-periodic Hopf bifurcation of the two-torus (instead of a
quasi-periodic period doubling), whereby an invariant three-torus
is created. In the Poincar\'e section (not shown), this
corresponds to an attracting two-torus.

The invariant objects involved in the transition to low-dynamical
chaos described in this section correspond to well-known fluid
flow patterns. In particular, the two-torus attractor in phase
space yields an \emph{amplitude vacillation} in the flow, whereas
the three-torus detected for $JT=8$ yields a \emph{modulated
amplitude vacillation}, see~\cite{GB83,RFMR05} and references
therein. However, a characterization of the strange attractors
occurring for large $T_E$ and of their relation to turbulence is
still lacking. Typically, low-dimensional nonhyperbolic strange
attractors, such as the Lorenz~\cite{Lor63} and H\'enon-Pomeau
attractors~\cite{HP,Sim79}, are the topological closure of a set
of unstable periodic orbits. Moreover, the H\'enon-Pomeau
attractor coincides with the closure of the unstable manifold
$W^u(p)$ of a fixed point $p$ of saddle type. See
\emph{e.g.}~\cite{BST98,Cao,MV,Viana} and references therein. To
the best knowledge of the authors, no similar properties has yet
been proved (or even formulated) with sufficient generality for
nonhyperbolic strange attractors of larger dimension.

We suspect that plenty of unstable periodic orbits and invariant
tori coexist with the attractor of
model~\hbox{\eqref{dotA1j}-\eqref{dotmj}} with $JT=32$, for
sufficiently large $T_E$. Indeed, from \figref{linearly} we deduce
that the Hadley equilibrium undergoes several other bifurcations
after the first one. Since the number of unstable eigenvalues of
the Hadley equilibrium increases at each Hopf bifurcation, the
periodic orbits that branch off have unstable manifolds of
increasingly high dimension. Moreover, these unstable periodic
orbits in turn undergo Hopf bifurcations (also called torus or
Ne\u\i mark-Sacker~\cite{Kuz}) where unstable two-tori branch off,
compare~\cite[Sec. 5]{SM}. It seems, therefore, that the phase
space quickly gets crowded with high-dimensional unstable
invariant manifolds. The question remains open whether such
complex dynamical characterizations of the system play a role in
the geometrical structure of the strange attractor and are
potentially useful for computing the statistical properties, let
it go for the time average fields considered in the classical
atmospheric circulations theories or the Hadley equilibrium of
most theories of atmospheric instability.

\subsection{Lyapunov Exponents and Dimension of the Strange Attractor}
\label{sec:lyap}

To characterize the dynamical properties of the strange attractors
of~\hbox{\eqref{dotA1j}-\eqref{dotmj}} we resort to the study of
the Lyapunov exponents~\cite{Ose,ER}. See \appref{Lyapunov} for a
description of the algorithm used to compute them. In what
follows, the Lyapunov exponents are denoted by $\lambda_1,
\lambda_2, \dots, \lambda_N$, with
$\lambda_1\ge\lambda_2\ge\dots\ge\lambda_N$, $N=6\times JT$.

\begin{figure}[htbp]
 \includegraphics[angle=270,width=0.49\textwidth]{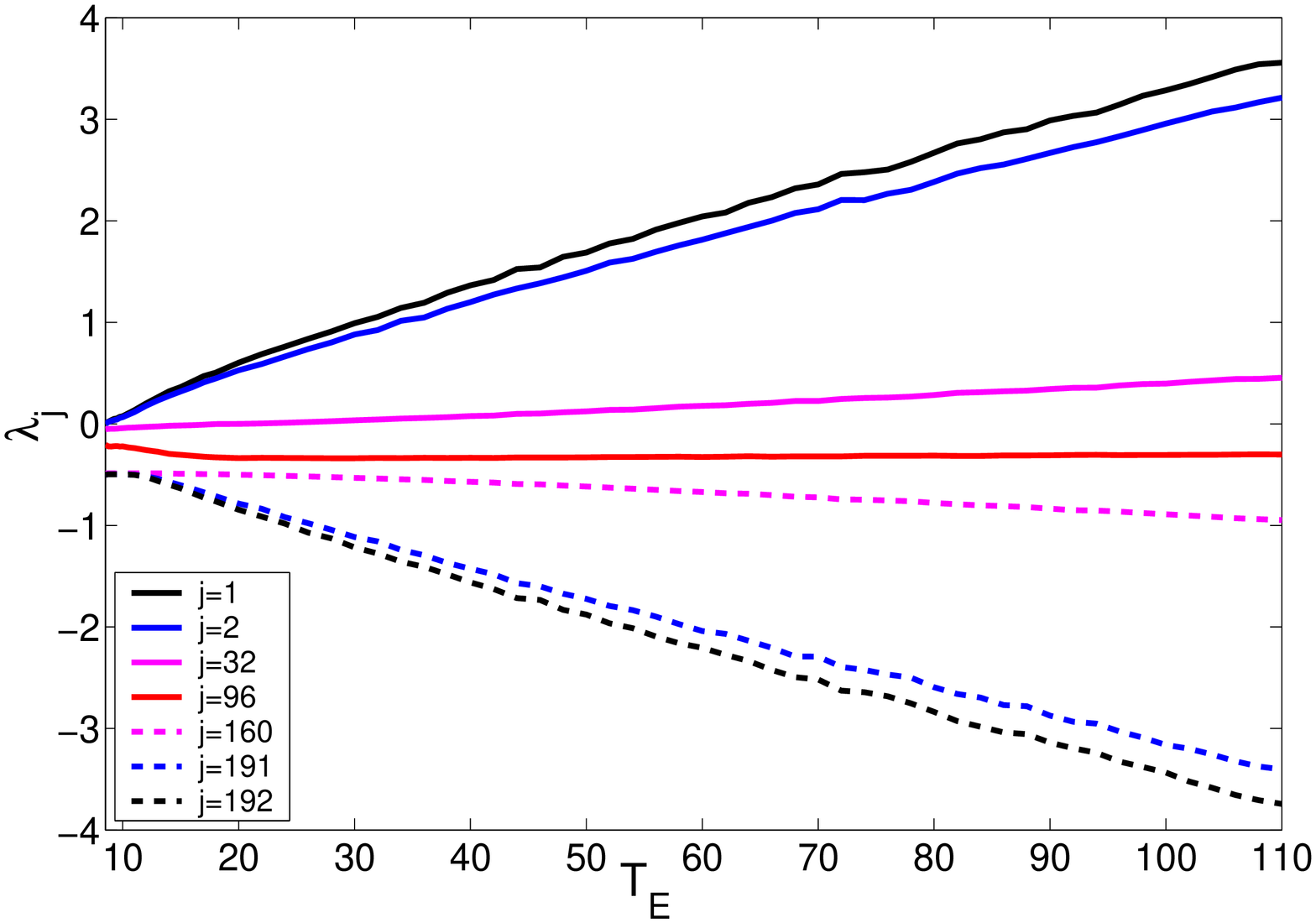}
  \includegraphics[angle=270,width=0.49\textwidth]{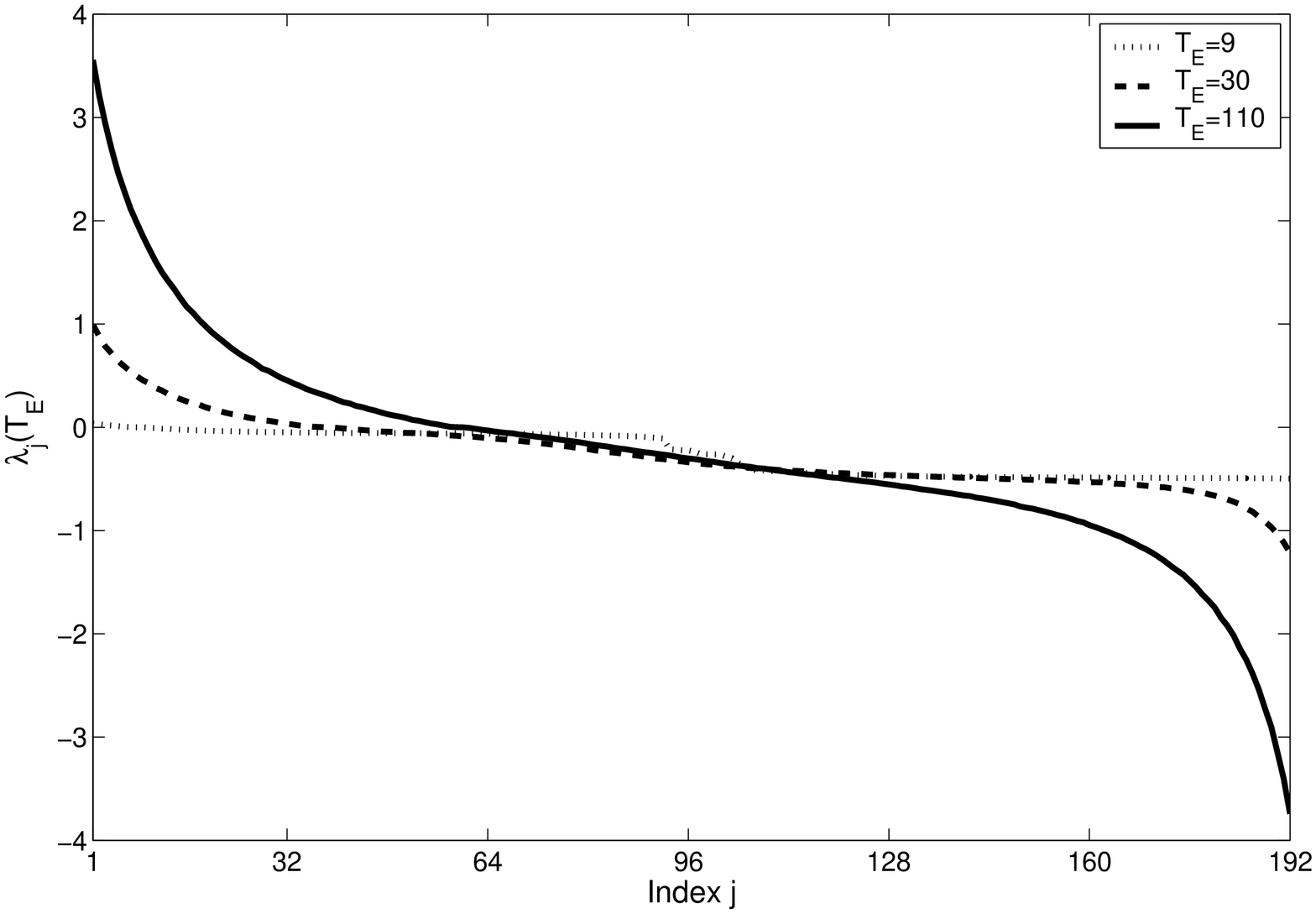}
  \caption{
    Left: Lyapunov exponents $\lambda_j$ for $JT=32$
    and for $j=1$, $2$, $32$, $96$, $160$, $191$, $192$ as a function of $T_E$.
    Right: Spectrum of the Lyapunov exponents for $T_E=9$, $T_E=30$, and $T_E=110$.
    Units as for $\lambda_j$ and $T_E$ as described in \tabref{tab1}.
  }
  \label{fig:Lyapunov32}
\end{figure}

In the left panel of \figref{Lyapunov32} we represent the
evolution of some of the $192$ Lyapunov exponents of the attractor
of~\hbox{\eqref{dotA1j}-\eqref{dotmj}} with $JT=32$ as $T_E$ is
increased. The maximal exponent $\lambda_1$ becomes positive as
$T_E$ crosses the torus breakdown value $T_E^{crit}$, and then
increases monotonically with $T_E$.

The spectrum of the Lyapunov exponents is plotted in the right
panel of \figref{Lyapunov32} for three different values of $T_E$,
again with $JT=32$. The distribution of the exponents approaches a
smooth shape for large $T_E$ and a similar shape is observed for
$JT=64$ (not shown). This suggests the existence of a well-defined
\emph{infinite baroclinicity} model obtained
from~\hbox{\eqref{tau}-\eqref{phi}} as a (possibly, singular
perturbation) limit for $T_E\to\infty$.

\subsubsection{Dimension of the Strange Attractor}

The Lyapunov exponents are used to compute the Lyapunov dimension
(also called Kaplan-Yorke dimension, see~\cite{ER,KY}) and metric
entropy (also known as Kolmogorov-Sinai entropy~\cite{ER}).

The Lyapunov dimension is defined by
\begin{equation}%%noref
  D_L=k+\frac{\sum_{j=1}^k\lambda_j}{\lvert \lambda_{k+1} \rvert},
\end{equation}%%noref
where $k$ is the unique index such that
$\sum_{j=1}^k\lambda_j\ge0$ and $\sum_{j=1}^{k+1}\lambda_j<0$.
Under general assumptions on the dynamical system under
examination, $D_L$ is an upper bound for the Hausdorff dimension
of an attractor.

We have also computed (not shown) other numerical estimates for
the dimension of an attractor: the correlation and information
dimensions~\cite{FarOttYor}. However, these estimates become
completely meaningless when the Lyapunov dimension increases
beyond, say, 20. In particular, the correlation and information
algorithms drastically underestimate the dimension. This is a
well-known problem: for large dimensions, prohibitively long time
series have to be used~\cite{GB83}. Ruelle~\cite{Rue90} suggests
the following rule of thumb: you need a time series of length
$10^{d/2}$ to estimate an attractor of dimension $d$. Therefore,
computational time and memory constraints in fact limit the
applicability of correlation-like algorithms to low-dimensional
attractors.

\begin{figure}[htbp]
  \includegraphics[angle=270,width=0.9\textwidth]{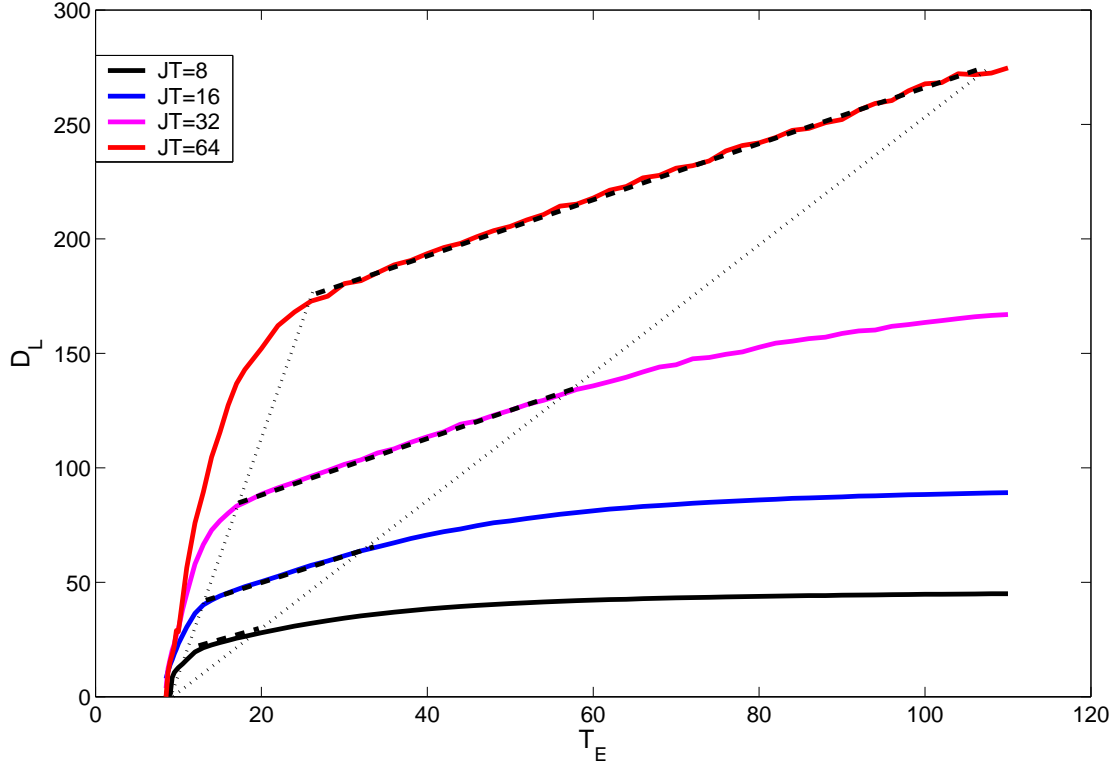}
  \caption{
    Lyapunov dimension of the attractor
    of~\hbox{\eqref{dotA1j}-\eqref{dotmj}} as a function of $T_E$
    for $JT=8$, $16$, $32$, and $64$. All the straight lines are
    parallel and the domain of validity of the linear fit is
    apparently homothetic.
  }
  \label{fig:dimly}
\end{figure}

The number of positive Lyapunov exponents (unstable
dimension~\cite{ER}) increases with $T_E$, which implies that the
Lyapunov dimension also does so. This is confirmed by a plot of
the Lyapunov dimension as a function of $T_E$ for four values of
the discretization order $JT=$ $8$, $16$, $32$, and $64$ (see
\figref{dimly}). For all the considered values of $JT$, it is
possible to distinguish three characteristic regimes in the
behavior of the function $D_L(T_E)$:
\begin{itemize}
    \item For small values of $(T_E-T_E^{crit})$, we have that $D_L
\propto(T_E-T_E^{crit})^\gamma$, with $\gamma$ ranging from
$\sim0.5$ ($JT=8$) to $\sim0.7$ ($JT=64$). The range of $T_E$
where this behavior can be detected increases with $JT$.
    \item For larger values of $T_E$ a linear scaling regime of
$D_L\sim \beta T_E+const.$ is found in all cases. The linear
coefficient is for all $JT$ remarkably close to $\beta\sim1.2$.
The domain of validity of the linear approximation is apparently
homothetic, as can be seen from the simple geometric construction
in figure \figref{dimly}.
    \item For $T_E$ larger than a $JT$-depending threshold, there occurs a
sort of \textrm{phase-space saturation} as the Lyapunov dimension
begins to increase sublinearly with $T_E$. Note that while for
$JT=8$ the model is in this regime in most of the explored
$T_E$-domain ($T_E\gtrsim20$), for $JT=64$ the threshold is
reached only for $T_E\gtrsim108$. In this latter regime of
parametric dependence the system is not able to provide an
adequate representation of the details of the dynamics of the
system. Further discussions on this point will be given in
\secref{bbox} and \secref{energy}.
\end{itemize}

\subsubsection{Entropy production}
The metric entropy $h(\rho)$ of an ergodic invariant measure
$\rho$ expresses the mean rate of information creation,
see~\cite{ER} for definition and other properties. If a dynamical
system possesses a SRB (Sinai-Ruelle-Bowen) invariant measure
$\rho$, then Pesin's identity holds:
\begin{equation}%%noref
  h(\rho)=\sum_{\lambda_j>0}\lambda_j.
\end{equation}%%noref
Existence of an SRB measure for is rather difficult to show for a
given nonhyperbolic attractor~\cite{ER}. It has been only proven
for low-dimensional cases such as the H\'enon~\cite{WY} or
Lorenz~\cite{Viana} strange attractors. More generally one has the
inequality $h(\rho)\ge\sum_{\lambda_j>0}\lambda_j$. We then simply
\emph{assume} the existence of a \emph{unique} SRB measure and
refer to the sum of the positive Lyapunov exponents as metric
entropy.

\begin{figure}[htbp]
  \includegraphics[angle=270,width=0.32\textwidth]{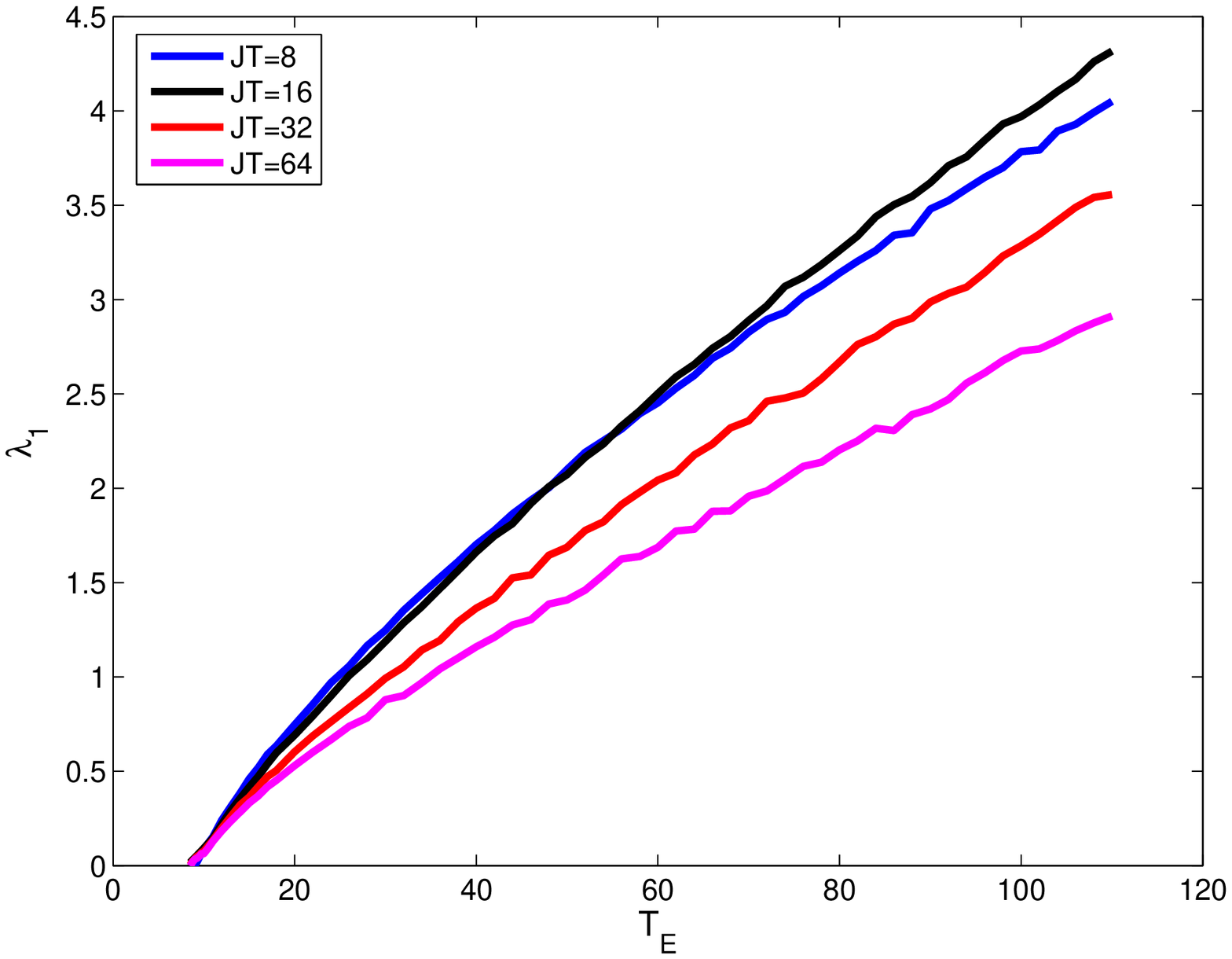}
  \includegraphics[angle=270,width=0.32\textwidth]{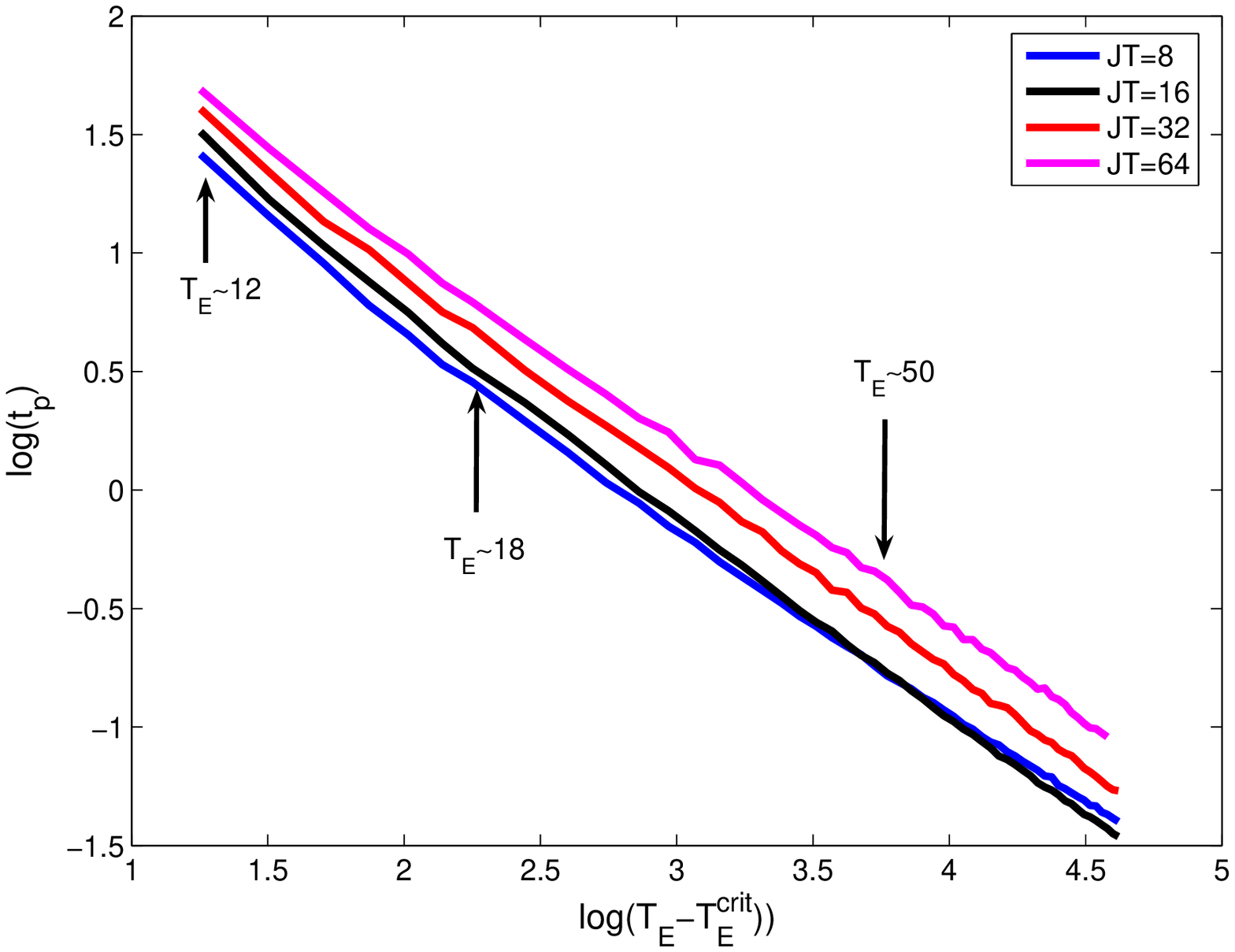}
  \includegraphics[angle=270,width=0.32\textwidth]{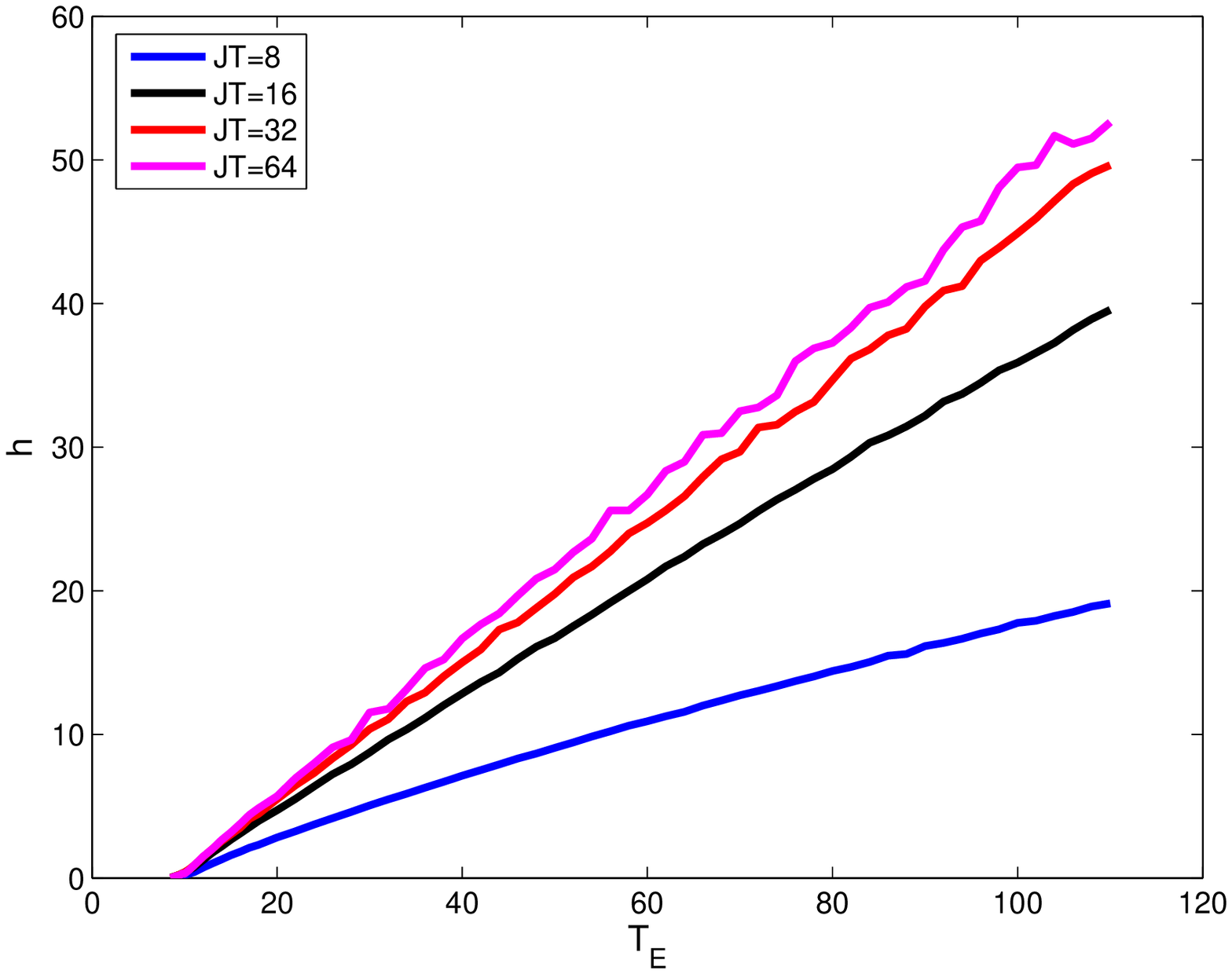}
  \caption{
    Left: maximal Lyapunov exponent on the attractor
    of~\hbox{\eqref{dotA1j}-\eqref{dotmj}} as a function of $T_E$
    for $JT=$ $8$, $16$, $32$, $64$.
    Center:  Log-Log plot of the predictability time of the system
    $t_p=\lambda_1^{-1}$ versus $T_E-T_E^{crit}$.
    Power laws ($t_p\propto(T_E-T_E^{crit})^{\gamma}$)
    are detected for all considered values of $JT$.
    Right: metric entropy.
    Linear dependences $h\sim\beta(T_E-T_E^{crit})$
    occur for all values of $JT$.
  }
  \label{fig:jt32-64}
\end{figure}

The maximal Lyapunov exponent, the predictability time
$t_p=\lambda_1^{-1}$, and the metric entropy as functions of $T_E$
are compared for $JT=$ $8$, $16$, $32$, and $64$ in
\figref{jt32-64}. It turns out that, for fixed $JT$, $\lambda_1$
increases sublinearly with $T_E$, whereas for $T_E$ fixed,
$\lambda_1$ decreases for increasing values of $JT$. Consequently,
for fixed $JT$ the predictability time decreases monotonically
with $T_E$. We note that, for all values of $JT$, if $T_E>14$ we
have that $t_p<10$, which corresponds in physical units to a
predictability time $t_p\lesssim 12$ days. Moreover, in the range
$T_E\gtrsim12$, $t_p$ is proportional to
$(T_E-T_E^{crit})^\gamma$, with $\gamma$ ranging between
$[-0.85,-0.8]$ depending on the considered value of $JT$. The
metric entropy has a marked linear dependence $h\sim \beta
(T_E-T_E^{crit})$, with $\beta$ ranging from $\sim0.15$ ($JT=8$)
to $\sim 0.5$ ($JT=64$). Moreover, for a given value of $T_E$, the
metric entropy increases with $JT$. From the dynamical viewpoint,
this means on one hand that the maximal sensitivity of the system
to variations in the initial condition along a \emph{single
direction} is largest for $JT=8$. On the other hand, there are
many more \emph{active degrees of freedom} for $JT=64$ and they
collectively produce a faster \textit{forgetting} of the initial
condition as time goes on.

\subsubsection{Parametric smoothness of the attractor properties with respect to $T_E$}

The dependence of the Lyapunov exponents and, consequently, of the
predictability time, of the Lyapunov dimension and metric entropy,
with respect to $T_E$ is remarkably smooth, especially if one
keeps in mind the paradigms of low-dimensional nonhyperbolic
strange attractors. For example, for the logistic mapping (see
\emph{e.g.}~\cite{ER}) the maximal Lyapunov exponent $\lambda_1$
is a discontinuous function of the parameter at every point where
$\lambda_1>0$. This is due to the fact that so-called
\emph{windows of periodicity}, that is, open parameter intervals
where the logistic mapping has a periodic attractor, are
\emph{dense} in the parameter axis. In the complement set of the
windows of periodicity, parameter values for which a strange
attractor occurs form a nowhere dense set of positive Lebesgue
measure. In fact, similar features seem to hold for many
low-dimensional mappings having strange attractors, such as the
H\'enon-like families~\cite{MV,Sim79,WY}, also
compare~\cite{BST98,BSV1,ER} and references therein.

No windows of periodicity were detected in the fully chaotic range
(say, $T_E>16$) for model~\hbox{\eqref{dotA1j}-\eqref{dotmj}},
independently of the truncation order $JT=$ $8$, $16$, $32$, $64$.
We have also tried slightly different spectral discretization
schemes and integration methods (such as leapfrog or Runge-Kutta
4), but this qualitative feature of smoothness and absence of
windows of periodicity persisted in all cases.

There are two possible explanations for this: either the windows
of periodicity are very narrow or there are no windows of
periodicity. A possible theoretical support for the latter case
might be provided by the concept of \emph{robust} strange
attractors. We refer the interested reader to~\cite{Viana} for a
discussion and more references. Also see~\cite{GOST05} for a class
of low-dimensional maps where strange attractors occur on open
parameter sets.

From the above it follows that, from the dynamical point of view,
the model~\hbox{\eqref{dotA1j}-\eqref{dotmj}} behaves in sensibly
different ways if the truncation order $JT$ is changed. For
example, in the earth-like regime $T_E=18$, the Lyapunov dimension
nearly doubles when passing from $JT=32$ to $JT=64$. However,
despite the quantitative differences, many qualitative features
remain the same for $JT\ge16$:
\begin{itemize}
\item
  the route for the creation of the strange attractor
  involves a Hopf bifurcation of the Hadley equilibrium,
  followed by quasi-periodic breakdown of the invariant torus;
\item
  a linear scaling regime exists for the Lyapunov dimension
  as a function of $T_E$;
\item
  the maximal Lyapunov exponent and the metric entropy
  increase monotonically with $T_E$;
\item
  the distribution of the Lyapunov exponents tends to a
  well-defined shape for $T_E$ large (\figref{Lyapunov32}~right);
\item
  the dependence of Lyapunov exponents, dimension and metric
  entropy with respect to $T_E$ is remarkably smooth.
\end{itemize}

\subsection{Bounding Box of the Attractor}
\label{sec:bbox} In this section we study the volume of the
bounding box $V_{BB}$ for the attractors of
model~\hbox{\eqref{dotA1j}-\eqref{dotmj}} previously described.
The bounding box of a set of points in an $N-$dimensional space is
defined as the smallest hyperparallelepiped containing the
considered set~\cite{Smith00,Smith02}. For clarity, in the
$N$-dimensional phase space, where $N=6\times JT$, the volume
$V_{BB}$ is computed as:
\begin{equation}
  \label{eq:bbox}
  V_{BB}=\prod_{k=1}^{N=6\times JT}
  \left[\max_{t_{tr}<t<t_{max}}
    \left(z_k\left(t\right)\right)-\min_{t_{tr}<t<t_{max}}
    \left(z_k\left(t\right)\right)
  \right].
\end{equation}
Here the $z_k$ denote the $6\times JT$ variables spanning the
phase space of the system, in our case the Fourier coefficients
$A^1_j$, $A^2_j$, $B^1_j$, $B^2_j$, $m_j$, and $U_j$, with
$j=1,\ldots,JT$. The condition $t>t_{tr}$ allows for the
transients to die out. Typically, $t_{tr}$ is rather safely fixed
to $1500$, which correspond to about five years.

When the Hadley equilibrium is the universal attractor, the volume
$V_{BB}$ is zero, while it is non-zero if the computed orbit is
attracted to a periodic orbit, a two-torus or a strange attractor.
In all cases $V_{BB}$, which represents the bulk size of the
attractor in phase space, grows with $T_E$. More precisely, each
of the factors in the product~\eqref{eq:bbox} increases with
$T_E$, so that expansion occurs in all directions of the phase
space. This matches the basic expectations on the behavior of a
dissipative system having a stronger energy input.

\begin{figure}[htbp]
  \includegraphics[angle=270,width=0.49\textwidth]{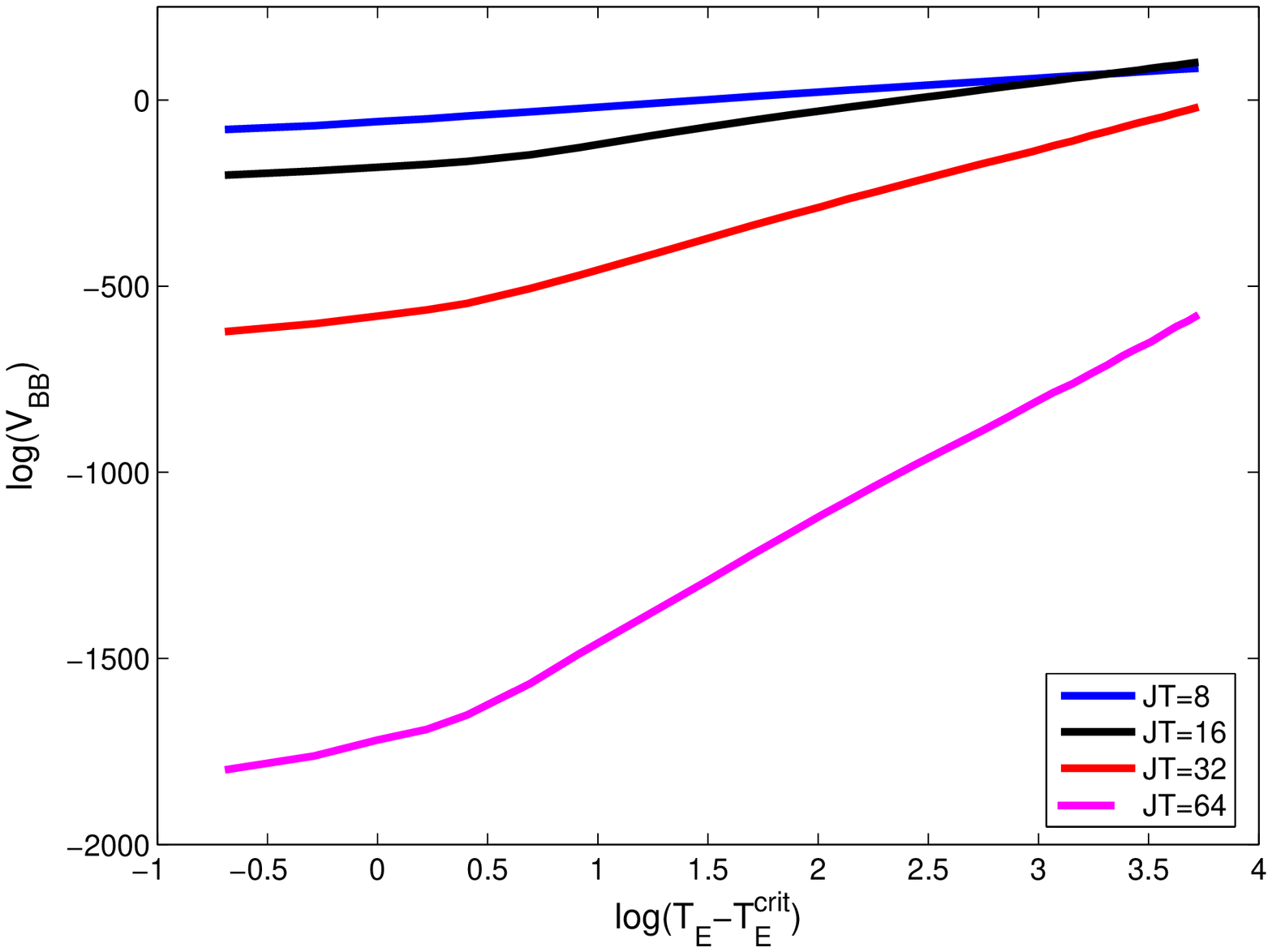}
  \includegraphics[angle=270,width=0.49\textwidth]{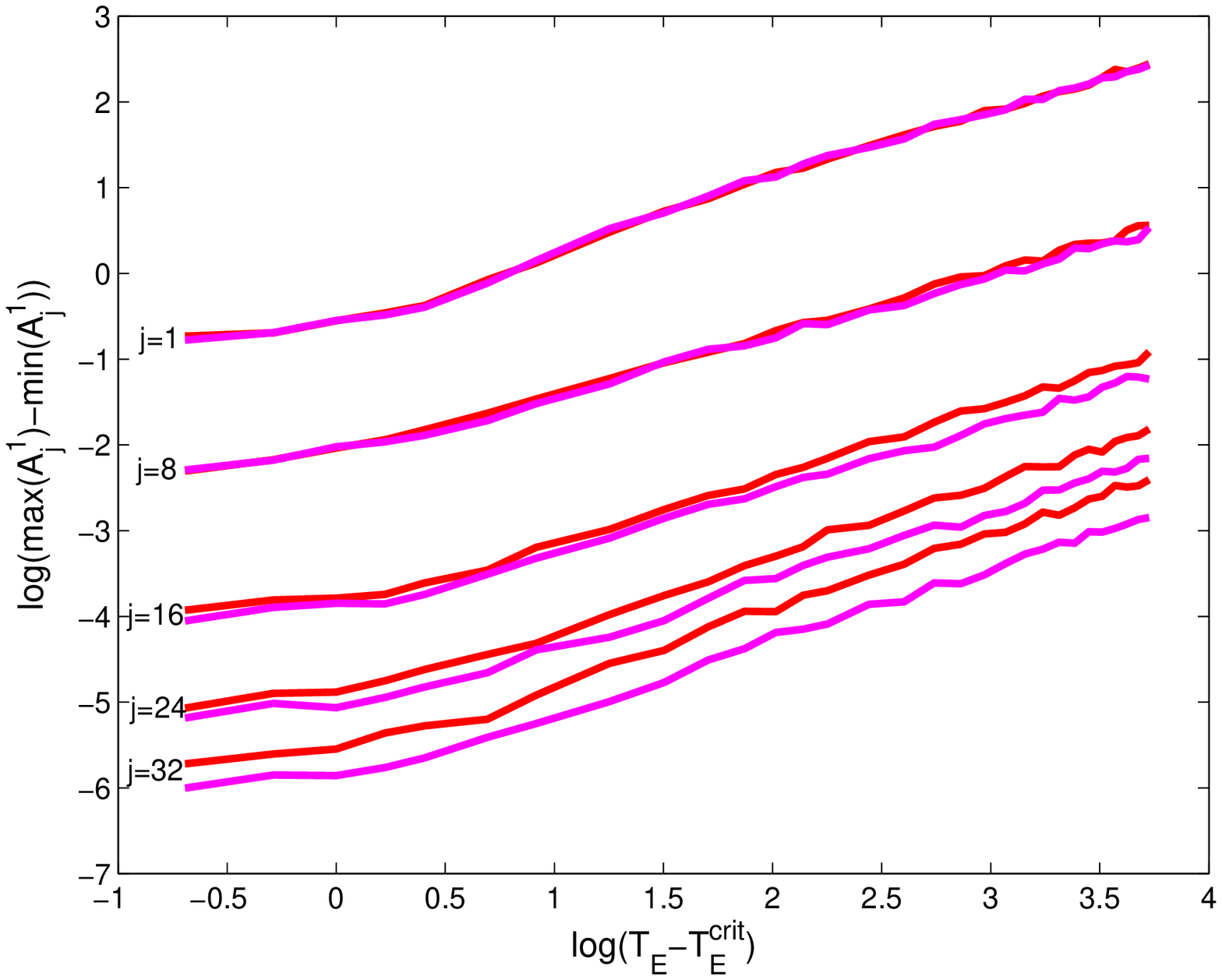}
  \caption{
    Left:
    Volume of the bounding box $V_{BB}$ of the attractor
    as a function of the detuning parameter $T_E-T_E^{crit}$
    for $JT=$ $8$, $16$, $32$, $64$.
    For description of the power law fits, see text and \tabref{bboxtable}.
    Right: Value of the corresponding sides of the bounding box
    pertaining to the variables $A_j^1$ for $JT=32$ (red lines)
    and $64$ (magenta lines).
    Notice the two power-law regimes mentioned in the text.
  }
  \label{fig:bbox}
\end{figure}

In the right panel of \figref{bbox} we present a plot of
$\log(V_{BB})$ as function of $T_E$ for the selected values of
$JT=$ $8$, $16$, $32$, and $64$. In the case $JT=8$, $V_{BB}$
obeys with great precision the power law
$V_{BB}\propto(T_E-T_{E}^{crit})^\gamma$ in the whole domain
$T_E\geq 9$. The best estimate for the exponent is $\gamma\sim40$.
Given that the total number of Fourier components is $6\times
JT=48$, this implies that the growth of the each side of the
bounding box is on the average proportional to about the
$5/6^{th}$ power of $(T_E-T_E^{crit})$.

For higher values of $JT$, two sharply distinct and well defined
power-law regimes occur. For $JT=16$, in the lower range of
$(T_E-T_{E}^{crit})$ - corresponding in all cases to $T_E\lesssim
T_{E}^{crit} + 1.5$ - the volume of the bounding box increases
with about the $35^{th}$ power of $(T_E-T_{E}^{crit})$, while in
the upper range of $(T_E-T_{E}^{crit})$ - for $T_E\gtrsim
T_{E}^{crit} + 1.5$ - the power-law exponent abruptly jumps up to
about $80$. For $JT=32$ the same regimes can be recognized, but
the values of the best estimates of the exponents are twice as
large as what obtained with $JT=16$. Similarly, for $JT=64$ the
best estimates of the exponents are twice as large as for $JT=32$.
The results on the power law fits of
$V_{BB}\propto(T_E-T_{E}^{crit})^\gamma$ are summarized in
Table~\ref{bboxtable}. We emphasize that in all cases the
uncertainties on $\gamma$, which have been evaluated with a
standard bootstrap technique, are rather low and total to less
than $3\%$ of the best estimate of $\gamma$. Moreover, the
uncertainty of the power-law fit greatly worsens if we detune the
value of $T_{E}^{crit}$ by as little as $0.3$, thus reinforcing
the idea that fitting a power law against the logarithm of
$(T_E-T_{E}^{crit})$ is a robust choice.

When considering separately the various sides of the bounding box
hyperparallelepiped (not shown), \textit{i.e.}, each of the
factors in the product~\eqref{eq:bbox}, we have that for $JT=8$
all of them increase as about $(T_E-T_E^{crit})^{5/6}$ in the
whole range. For $JT=$ $16$, $32$, and $64$, in the lower range of
$T_E$ each side of the bounding box increases as about the
$1/3^{rd}$ power of $(T_E-T_E^{crit})$, while in the upper range
of $T_E$ each side of the bounding box increases as about the
$5/6^{th}$ power of $(T_E-T_E^{crit})$. Selected cases are
depicted in the right panel of \figref{bbox}. So for a given value
of truncation order $JT$, the ratios between the ranges of the
various degrees of freedom are essentially unchanged when varying
$T_E$, so that the system obeys a sort of self-similar scaling
with $T_E$.

Summarizing, for sufficiently high truncation order ($JT\ge16$) a
robust parametric dependence is detected for the volume of the
bounding box as a function of $T_E$:
\begin{equation}%%noref
  V_{BB}\propto(T_E-T_E^{crit})^{\gamma}, \quad \gamma=\epsilon N
  \quad \epsilon\sim\begin{cases}1/3, \quad T_E-T_E^{crit}\lesssim
    1.5,
    \\5/6, \quad T_E-T_E^{crit}\gtrsim
    1.5,\end{cases}
\end{equation}%%noref
where $N=6\times JT$ is the number degrees of freedom.

The comparison, for, say, $JT=$ $16$ and $32$, of factors
in~\eqref{eq:bbox} having the same order for the same value of
$T_E-T_E^{crit}$ provides insight about the sensitivity to model
resolution. In the following discussion, we examine the variables
$A_j^1$ but similar observations apply to all other variables
$A_j^2$, $B_j^1$, $B_j^2$, $U_j$, and $m_j$. The factors related
to the the gravest modes, such as
$\left[\max\left(A_j^1\left(t\right)\right)-\min\left(A_1^1\left(t\right)\right)\right]$,
agree with high precision, thus suggesting that the large scale
behavior of the system is only slightly affected by variation of
model resolution. When considering the terms related to the
fastest latitudinally varying modes allowed by both truncation
orders, such as
$\left[\max\left(A_j^1\left(t\right)\right)-\min\left(A_j^1\left(t\right)\right)\right]$
with $17\leq j\leq 32$, we have that those obtained for $JT=32$
are \emph{larger} than the corresponding factors obtained for
$JT=64$, and the distance between pairs of the same order
increases with $j$. See the right panel of \figref{bbox}. This is
likely to be the effect of spectral aliasing~\cite{Boy}: the
fastest modes of the model with lower resolution \textit{absorb}
the dynamics contained in the scales which are instead resolved in
the higher-resolution model. The same effect is observed when
comparing, for $JT=16$ and $32$, coefficients of the same order
such as
$\left[\max\left(A_j^1\left(t\right)\right)-\min\left(A_j^1\left(t\right)\right)\right]$
with $9\leq j\leq 16$. The $JT=8$ case does not precisely match
this picture.

\clearpage
\section{Statistical properties of the total energy and latitudinally averaged
  zonal wind}\label{sec:energy}

In this section the model~\hbox{\eqref{dotA1j}-\eqref{dotmj}} is
studied by means of observables (functions of state space
variables) of physical significance, as opposed to the quantities
derived from the Lyapunov exponents and the volume of the bounding
box used in \secref{lyap}, which are more typical indicators used
in dynamical systems analysis.

\subsection{Total energy}

The total energy of the system $E(t)$ defined in~\eqref{eq:energy}
isa very relevant observable of physical significance for the
system. In \tabref{tab1} we report its conversion factor between
the non-dimensional and dimensional units. For the Hadley
equilibrium, the time-independent expression for the total energy
is derived by plugging~\eqref{elim1} into~\eqref{energydensity}
and then computing the integral~\eqref{eq:energy}:
\begin{equation}\label{eq:energyhad}
\overline{E(t)}=\frac{\delta
p}{g}L_xL_y\left(\frac{RT_E}{4f_0}\frac{1}{1+\frac{\kappa}{\nu_E}\left(\frac{\pi}{L_y}\right)^2}\right)^2\left(\frac{\pi^2}{L_y^2}+\frac{1}{H_2^2}\right).
\end{equation}
The total energy is proportional to $T_E^2$ and is mostly stored
as potential energy~\cite{Peix}, which is described by the second
term of the sum in~\eqref{eq:energyhad}.

\begin{figure}[p]
  \includegraphics[angle=270,width=0.49\textwidth]{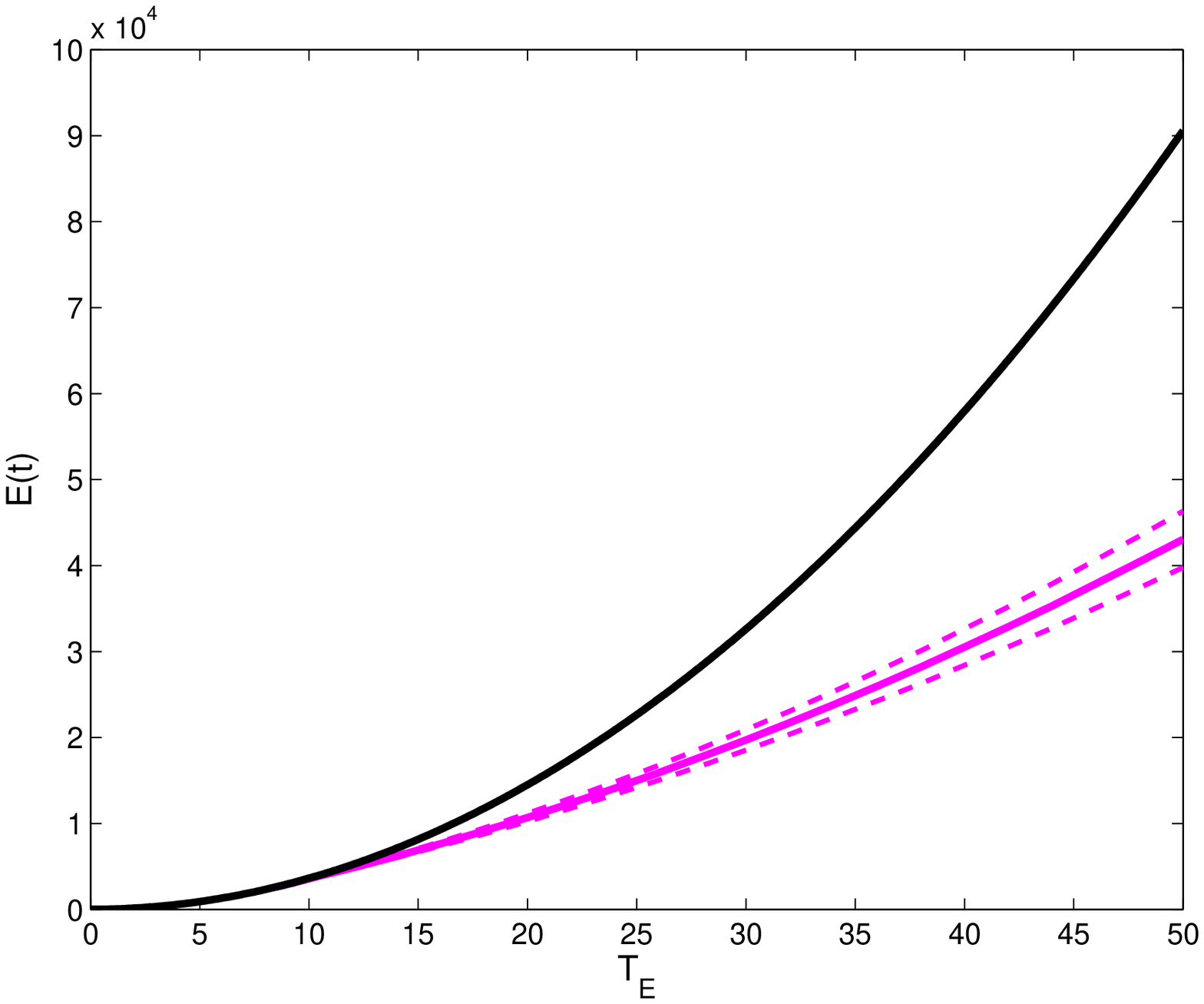}
  \includegraphics[angle=270,width=0.49\textwidth]{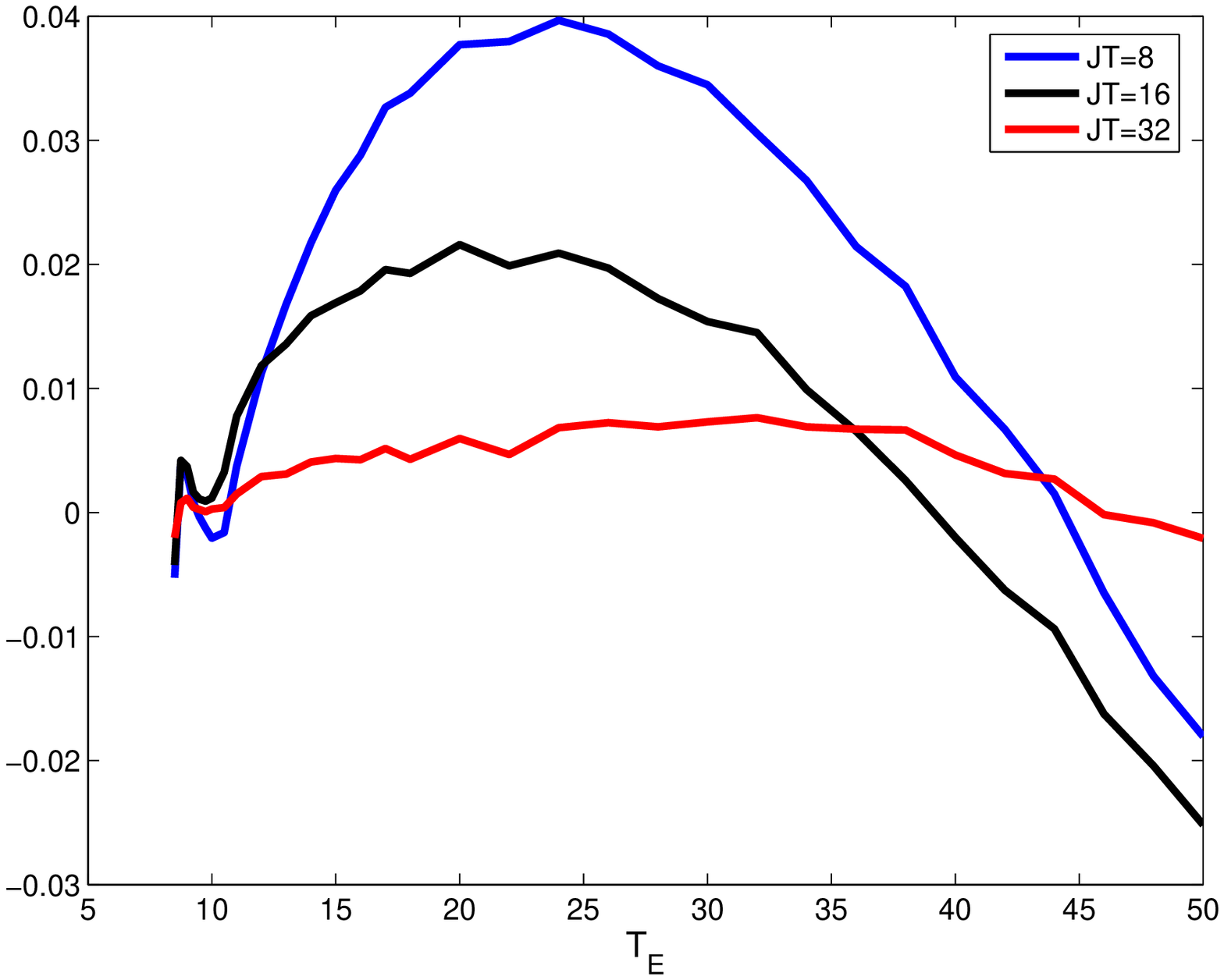}\\
  \caption{
    Left: $\overline{E(t)}$ for the Hadley equilibrium (black line)
    and deduced from the observed fields in the chaotic regime
    for $JT=64$ (magenta line);
    the magenta dashed line delimit the $\sigma$-confidence interval.
    Right: fractional deviations of $\overline{E(t)}$,
    for $JT=$ $8$, $16$, and $32$, with respect to $JT=64$.
    See text for details.
  }
  \label{fig:emean}
\end{figure}

In \figref{emean} we present the results obtained for the various
values of $JT$ used in this work. In the left panel we present the
$JT=64$ case, which is representative of what obtained also in the
other cases. The time-averaged total energy is monotonically
increasing with $T_E$, but when the system enters the chaotic
regime, $\overline{E(t)}$ is much lower than the value at the
coexisting Hadley equilibrium. This behavior may be related to the
much larger dissipation fuelled by the chaos-driven activation of
the smaller scales. In the chaotic regime $E(t)$ is characterized
by temporal variability, which becomes more and more pronounced
for larger values of $T_E$.

In the right panel of \figref{emean} we compare the cases
$JT=8,16,32$ with respect to $JT=64$. The overall agreement of
$\overline{E(t)}$ is good but progressively worsens when
decreasing $JT$: for $JT=32$, the maximal fractional difference is
less than $0.01$, while for $JT=8$ it is about one order of
magnitude larger. Differences between the representations given by
the various truncations levels also emerge in power law fits such
as $\overline{E(t)}\propto T_E^\gamma$. In the regime where the
Hadley equilibrium is attracting, this fit is exact, with exponent
$\gamma=2$. For $T_E-T_E^{crit}\lesssim 1.5$ and $T_E>T_E^H$ (the
value of the first Hopf bifurcation, see \tabref{hopf1}), for all
the values of $JT$ the power law fit is good, with
$\gamma=1.9\pm0.03$, so that a weakly subquadratic growth is
realized. For $T_E-T_E^{crit}\gtrsim 1.5$, only the $JT=$ $32$ and
$64$ simulations of $\overline{E(t)}$ obey with excellent
approximation a weaker power law, with $\gamma=1.52\pm0.02$ in
both cases, while the cases $JT=$ $8$ and $16$ do not
satisfactorily fit any power law.

The agreement worsens in the upper range of $T_E$, which points at
the criticality of the truncation level when strong forcings are
imposed. Nevertheless, the observed differences are strikingly
small between the cases, say, $JT=8$ and $JT=64$, with respect to
what could be guessed by looking at the Lyapunov dimension,
entropy production, and bounding box volume diagnostics analyzed
in the previous sections, where essentially only $JT=32$ and
$JT=64$ had a satisfactory agreement. This suggests that when
analyzing global observables, the resolution requirements for
obtaining good statistical indicators are much more relaxed.

\subsection{Zonal wind}
We here examine the latitudinal average, denoted by $\langle
\bullet \rangle$, of $U$ and $m$:
\begin{align}%%noref
  \langle U(y,t)\rangle&=\frac{1}{L}\int_0^L U(y,t)dy
  =\frac{2}{\pi}\sum_{j=1,\;\text{$j$ odd}}^{JT}\frac{U^j}{j},\\
  \langle m(y,t)\rangle &=\frac{1}{L}\int_0^L m(y,t)dz
  =\frac{2}{\pi}\sum_{j=1,\;\text{$j$ odd}}^{JT}\frac{m^j}{j}.
\end{align}%%noref
Since $U(y,t)$ represents the zonal average of the mean of the
zonal wind at the two pressure levels $p_1$ and $p_3$ at latitude
$y$, $\langle U(y,t)\rangle$ is proportional to the total zonal
momentum of the atmosphere. Instead, $\langle m(y,t)\rangle$
represents the spatially averaged halved difference between the
the zonal wind at the two pressure levels $p_1$ and $p_3$.
Computation of such space averages at the time-independent Hadley
equilibrium~\eqref{mTAU} is straightforward:
\begin{equation}%%noref
  \left\langle  m\left(y\right) \right\rangle=\left\langle U\left(y\right) \right\rangle=\frac{R}{f_0
  L_y}\frac{T_E}{2}\frac{1}{1+\frac{\kappa}{\nu_N}\left(\frac{\pi}{L_y}\right)^2}.
\end{equation}%%noref
Since we cannot have net, long-term zonal forces acting on the
atmosphere at the surface interface, the spatial average of the
zonal wind at the pressure level $p_3$ must be zero. Therefore,
the outputs of the numerical integrations must satisfy the
following constraint:
\begin{equation}
  \label{mUmeanaverage}
  \overline{\left\langle  m\left(y,t\right) \right\rangle}=\overline{\left\langle U\left(y,t\right)
  \right\rangle},
\end{equation}
where $\overline{X}$ denotes the time-average of the field $X$.
The constraint~\eqref{mUmeanaverage} is automatically satisfied at
the Hadley equilibrium.

\begin{figure}[p]
  \includegraphics[angle=270,width=0.49\textwidth]{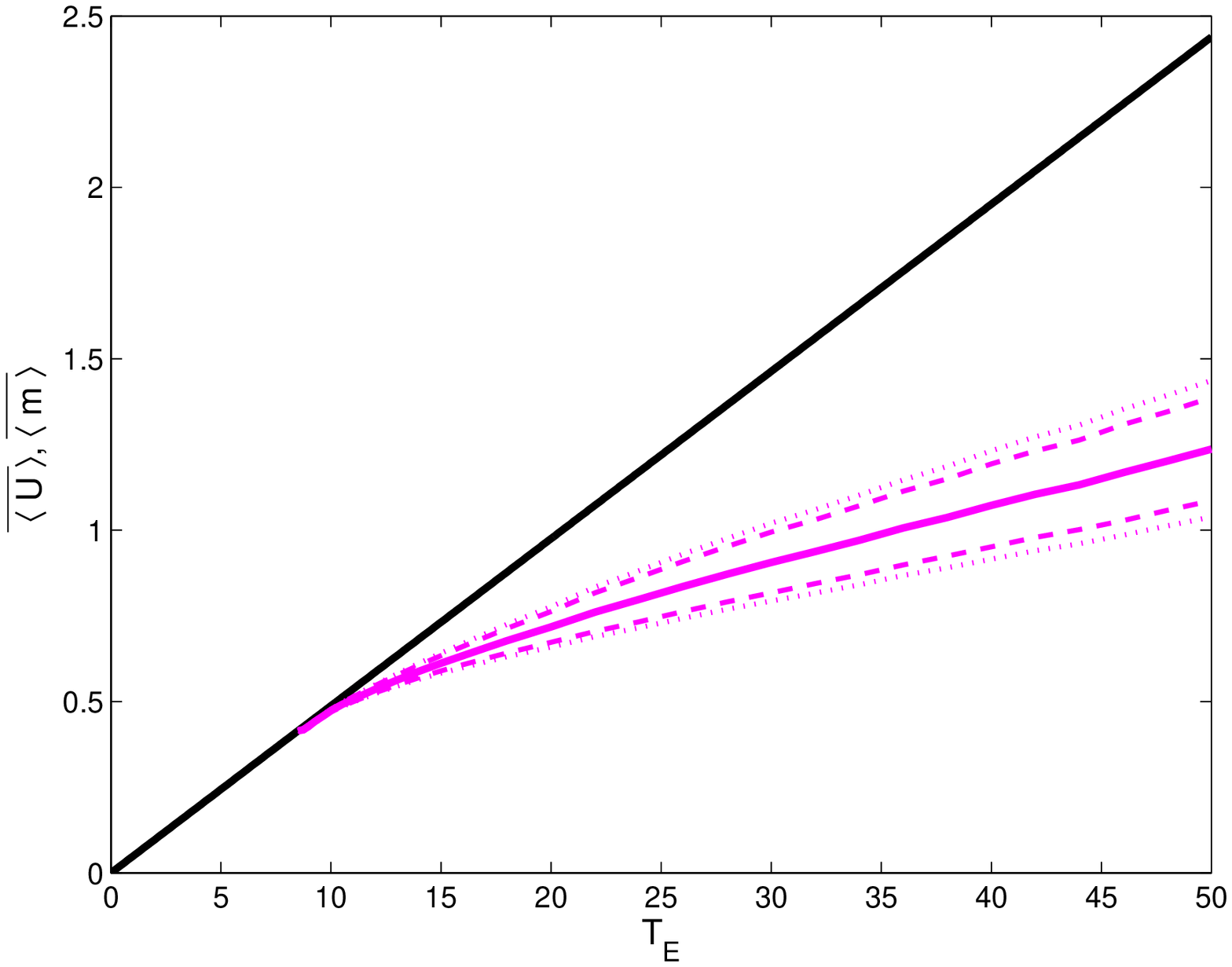}
  \includegraphics[angle=270,width=0.49\textwidth]{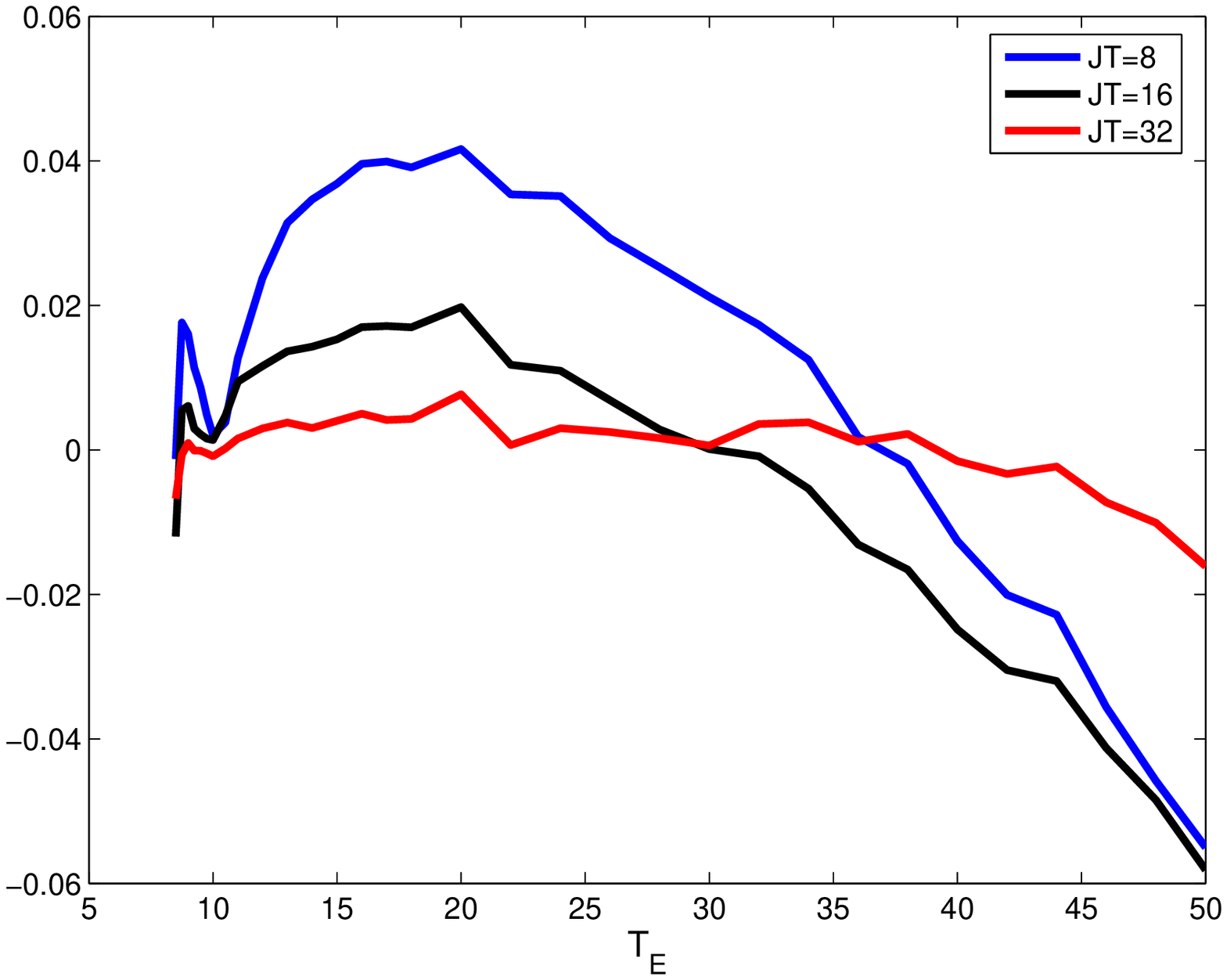}\\
  \caption{
    Left: $\overline{\langle U \rangle}=\overline{\langle m \rangle}$
    for the Hadley equilibrium (black line)
    and deduced from the observed fields in the chaotic regime
    for $JT=64$ (magenta line);
    the magenta dashed and dotted lines delimit
    the $\sigma$-confidence interval for $\langle U \rangle$
    and $\langle m \rangle$, respectively.
    Right: fractional deviations of
    $\overline{\langle U \rangle}=\overline{\langle m \rangle}$
    for $JT=$ $8$, $16$, and $32$ with respect to $JT=64$.
    See text for details.
  }
  \label{fig:umean}
\end{figure}

The results are presented in \figref{umean}. In the left panel we
plot the outputs for $JT=64$, which, similarly to the total energy
case, is well representative of all the $JT$ cases. We first note
that the constraint~\eqref{mUmeanaverage} is obeyed within
numerical precision. The average winds are monotonically
increasing with $T_E$, but, when the system enters the chaotic
regimes, the averages $\overline{\langle
m(y,t)\rangle}=\overline{\langle U(y,t)\rangle}$ have a much
smaller value than at the Hadley equilibrium, and they display
sublinear growth with $T_E$. Moreover, for $T_E>T_E^{crit}$ the
temporal variability of the time series $\langle m(y,t)\rangle$
and $\langle U(y,t)\rangle$ increases with $T_E$. The variability
of $\langle m(y,t)\rangle$ results to be slightly larger than that
of $\langle U(y,t)\rangle$, probably because the latter is related
to a \textit{bulk} mechanical property of the system such as the
total zonal momentum.

Since we are dealing with a quasi-geostrophic system, these
observations on the wind fields imply that while the time-averaged
meridional temperature difference between the northern and
southern boundary of the system increases monotonically with
$T_E$, as to be expected, the realized value is greatly reduced by
the onset of the chaotic regime with respect to the corresponding
Hadley equilibrium. This is the signature of the negative feedback
due to a mechanism similar to the baroclinic
adjustment~\cite{Stone}: when the poleward eddy transport of heat
is realized, it causes the reduction of the meridional temperature
gradient, thus limiting the wind shear. Note that in this model
the adjustment, as opposed to the general case, is essentially
correct in a variational context, since only one zonal wave is
considered, and so the fastest growing unstable wave is also the
wave transporting northward the largest amount of
heat~\cite{SP,SM}. Nevertheless, the adjustment mechanism does not
keep the system \textit{close to marginal stability}, as
envisioned in some baroclinic adjustment theories, since for
$T_E>T_E^{crit}$ both the instantaneous and the time-averaged
fields of the system are completely different from those realized
at the Hadley equilibrium.

The effects of lowering $JT$ are illustrated in
\figref{umean}~right. The overall agreement, expressed by a small
value of the fractional differences, progressively worsens for
smaller $JT$. Notice the similarity of the functional shapes with
\figref{emean}~right. The results in \figref{umean}~right can be
summarized as follows: the coarser-resolution models have higher
total temperature difference between the two boundaries for values
of $T_E$ up to about $30$ and lower temperature differences for
higher values of $T_E$. This implies that while for $T_E\lesssim
30$ the latitudinal heat transport increases with $JT$ as a
positive trade-off between the higher number of unstable
baroclinic modes (within a sloppy linear thinking) or, better,
smaller scale baroclinic conversion processes taking place in a
higher-dimensional attractor, and the enhancement of the
barotropic and viscous stabilizing effects, for $T_E \gtrsim 30$
the converse is true.

Again, differences between the various truncations levels emerge
as one attempts power law fits of the form $\overline{\langle
m(y,t)\rangle}=\overline{\langle U(y,t)\rangle}\propto
T_E^\gamma$. For the Hadley equilibrium regime we have $\gamma=1$.
For $T_E\lesssim10$ and above the first Hopf bifurcation, for all
values of $JT$ the power law fit is good, with
$\gamma=0.875\pm0.05$. For $T_E-T_E^{crit}\gtrsim 1.5$, only the
simulations with $JT=$ $32$ and $64$ obey a power law (with
$\gamma=0.58\pm0.02$) with excellent approximation, while the
realizations of the $JT=$ $8$ and $16$ cases do not fit any power
law.

\begin{figure}[p]
  \includegraphics[angle=270,width=0.49\textwidth]{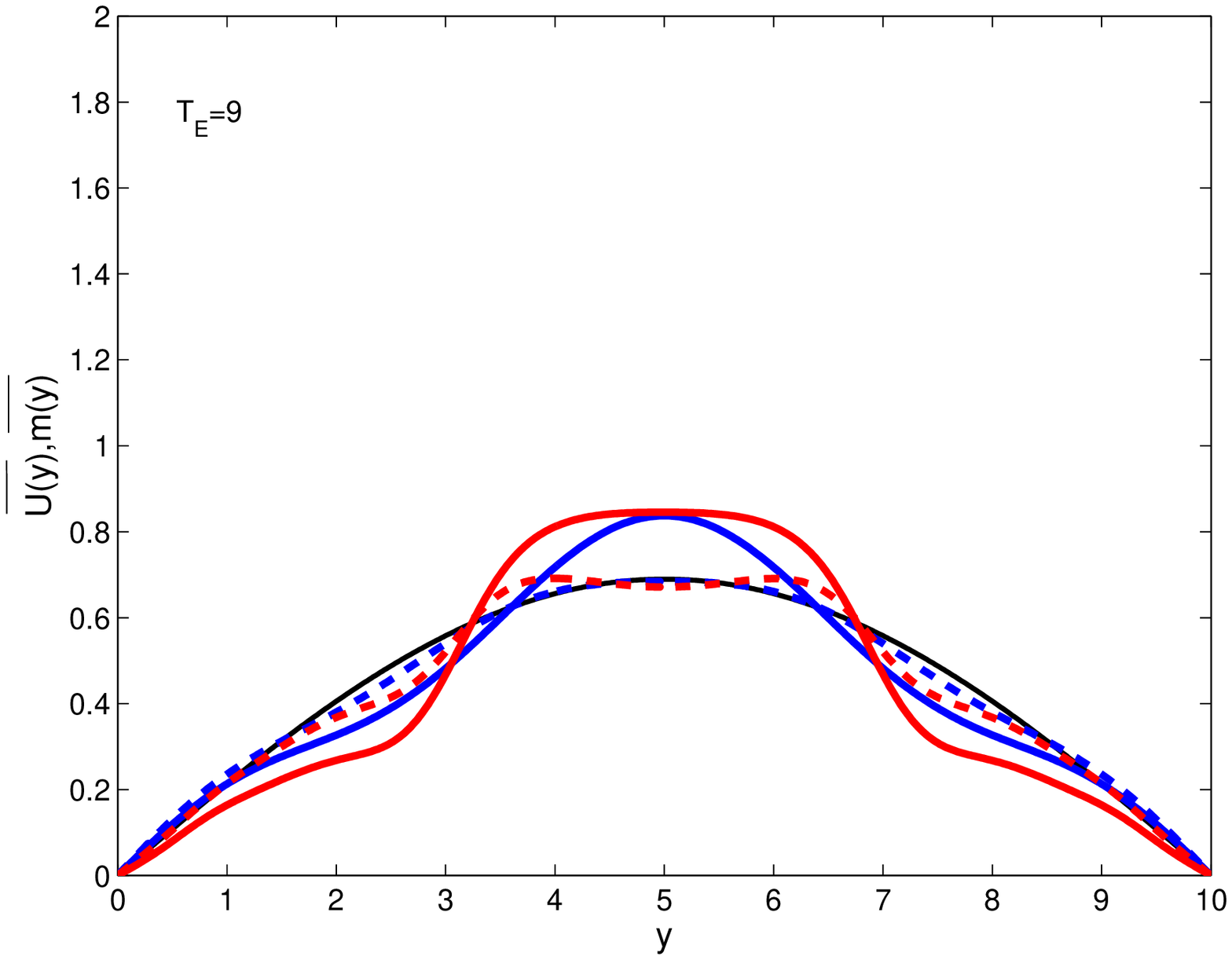}
  \includegraphics[angle=270,width=0.49\textwidth]{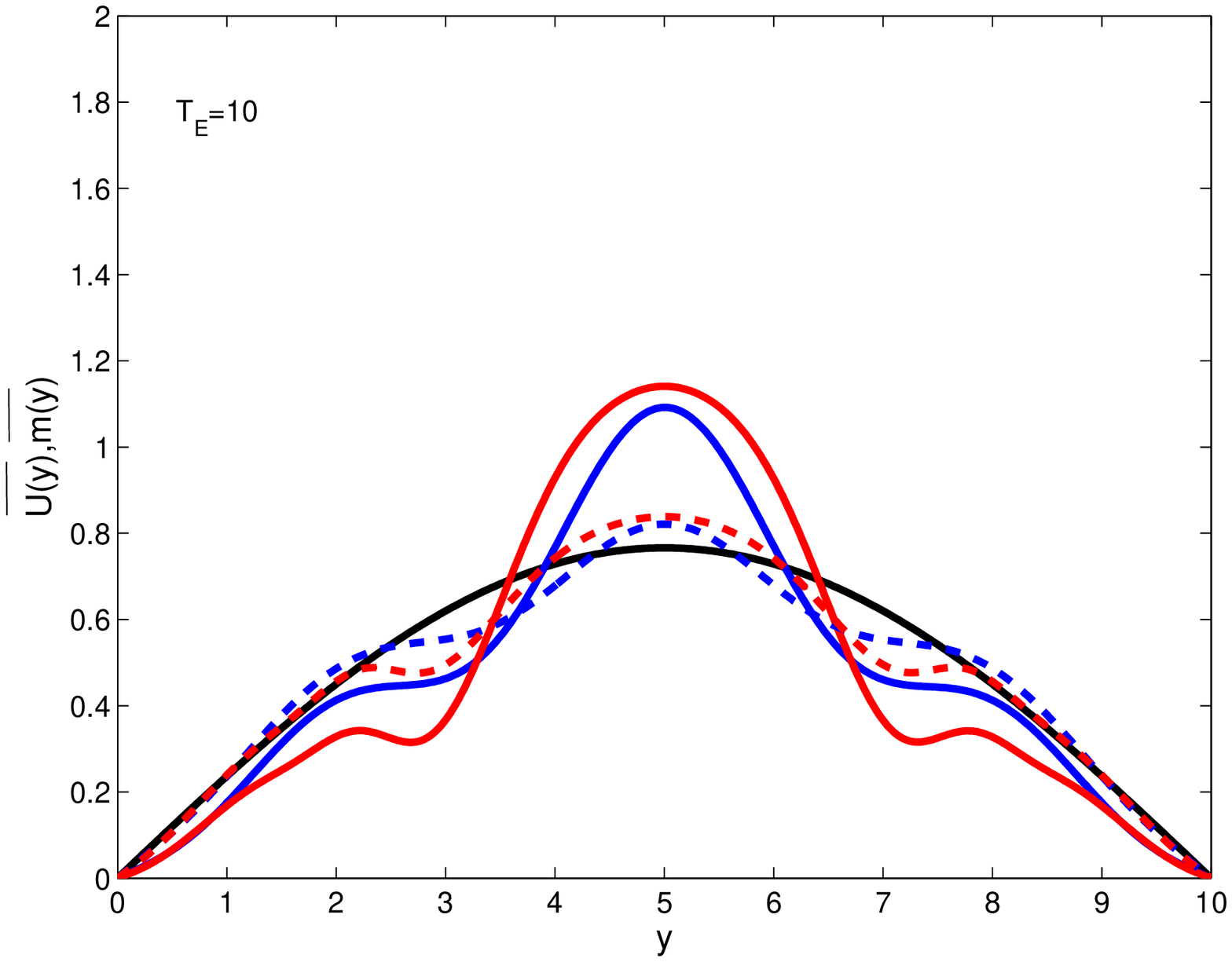}\\
  \includegraphics[angle=270,width=0.49\textwidth]{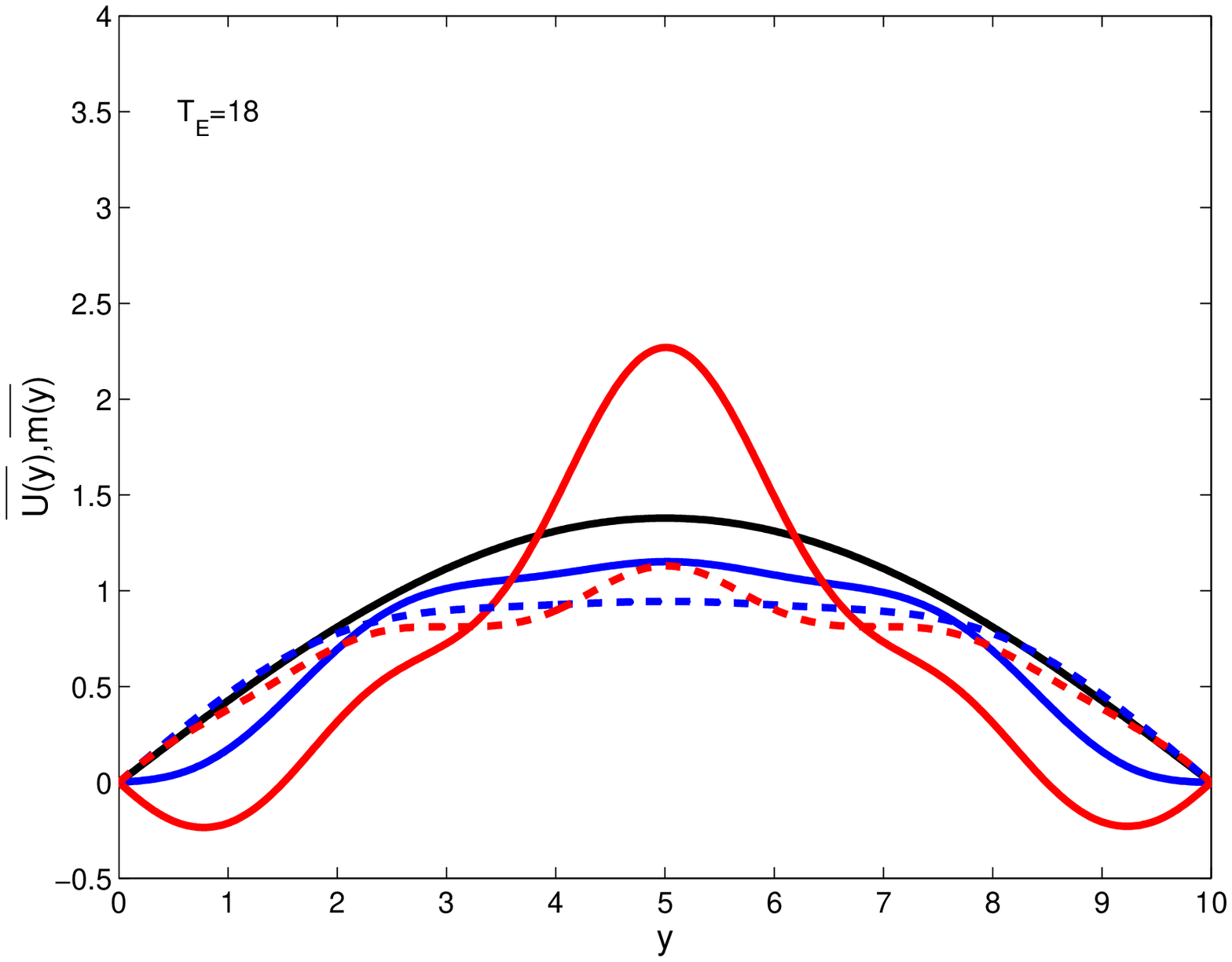}
  \includegraphics[angle=270,width=0.49\textwidth]{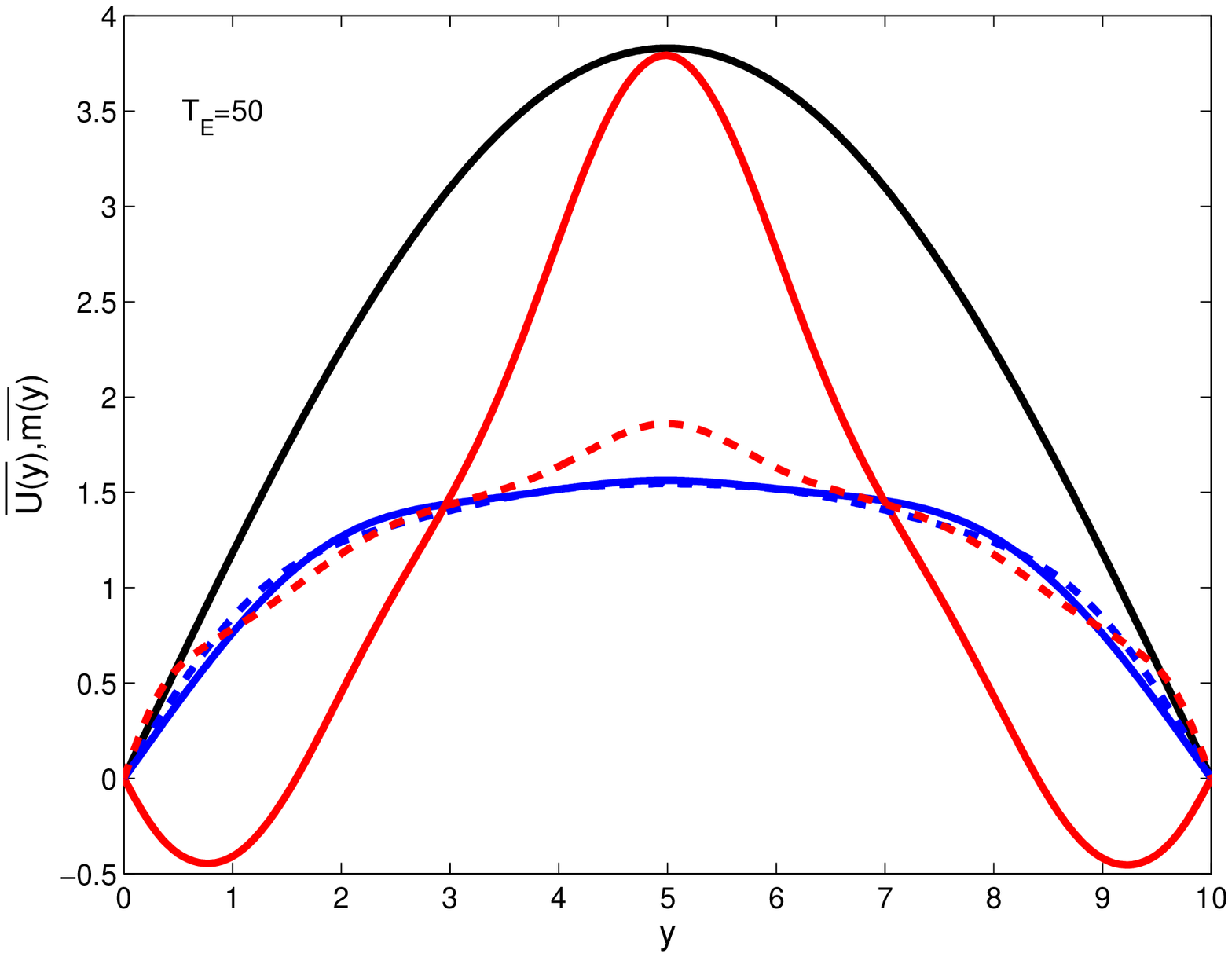}
 \caption{
   Time-averaged latitudinal profiles $\overline{U(y)}$
   (solid lines) and $\overline{m(y)}$ (dashed lines).
   In all figures the black solid line indicates
   the $U(y)=m(y)$ profile of the Hadley equilibrium,
   the blue and red lines refer to the cases $JT=8$ and $JT=32$, respectively.
   The values of $T_E$ are indicated.
   Note that the vertical scale for $T_E=9$ and $10$ is about $1/2$
   as for the other two figures.
  }
  \label{fig:profu}
\end{figure}

By examining more detailed diagnostics on the winds, such as the
time-averaged latitudinal profiles of $U(y)$ and of $m(y)$
(\figref{profu}), relevant differences are observed between $JT=8$
and the other three cases. Results are presented for $JT=8$ and
$JT=32$, the latter being representative also of $JT=16$ and $64$.
We first note that already for $T_E=$ $9$ and $10$, such that only
a weakly chaotic motion is realized, the $\overline{U(y)}$ and
$\overline{m(y)}$ profiles feature in both resolutions relevant
qualitative differences with respect to the corresponding Hadley
equilibrium profile, although symmetry with respect to the center
of the channel is obeyed. The $\overline{U(y)}$ and
$\overline{m(y)}$ profiles are different (the
constraint~\eqref{mUmeanaverage} being still satisfied), with
$\overline{U(y)}>\overline{m(y)}$ at the center and
$\overline{U(y)}<\overline{m(y)}$ at the boundaries of the
channel. Nevertheless, like for the Hadley equilibrium, both
$\overline{U(y)}$ and $\overline{m(y)}$ are positive and are
larger at the center of the channel than at the boundaries.
Consequently, at pressure level $p_1$ there is a westerly flow at
the center of the channel and easterly flows at the two
boundaries, and that at pressure level $p_3$ the wind is
everywhere westerly and peaks at the center of the channel. Such
features are more pronounced for the $JT=32$ case, where the
mechanism of the convergence of zonal momentum is more accurately
represented.

For larger values of $T_E$, the differences between the two
truncation levels become more apparent. For $JT=8$, the observed
$\overline{U(y)}$ and $\overline{m(y)}$ profiles tend to flatten
in the center of the channel and to become more similar to each
other. Therefore, somewhat similarly to the Hadley equilibrium
case, the winds at the pressure level $p_1$ tend to vanish and all
the dynamics is restricted to the pressure level $p_3$. The
$\overline{m(y)}$ profiles for $JT=32$ are quite similar to those
of $JT=8$, even if they peak and reach higher values in the center
of the channel and are somewhat smaller at the boundaries. So when
a finer resolution is used, a stronger temperature gradient is
realized in the channel center. The $\overline{U(y)}$ profiles
obtained for $JT=32$ are instead very different. They feature a
strong, well-defined peak in the channel center and negative
values near the boundaries. Therefore, the winds in the upper
pressure level are strong westerlies, and peak in the center of
the channel, while the winds in the lower pressure level feature a
relatively strong westerly jet in the center of the channel and
two compensating easterly jets at the boundaries. The fact that
for higher resolution the wind profiles are less smooth and have
more evident jet-like features is related to the more efficient
mechanism of barotropic stabilization, which, through zonal wind
convergence, \textit{keeps the jet together}
\cite{Kuo73,SIMM,RAND,JAMES}.

Examination of the latitudinal profiles in \figref{profu}
clarifies our choice to extend the latitudinal domain of the model
beyond the geometrically and geographically realistic mid-latitude
channel. Thanks to this, the wind fields in the central portion of
the domain (the latter corresponds to mid-latitudes and is of
primary interest in this work), are rather different than at the
boundary regions. The observed features, and especially the
presence of a jet, are in qualitative agreement with the real
atmosphere if models having truncation order of $JT\ge16$ are
used.

Summarizing, by considering the latitudinal average of the wind
fields in the mid-latitudes range $[0.25 L_y,0.75 L_y]$, the
$JT=8$ model greatly differs from the higher resolution models,
since $\langle m\rangle $ and especially $\langle U\rangle $ are
underestimated. Indeed, these diagnostics do not only rely on a
global balance, which is relatively weakly resolution-dependent
(see previous section), but also on the resolution-sensitive
representation of internal processes such as the zonal wind
convergence.

\newpage
\section{Summary and Conclusions}
\label{sec:conclusions}

We have described the construction and the dynamical behavior of
an intermediate complexity model of the atmospheric system. The
\textit{ab-initio} equations of dynamics and thermodynamics of a
stratified fluid are specialized to the quasi-geostrophic motion
and a new detailed derivation of the quasi-geostrophic two-layer
model of the planetary scale atmospheric flow in a mid-latitudes
beta-plane is provided. The derivation is performed by retaining,
at each step, the variables as expressed in physical units, while
the non-dimensionalization procedure, useful for the numerical
integrations, is introduced at last.

A single zonal wave solution is assumed and a partial differential
equation is derived for its coefficients. By a spectral
discretization in the latitudinal direction (using a Fourier
half-sine expansion), the latter equation is reduced to a system
of $N=6 \times JT$ ordinary differential equations, where $JT+1$
is the number of nodes of the (latitudinally speaking) fastest
varying base function. We have considered the cases $JT=$ $8$,
$16$, $32$, and $64$.

By increasing the parameter $T_E$, corresponding to the imposed
equator-to-pole temperature gradient, the system develops a
strange attractor in phase space. The route leading to the
formation of this strange attractor involves:
\begin{itemize}
\item
  a Hopf bifurcation at $T_E=T_E^H$
  responsible for the loss of stability of the Hadley equilibrium
  (corresponding to corresponding to baroclinic instability),
  where a periodic orbit branches off;
\item
  a Hopf bifurcation where a two-torus is created;
\item
  a finite number of quasi-periodic period doublings
  of the invariant two-torus;
\item
  two-torus breakdown at $T_E=T_E^{crit}$.
\end{itemize}
Statistical indicators, such as lagged autocorrelations, have been
used to characterize the observed quasi-periodic or strange
attractors for various values of $T_E$. To generate the required
time series, a physically relevant observable has been computed,
the total energy of the system. For $T_E$ close to $T_E^{crit}$
quasi-periodic intermittency and very weak chaoticity are
detected. The corresponding flow pattern might be classified as an
amplitude vacillation, like for the two-torus dynamics. For larger
$T_E$ the lagged autocorrelation typically decays (exponentially)
fast. The observed route to chaos is qualitatively the same for
$JT=$ $16$, $32$, and $64$, and the values $T_E^H$ and
$T_E^{crit}$ weakly depend on $JT$ (\tabref{hopf1}). Structural
differences occur for $JT=8$: a transition to a three-torus,
yielding a modulated amplitude vacillation, is involved.

The strange attractor is further studied by means of the Lyapunov
exponents, where we have varied both $T_E$ and model resolution
$JT$. Although the system qualitative behavior is analogous for
different values of $JT$, there are significant quantitative
differences. In all cases, the maximal Lyapunov exponent
$\lambda_1$ increases with $T_E$, and it is possible to robustly
fit a power law of the form
$\lambda_1\propto(T_E-T_E^{crit})^\gamma$. For $T_E$ fixed, the
maximal Lyapunov exponent decreases with $JT$ (so that the
predictability time increases). On the contrary, the metric
entropy increases linearly with $T_E$ for all examined values of
$JT$, and is larger for larger values of $JT$. In other words, the
fastest (the total) dynamical instability of the system is smaller
(larger) for larger $JT$, where the dynamics is more accurately
represented.

The Lyapunov dimension $D_L$ increases with both $T_E$ and $JT$.
The dependence of $D_L$ on $T_E$ is qualitatively the same for all
values of $JT$: by increasing $T_E$ there is an initial phase
where the dimension quickly grows with a power law
$D_L\propto(T_E-T_E^{crit})^\gamma$, followed by a linear scaling
regime. For large $T_E$, the dimension \textit{saturates} and
depends sublinearly on $T_E$. The latter effect is, of course,
more evident for small values of $JT$. It provides a measure of
accuracy of the spectral discretization (as far as the details of
the dynamics are concerned), which turns out to depend on $T_E$.

When considering the bounding box of the system, \textit{i.e.} the
minimal hyperparallelepiped containing the attractor in phase
space, for sufficiently high truncation order $JT$ each side of
the box increases as $\propto(T_E-T_E^{crit})^{1/3}$ for
$T_E-T_E^{crit}\lesssim 1.5$ and as
$\propto(T_E-T_E^{crit})^{5/6}$ for larger values of $T_E$. So for
a given value of $JT$ the ratios of the ranges of the various
degrees of freedom remain essentially unchanged when varying
$T_E$, yielding a self-similar scaling property. The volume of the
bounding box $V_{BB}$ then results to increase as
$\propto(T_E-T_E^{crit})^{N/3}$ and as
$\propto(T_E-T_E^{crit})^{5N/6}$ in the mentioned domains of
$T_E$.

A peculiar feature of this dynamical system is the rather smooth
dependence on the parameter $T_E$ of all the examined properties
of the strange attractor. No windows of periodicity have been
detected in the chaotic range and this is quite uncommon
especially when comparing with low-dimensional chaotic systems
such as the H\'enon-Pomeau mapping~\cite{HP,Sim79} or the Lorenz
flow~\cite{Lor63} (also see~\cite{BST98,BSV1,ER}). Although
\emph{structural stability}~\cite{ER} is out of the question,
other stability concepts (such as robustness~\cite{Viana}) might
provide an alternative and more practical theoretical basis for
the explanation of the observed parametric smoothness, perhaps
also for other systems of intermediate and high dimensionality.

Despite the sensitivity of Lyapunov exponents and dimension to
model resolution $JT$, certain observables of physical interest,
such as the time-averaged total energy of the system, or the
time-averaged spatially averaged zonal wind fields, are in
\emph{quantitative} agreement for all values of $JT$, except for
high values of $T_E$. Indeed these quantities are representative
of global balances, which turn out to be only slightly affected by
model resolution. When the system enters the chaotic regime, the
average total energy and average zonal winds have lower values
than those of the coexisting - and unstable - Hadley equilibrium,
because the chaos-driven occupation of the faster-varying
latitudinal modes fuels viscous dissipation, which acts
preferentially on the small scales. Other mechanisms which are
present in the real atmosphere, such as the \textit{barotropic
governor} \cite{NAKA}, are not represented in this schematic
model. Moreover, the total energy and the average wind field at
the Hadley equilibrium depend quadratically and linearly on $T_E$,
respectively, in the chaotic regimes such quantities obey a
subquadratic and sublinear power law $\propto T_E^\gamma$,
respectively. For both quantities, the exponents of the power laws
decrease abruptly as $T_E-T_E^{crit}$ crosses $1.5$. An analogous
sharp change is observed for $V_{BB}$, which suggests the onset of
a self-similar scaling law.

Nevertheless, when analyzing more detailed diagnostics on the
winds at the two pressure levels, relevant differences emerge
between the model with $JT=8$ and those with the higher
resolutions. For $JT=8$ the wind profiles are rather flat in
latitude and very weak in the lower pressure level. For the higher
resolution models the winds in the upper pressure level are strong
westerlies and peak in the center of the channel (corresponding to
mid-latitudes), in qualitative agreement with reality. The winds
in the lower pressure level feature a relatively strong westerly
jet in the center of the channel and two compensating easterly
jets at the boundaries. The fact that for higher resolution the
wind profiles are less smooth and have more clear-cut jet-like
features is related to the more efficient mechanism of barotropic
stabilization, which, through zonal wind convergence,
\textit{keeps the jet together}.

The model we study, although admittedly very schematic, is
Earth-like in that it features some fundamental processes
determining the general circulation of the Earth atmosphere, in
particular:
\begin{itemize}
\item
  the complex process of atmospheric baroclinic conversion, transforming
  available potential energy associated with (latitudinally) differential
  Sun heating into kinetic energy of synoptic scale motions
  of the mid-latitudes atmosphere;
\item
  nonlinear stabilization by eddy momentum convergence from non-symmetric
  baroclinic disturbances into the zonal jet;
\item
  viscous dissipation.
\end{itemize}
While the baroclinic conversion process is essentially well
represented in all models (even if those with higher resolution
are more efficient in the conversion for large $T_E$, since
conversion can take place also on smaller spatial scales), the
descriptions of the barotropic zonal wind convergence and of the
viscous dissipation are much more critically dependent on the
latitudinal truncation order, since the latter processes are
represented by terms involving the latitudinal derivatives of the
fields.

When a larger pool of available energy is provided, the dynamics
of the system is richer, since the baroclinic conversion process
can transfer larger amounts of energy to the disturbances: for
each given value of $JT$, the largest Lyapunov exponent, the
metric entropy, the Lyapunov dimension of the attractor, the mean
and the variability of the total energy and of the latitudinally
averaged zonal wind fields all increase with $T_E$. The
enhancement of the efficacy of the baroclinic conversion process
when higher resolution is adopted is highlighted by the increase
with $JT$, for a fixed value of $T_E$, of the number of linearly
unstable modes of the Hadley equilibrium, of the Lyapunov
dimension of the attractor, and of the metric entropy.

The critical dependence of the efficiency of the two mentioned
stabilizing processes on the model resolution is illustrated by
several results, \textit{e.g.} in the dependence of the parameters
$T_E^H$ and $T_E^{crit}$ on $JT$ (it is easier to destabilize a
system with lower resolution), in the fact that there are fewer
unstable modes of the Hadley equilibrium for larger values of $JT$
in the vicinity of $T_E^H$, in the fact that the predictability
time increases with $JT$ for a given value of $T_E$ and in the
features of the latitudinal profiles of the winds.

Although relevant ingredients of geometrical (horizontal
convergence due to the Earth curvature, latitudinal boundary
conditions at the margins of the middle latitude circumpolar
vortex, etc.) and dynamical (stabilization mechanisms other then
momentum convergence such as the so-called \textit{barotropic
governor}~\cite{NAKA}) nature of the real atmospheric circulation
are still missing in this, very preliminary, theoretical
representation, some important general conclusions are drawn from
the described results.

Pessimistic conclusions (in increasing order of pessimism):

\begin{itemize}
\item
  No simple \textit{mean field} or \textit{macroscopic adjustment}
  theory can be formulated for such complex nonlinear systems,
  even for relatively simple models as those proposed in this work.
\item
  It is, in general, doubtful whether \textit{invariant manifolds}
  in phase space - such as fixed points, periodic orbits -
  carry any useful information concerning the
  general circulation of the system.
\item
  Beyond the time of deterministic predictability,
  ``averaging'' is of no practical use;
  it is not clear what else should be done in order to produce
  \textit{useful} - in a statistical sense - \textit{predictions}.
\end{itemize}

Optimistic conclusions:
\begin{itemize}
\item
  Although some dynamical system properties, such as Lyapunov exponents
  and dimension, are strongly model-dependent,
  some other - of great physical interest - are not.
\item
  Increasing refinement (number of degrees of freedom) of models may
  produce smoother dependence on macroscopic parameters.
\item
  It is not outside the range of practically feasible, although possibly
  challenging, projects to put together an intermediate dimensionality model
  - with hundreds of (well chosen!) degrees of freedom -
  with \emph{stable properties}
  which is relevant for a theory of general atmospheric circulation.
\end{itemize}

\begin{acknowledgments}
  We wish to thank Mara Felici for technical and scientific help.
\end{acknowledgments}

\newpage

\appendix

\section{On the numerical methods}
\label{app:numerical} We begin by describing the projection
operator used in the definition of the vector
field~\hbox{\eqref{dotA1j}-\eqref{dotmj}}. As it is customary with
climatological spectral models~\cite{Holton}, a pseudospectral
method is used, also known as Fourier collocation~\cite{Boy,GO}.

The fields $A$, $B$, $U$, $m$, appearing in the nonlinear terms
of~\hbox{\eqref{dotA1j}-\eqref{dotmj}}, are first evaluated at
$JT$ collocation points $y_1,\dots,y_{JT}$, equally spaced in the
$y$-domain $(0,L_y)$. This is achieved by a Discrete Sine
Transform of $A^1_j$, $A^2_j$, $B^1_j$, $B^2_j$, $U_j$, $m_j$,
with $j=1,\dots,JT$. The terms involving second derivatives with
respect to $y$ are also computed in this way, by premultiplying
for a suitable coefficient involving the wave numbers $w_j$. Then
all the nonlinear terms are evaluated pointwise, at each of the
collocation points $y_1,\dots,y_{JT}$. Lastly, an inverse Discrete
Sine Transform is carried out, yielding the Fourier coefficients
of the nonlinear terms. The software library
\texttt{fftw3}~\cite{fftw}, publicly available at
\texttt{www.fftw.org}, has been used for the Discrete Sine
Transform.

The numerical solution of the system of ordinary differential
equations~\hbox{\eqref{dotA1j}-\eqref{dotmj}} is computed by means
of a standard Runge-Kutta-Fehlberg(4,5) algorithm~\cite{SWD} with
adaptive stepsize, where the approximated solution is carried by
the order five method. The local truncation error is kept below
$1.e-6$. The stepsize adjustment procedure is similar to that of
\texttt{DOPRI5}, available at
(\texttt{www.unige.ch/}$\sim$\texttt{hairer}).

The total energy of~\hbox{\eqref{dotA1j}-\eqref{dotmj}}, is
computed according to~\eqref{eq:energy}. In terms of the Fourier
coefficients $A^1_j$, $B_1^j$, \dots, this yields the expression
\begin{equation}%%noref
  \begin{split}
    E(t)=L_xL_y\Biggl\{\Biggr.
      &\left(\frac{12\pi}{L_x}\right)^2\sum_{j=1}^{JT}   \left(\left(A^1_j\right)^2+\left(B^1_j\right)^2+
      \left(\frac{\pi}{L_y}\right)^2 j^2\left(\left(A^2_j\right)^2+\left(B^2_j\right)^2\right)+
      \frac{2}{H_2^2}\left(B^2_j\right)^2\right)+\\
   & \left(\frac{12\pi}{L_x}\right)^2\sum_{j=1}^{JT}   \left(\left(A^2_j\right)^2+\left(B^2_j\right)^2+
      \left(\frac{\pi}{L_y}\right)^2 j^2\left(\left(A^1_j\right)^2+\left(B^1_j\right)^2\right)+
      \frac{2}{H_2^2}\left(B^1_j\right)^2\right)+\\
      &\frac12 \sum_{j=1}^{JT}  \left(U_j^2+m_j^2+
      \frac{2}{H_2^2}\left(\frac{L_y}{\pi}\right)^2\left(\frac{m_j}{j}\right)^2\right)\Biggl.\Biggr\}.
  \end{split}
\end{equation}%%noref
For the computation of the averages in \secref{energy}, time
series of 315360 adimensional time units (1000 years in natural
units) have been computed for all values of $JT$, preceded by a
transient of five years (time is expressed in the scale of the
system, see \tabref{tab1}). The observables $E(t)$, $U(y,t)$, and
$m(y,t)$ have been sampled every 0.216 time units (four times a
day), thereby obtaining time series of $1460000$ elements. The
sample mean and sample standard deviation have been computed
according to the usual formulas:
\begin{equation}%%noref
  \overline{E(t)}=\frac{1}{n}\sum_{i=1}^nE(t_i),
  \qquad
  \sigma_E^2=  \frac{1}{n-1} \left( \sum_{i=1}^nE^2(t_i) - n\overline{E(t)}^2\right).
\end{equation}%%noref
The initial condition used for all computations is $A^1_1 = -0.8,
A^1_2 = 0.65, B^1_1 =  0.2, B^1_2 = 0.2, B^2_1 = 0.4, B^2_2 = 0.1,
U_1  =  1.26, m_1  =  1.1$, with the remaining coefficients set to
$0$, as in~\cite{SM}.

\newpage
\section{Lyapunov exponents}
\label{app:Lyapunov} The Lyapunov exponents of
system~\hbox{\eqref{dotA1j}-\eqref{dotmj}} are estimated according
to the algorithm described by Galgani, Giorgilli, Benettin and
Strelcyn~\cite{BGGS}. The first variational equations
of~\hbox{\eqref{dotA1j}-\eqref{dotmj}} are integrated during a
period of time $T$, with the identity matrix as initial condition.
During integration, at time $t$  the canonical orthonormal basis
is mapped onto a new set of vectors
$(\mathbf{v}_1^t,\mathbf{v}_2^t,\dots,\mathbf{v}_{N}^t)$, where
$N=6\times JT$ is the dimension of the phase space. Each vector
tends to align itself along the direction of maximal expansion (or
of minimal compression). Thus all $\mathbf{v}_j^t$'s tend to
collapse onto one direction. To prevent this, the Gram-Schmidt
process is applied to
$(\mathbf{v}_1^{t_1},\mathbf{v}_2^{t_1},\dots,\mathbf{v}_{N}^{t_1})$
at $t=t_1$, yielding a set
$(\tilde{\mathbf{v}}_1^{t_1},\dots,\tilde{\mathbf{v}}_N^{t_1})$ of
orthogonal vectors. The vectors are normalized by putting
$\mathbf{w}_j^{t_1}=\tilde{\mathbf{v}}_j^{t_1}/{\lVert\tilde{\mathbf{v}}_j^{t_1}\rVert}$
for $j=1,\dots,N.$ Then a new frame of vectors
$(\mathbf{v}_1^{t_2},\dots,\mathbf{v}_N^{t_2})$, with $t_2=2t_1$,
is computed by integrating the first variational equations taking
as initial condition the orthonormal vectors
$(\mathbf{w}_1^1,\dots,\mathbf{w}_N^1)$ from the previous step,
and the whole process is repeated. At iteration step $k$, define
$t_k=kt_1$ and
$$
c_j^k=\prod_{i=1}^k{\lVert\tilde{\mathbf{v}}_j^{t_k}\rVert}
\quad\text{ and }\quad
\mathbf{w}_j^{t_k}=\frac{\tilde{\mathbf{v}}_j^{t_k}}{\lVert{\tilde{\mathbf{v}}_j^{t_k}}\rVert}
\quad\text{ for }\quad j=1,\dots,N.
$$
The orthonormalization process does not change the direction of
$\mathbf{v}_1^{t_k}$, so that $\mathbf{w}_1^{t_k}$ still points to
the direction of maximal stretch. Denoting by $\lambda_j$,
$j=1,\dots,N$, the Lyapunov exponents in decreasing order of
magnitude, the length $c_1^k$ of $\mathbf{v}_1^{t_k}$ is
approximately proportional to $e^{k\lambda_1}$. The plane spanned
by $\mathbf{v}_1^{t_k}$ and $\mathbf{v}_2^{t_k}$ is not changed by
the Gram-Schmidt process and tends to adjust to the subspace of
maximal growth of surfaces. The rate of growth of areas is
proportional to $e^{k(\lambda_1+\lambda_2)}$. In particular, since
$\mathbf{v}_1^{t_k}=\mathbf{w}_1^{t_k}$ and $\mathbf{w}_2^{t_k}$
are orthonormal, the length of the projection of
$\mathbf{v}_2^{t_k}$ upon $\mathbf{w}_2^{t_k}$ is proportional to
$e^{k\lambda_2}$. A similar argument for growth of volumes yields
that $c_j^k$ is proportional to $e^{k\lambda_j}$. Therefore, the
Lyapunov exponent $\lambda_j$ is estimated by the averages
\begin{equation}%%noref
  \lambda_j\approx\frac{1}{k}\log(c^k_j),
\end{equation}%%noref
where $k=T/t_1$. We have chosen $T$ of the order of 3150
adimensional time units (about 10 years in natural units) for all
values of $JT$, while $t_1$ has been chosen as 0.864 adimensional
time units (1 day), which allow for an excellent convergence of
the exponents.

Actually, we have used a version of the algorithm~\cite{BGGS} in
which the variational equations are not integrated explicitly, but
approximated by means of numerical differentiation: $N$
trajectories are simultaneously integrated, starting from points
nearby a reference orbit. The distances from the reference orbit
are normalized at regular time steps~\cite{Si1}.

The library \texttt{LAPACK} (\texttt{www.netlib.org}) has been
used for Gram-Schmidt orthogonalization and for other computations
in this work.

\clearpage
\newpage

\begin{table}[htb]
  \centering
  \begin{minipage}[c]{\textwidth}
    \begin{tabular}{|c|c|c|}
      \hline
      Variable &  Scaling factor & Value of scaling factor \\
      \hline
      $x$ & $l$ & $10^6m$\\
      $y$ &  $l$ & $10^6m$\\
      $t$ &  $u^{-1}l$ & $10^5s$\\
      $\psi_1$ &  $ul$ & $10^7m^2s^{-1}$\\
      $\psi_3$ &  $ul$ & $10^7m^2s^{-1}$\\
      $\phi$ &  $ul$ & $10^7m^2s^{-1}$\\
      $\tau$ &  $ul$ & $10^7m^2s^{-1}$\\
      $A^1_n$ &  $ul$ & $10^7m^2s^{-1}$\\
      $B^1_n$ &  $ul$ & $10^7m^2s^{-1}$\\
      $A^2_n$ &  $ul$ & $10^7m^2s^{-1}$\\
      $B^2_n$ &  $ul$ & $10^7m^2s^{-1}$\\
      $m$ &  $u$ & $10ms^{-1}$\\
      $U$ &  $u$ & $10ms^{-1}$\\
      $m_n$ &  $u$ & $10ms^{-1}$\\
      $U_n$ &  $u$ & $10ms^{-1}$\\
      $w_n$ &  $l^{-1}$ & $10^{-6}m^{-1}$ \\
      $Lag$ &  $u^{-1}l$ & $10^5s$\\
      $\lambda_j$ &  $ul^{-1}$ & $10^{-5}s^{-1}$\\
      $t_p$ &  $u^{-1}l$ & $10^{5}s$\\
      $T$ &  $ulf_0R^{-1}$ & $3.5K$\\
      $E$ &  $u^2l^2(\delta p) g^{-1}$ & $5.1 \times 10^{17} J$\\
      \hline
    \end{tabular}
    \caption{
      Variables of the system and non-dimensionalization factors.
      For $A^1_n$, $B^1_n$, $A^2_n$, $B^2_n$, $m_n$, $U_n$, and $w_n$,
      the index $n$ ranges from $1$ to $JT$.
      For $\lambda_j$, $n=1,\ldots,6\times JT$.
    }
    \label{tab1}
  \end{minipage}
\end{table}

\begin{table}[htb]
  \centering
  \begin{minipage}[c]{\textwidth}
    \begin{tabular}{|c|c|c|c|c|}
      \hline
      Parameter &  Dimensional Value & Non-dimensional value & Scaling factor & Value of scaling factor \\
      \hline
      $L_x$ & $3\times10^7m$ & $29$ & $l$ & $10^6m$ \\
      $L_y$ & $10^7m$ & $10$ & $l$ & $10^6m$ \\
      $\chi$ & $2\pi/\left(4.833\times10^6\right)m^{-1}$ & $1.3$ & $l^{-1}$ & $10^{-6}m^{-1}$ \\
      $H_2$ & $7.07\times10^5m$ & $7.07\times10^{-1}$ & $l$ & $10^6m$ \\
      $f_0$& $10^{-4}s^{-1}$& $10$ &  $ul^{-1}$ & $10^{-5}s^{-1}$\\
      $\beta$& $1.6\times10^{-11}m^{-1}s^{-1}$& $1.6$ &  $ul^{-2}$ & $10^{-11}m^{-1}s^{-1}$\\
      $\nu_E$ & $5.5\times10^{5}m^2s^{-1}$ & $5.5\times10^{-2}$ &  $ul$ &  $10^{7}m^2s^{-1}$\\
      $\kappa$ &  $2.8\times10^{5}m^2s^{-1}$ & $2.8\times10^{-2}$ & $ul$  &  $10^{7}m^2s^{-1}$ \\
      $\nu_N$ & $1.1\times10^{-6}s^{-1}$ & $1.1\times10^{-1}$ & $ul^{-1}$ &$10^{-5}s^{-1}$ \\
      $T_E$ & $28K$ to $385K$ &  $8$ to $110$ & $ulf_0R^{-1}$  & $3.5K$\\
      \hline
    \end{tabular}
    \caption{
      Values of the parameters used in this work and non-dimensionalization factors. }
    \label{tab2}
  \end{minipage}
\end{table}

\begin{table}[htb]
  \centering
  \begin{minipage}[c]{\textwidth}
    \begin{tabular}{|c|c|c|}
      \hline
      $JT$ &  $T_E^H$ & $T_E^{crit}$ \\
      \hline
      $8$ & $7.83$ & $9.148$ \\
      $16$ &  $8.08$ & $8.415$ \\
      $32$ & $8.28$ & $8.522$ \\
      $64$ & $8.51$  &  $8.663$\\
      \hline
    \end{tabular}
  \caption{
    Approximate values of the parameter $T_E$
    where the Hadley equilibrium loses stability via Hopf bifurcation ($T_E^H$)
    and where the onset of the chaotic regime occurs ($T_E^{crit}$)
    for each of the considered orders of truncation $JT$.
    See text for details.
  }
  \label{hopf1}
  \end{minipage}
\end{table}

\begin{table}[htb]
  \centering
  \begin{minipage}[c]{\textwidth}
    \begin{tabular}{|c|c|c|}
      \hline
      $JT$ &  $\gamma[\log(T_E-T_E^{crit})\leq0.5]$ & $\gamma[\log(T_E-T_E^{crit})\geq0.5]$ \\
      \hline
      $8$ & $40\pm1$ & $40\pm1$ \\
      $16$ & $33\pm3$ & $80\pm1$ \\
      $32$ & $66\pm2$ & $160\pm1$ \\
      $64$ & $133\pm4$  &  $320\pm1$\\
      \hline
    \end{tabular}
    \caption{
      Power-law fits of the volume of the bounding box as
      $V_{BB}\propto(T_E-T_E^{crit})^\gamma$
      in two different ranges of $T_E-T_E^{crit}$
      for each of the considered orders of truncation $JT$.
      See text and \figref{bbox} for details.
    }
  \label{bboxtable}
  \end{minipage}
\end{table}

\clearpage
\newpage

\clearpage

\begin{thebibliography}{99}

\bibitem{Peix}
  J.P.~Peixoto, A.H.~Oort:
  \textit{Physics of Climate}, Am. Inst. of Phys., College Park, 1992.

\bibitem{Luc02}
  V. Lucarini:
  Towards a definition of climate science,
  \textit{Int. J. Environment and Pollution} \textbf{18} (2002), 409--414.

\bibitem{Lor69}
  E.N.~Lorenz:
  The predictability of a flow which possesses many scales of motion,
  \textit{Tellus} \textbf{21}, (1969), 289--307.


\bibitem{Lor76}
  E.N.~Lorenz:
  Nondeterministic theories of climatic change,
  \textit{Quaternary Res.} \textbf{6} (1976), 495--506.

\bibitem{Char50}
  J.G.~Charney, R.~Fj\"ortoft, J.~von~Neumann:
  Numerical integration of the barotropic vorticity equation,
  \textit{Tellus} \textbf{2} (1950), 237--254.


\bibitem{Lor67}
  E.N.~Lorenz:
  \textit{The Nature and Theory of the General
    Circulation of the Atmosphere},
  World Meteorol. Organ., Geneva, 1967.


\bibitem{Lor83}
  E.N.~Lorenz:
  A History of Prevailing Ideas about the
  General Circulation of the Atmosphere,
  \textit{Bull. Am. Met. Soc.} \textbf{64} (1983), 730--769.

\bibitem{Jef24}
  H.~Jeffreys:
  On the Formation of Waves by Wind,
  \textit{Proc. Roy. Soc. Lond.} \textbf{107} (1924), 189--206.


\bibitem{Jef25}
  H.~Jeffreys:
  On the Formation of Waves by Wind,
  \textit{Proc. Roy. Soc. Lond.}, \textbf{110A} (1925), 341--347.


\bibitem{Palmen}
  E.~Palmen:
  The Role of Atmospheric Disturbances in the General Circulation,
  \textit{Quart. J. Roy. Meteor. Soc.} \textbf{77} (1951), 337--354.


\bibitem{Mar}
  M.~Margules:
  Die energie der St\"{u}rme,
  \textit{Jahrb. Zentralanst. Meteor. Wien} \textbf{40} (1903), 1--26.


\bibitem{Lor55}
  E.N.~Lorenz:
  Available potential energy and the maintenance of the
  general circulation,
  \textit{Tellus} \textbf{7}, (1955), 157--167.


\bibitem{Lor60}
  E.N.~Lorenz:
  Generation of available potential energy and the
  intensity of the general circulation,
  in \textit{Dynamics of Climate}, R.L. Pfeffer ed.,
  Pergamon, Tarrytown (1960), 86--92.

\bibitem{Black}
  M.L.~Blackmon:
  A climatological spectral study of the 500 mb geopotential height
  of the Northern Hemisphere,
  \textit{J. Atmos. Sci.} \textbf{33} (1976), 1607--1623

\bibitem{SP}
  A.~Speranza:
  Deterministic and statistical properties of the westerlies,
  \textit{Paleogeophysics} \textbf{121} (1983), 511--562


\bibitem{DellAquila}
  A.~\hbox{dell'Aquila}, V.~Lucarini, P.M.~Ruti, S.~Calmanti:
  Hayashi spectra of the northern hemisphere mid-latitude atmospheric
  variability in the {NCEP--NCAR} and ECMWF reanalyses,
  \textit{Clim. Dyn.} (2005), DOI: 10.1007/s00382-005-0048-x.

\bibitem{Char47}
  J.G.~Charney:
  The Dynamics of Long Waves in a Baroclinic Westerly Current,
  \textit{J. Atmos. Sci.} \textbf{4} (1947), 136--162.

\bibitem{Eady49}
  E.T.~Eady:
  Long waves and cyclone waves,
  \textit{Tellus} \textbf{1} (1949), 33--52.


\bibitem{Lor63}
  E.N.~Lorenz:
  Deterministic Nonperiodic Flow,
  \textit{J. Atmos. Sci.} \textbf{20} (1963), 130--141.


\bibitem{IPCC}
  Intergovernmental Panel on Climate Change 2001, Working Group I:
  \textit{Climate Change 2001: The Scientific Basis},
  Cambridge University Press, Cambridge, 2001.


\bibitem{HAL}
  N.M.J.~Hall, P.D.~Sardeshmukh:
  Is the time-mean Northern Hemisphere flow baroclinically unstable?,
  \textit{J. Atmos. Sci.}, \textbf{55} (1998), 41--56.

\bibitem{Kuo}
  H.L.~Kuo:
  On Production of Long-term Mean Zonal Current and Eddy
  Momentum and Heat Transports in Atmosphere,
  \textit{Pure Appl. Geophys.} \textbf{158} (2001), 1047--1064.



\bibitem{Far}
  J.D.~Farmer:
  Chaotic attractors of an infinite-dimensional dynamic system,
  \textit{Physica D} \textbf{4} (1982), 366--393.


\bibitem{SM}
  A.~Speranza, P.~Malguzzi:
  The statistical properties of a zonal jet in a baroclinic atmosphere:
  a semilinear approach. Part I: two-layer model atmosphere,
  \textit{J. Atmos. Sci.} \textbf{48} (1988), 3046--3061.


\bibitem{MTS}
  P.~Malguzzi, A.~Trevisan, A.~Speranza:
  Statistics and predictability for an intermediate dimensionality
  model of the baroclinic jet,
  \textit{Ann. Geoph.} \textbf{8} (1990), 29--35.

\bibitem{Fred}
  J.S.~Frederiksen:
  Instability theory and nonlinear evolution of blocks and mature anomalies,
  \textit{Advances in Geophysics} \textbf{29} (1986), 277--303.

\bibitem{Lor80}
  E.N.~Lorenz:
  Attractor sets and quasi-geostrophic equilibrium,
  \textit{J. Atmos. Sci.} \textbf{37} (1980), 1685--1699.

\bibitem{Ped}
  J.~Pedlosky:
  \textit{Geophysical Fluid Dynamics} (2nd ed.),
  Springer-Verlag, New York, 1987.



\bibitem{Phil54}
  N.A.~Phillips:
  Energy transformations and meridional circulations
  associated with simple baroclinic waves in a two-level,
  quasi-geostrophic model, \textit{Tellus} \textbf{6} (1954), 273--286.



\bibitem{ER}
  J.-P.~Eckmann, D.~Ruelle:
  Ergodic theory of chaos and strange attractors,
  \textit{Rev. Mod. Phys.} \textbf{57} (1985), 617--655.



\bibitem{SPELUC}
  A.~Speranza, V.~Lucarini:
  {Environmental Science:} physical principles and applications,
  in \textit{Encyclopedia of Condensed Matter Physics},
  F.~Bassani, J.~Liedl, P.~Wyder eds.,
  Elsevier, Amsterdam, in press (2005).


\bibitem{Holton}
  J.R.~Holton:
  \textit{An Introduction to Dynamic Meteorology},
  Academic Press, San Diego, 1992.


\bibitem{Hosk}
  B.J.~Hoskins,  M.E.~McIntyre,  A.W.~Robertson:
  On the use and significance of isentropic potential vorticity maps,
  \textit{Quart. J. R. Met. Soc.} \textbf{111} (1985), 877--946.

\bibitem{Luc05}
  V.~Lucarini, J.J.~Saarinen, K.-E.~Peiponen, E.~Vartiainen:
  \textit{Kramers-Kronig Relations in Optical Materials Research},
  Springer, Heidelberg, 2005.

\bibitem{HH}
  I.M.~Held, A.Y.~Hou:
  Nonlinear Axially Symmetric Circulations in a Nearly Inviscid Atmosphere,
  \textit{J. Atmos. Sci.} \textbf{37} (1980), 515--533


\bibitem{Kuz}
  Yu.~Kuznetsov:
  \textit{Elements of Applied Bifurcation Theory} (2nd ed.),
  Springer--Verlag (1998).

\bibitem{Kuo73}
  H.L.~Kuo:
  Dynamics of quasigeostrophic flows and instability theory,
  \textit{Adv. Appl. Mech.} \textbf{13} (1973), 247--330.


\bibitem{SIMM}
  A.J.~Simmons, B.J.~Hoskins:
  The life cycles of some nonlinear baroclinic waves,
  \textit{J. Atmos. Sci.} \textbf{35} (1978), 414--432.


\bibitem{RAND}
  W.J.~Randel, J.L.~Stanford:
  The observed life cycle of a baroclinic instability,
  \textit{J. Atmos. Sci.} \textbf{42} (1985), 1364--1373

\bibitem{JAMES}
  I.N.~James, L.J.~Gray:
  Concerning the effect of surface drag on the circulation
  of a baroclinic planetary atmosphere,
  \textit{Quart. J. Roy. Meteor. Soc.} \textbf{112} (1986), 1231--1250.


\bibitem{DraReid81}
  P.G.~Drazin,  W.H.~Reid:
  \textit{Hydrodynamic stability},
  Cambridge University Press, Cambridge, 1981.




\bibitem{PM}
  Y.~Pomeau, P.~Manneville:
  Intermittent transition to turbulence in dissipative dynamical systems,
  \textit{Comm. Math. Phys.} \textbf{74} (1980), 189--197.

\bibitem{BS}
  A.~Brandstater, H.~L.~Swinney:
  Strange attractors in weakly turbulent Couette-Taylor flow,
  \textit{Phys. Rev. A} \textbf{35} (1987), 2207--2220.

\bibitem{BSV1}
  H.W.~Broer, C.~Sim\'o, R.~Vitolo:
  Bifurcations and strange attractors in the \hbox{Lorenz-84}
  climate model with seasonal forcing,
  \textit{Nonlinearity} \textbf{15} (2002), 1205--1267.

\bibitem{FHW}
  J.D.~Farmer, J.~Hart, P.~Weidman:
  A Phase Space Analysis of Baroclinic Flow,
  \textit{Physics Letters A} \textbf{91} (1982), 22--24.

\bibitem{RFMR05}
  A.~Randriamampianina, W.-G.~Fr\"uh, P.~Maubert, P.L.~Read:
  DNS of bifurcations to low-dimensional chaos in an air-filled
  rotating baroclinic annulus,
  \textit{preprint at}  \texttt{http://www-atm.physics.ox.ac.uk/user/read/} (2005).

\bibitem{BHS}
  H.W.~Broer, G.B.~Huitema, M.B.~Sevryuk:
  \textit{Quasi-periodic Motions in Families of Dynamical Systems,
  Order amidst Chaos},
  Springer LNM \textbf{1645} (1996).

\bibitem{BSV2}
  H.W.~Broer, C.~Sim\'o, R.~Vitolo:
  Chaos and quasi-periodicity in diffeomorphisms
  of the solid torus,
  \textit{preprint}  \verb|mp_arc| \#05-107 (2005).


\bibitem{BST98}
  H.W.~Broer, C.~Sim\'o, J.C.~Tatjer:
  Towards global models near homoclinic tangencies
  of dissipative diffeomorphisms,
  \textit{Nonlinearity} \textbf{11} (1998), 667--770.


\bibitem{GB83}
  J.~Guckenheimer, G.~Buzyna:
  Dimension measurements for Geostrophic Turbulence,
  \textit{Phys. Rev. Lett.} \textbf{51(16)} (1983), 1438--1441.

\bibitem{HP}
  M.~H\'enon, Y.~Pomeau:
  Two strange attractors with a simple structure,
  in \textit{Turbulence and Navier-Stokes equations} \textbf{565} (1976),
  Springer-Verlag, 29--68.


\bibitem{Sim79} C.~Sim\'o:
  On the H\'enon--Pomeau attractor,
  \textit{J. Stat. Phys.} \textbf{21} (1979), 465--494.



\bibitem{Cao}
  Y.~Cao:
  The transversal homoclinic points are dense in the codimension-1
  H\'enon-like strange attractors,
  \textit{Proc. Amer. Math. Soc.} \textbf{127} (1999), 1877--1883.


\bibitem{MV}
  L.~Mora, M.~Viana:
  Abundance of strange attractors,
  \textit{Acta Math.} \textbf{171} (1993), 1--71.


\bibitem{Viana}
  M.~Viana:
  What's new on Lorenz strange attractors?,
  \emph{Math. Intelligencer} \textbf{22-3} (2000), 6--19.


\bibitem{Ose}
  V.I.~Oseledec:
  A multiplicative ergodic theorem. Lyapunov characteristic numbers
  for dynamical systems,
  \textit{Trudy Mosk. Mat. Obsc. (Moscow Math. Soc.)}
  \textbf{19} (1968), 19.



\bibitem{KY}
  J.~Kaplan, J.~Yorke:
  Chaotic behaviour of multidimensional difference equations, in
  Functional Differential Equations and Approximations of Fixed Points,
  \textit{Springer LNM} (1979), 204--227.



\bibitem{FarOttYor}
  J.D.~Farmer, E.~Ott, and J.A.~Yorke:
  The dimension of chaotic attractors,
  \textit{Physica D} \textbf{7} (1983), 153--180.


\bibitem{Rue90}
  D.~Ruelle:
  Deterministic chaos: the science and the fiction,
  \emph{Proc. R. Soc. London A} \textbf{427} (1990), 241--248.



\bibitem{WY}
  Q.~Wang, L.-S.~Young:
  Strange Attractors with One Direction of Instability,
  \textit{Comm. Math. Phys.} \textbf{218} (2001), 1--97.

\bibitem{GOST05}
  S.V.~Gonchenko, I.I.~Ovsyannikov, C.~Sim\'o, D.~Turaev:
  Three-dimensional H\'enon-like maps and wild Lorenz-like attractors,
  \textit{preprint at} \texttt{http://www.maia.ub.es/dsg/2005}.


\bibitem{Smith00}
  L.A.~Smith:
  Disentangling Uncertainty and Error:
  On the Predictability of Nonlinear Systems,
  in \emph{Nonlinear Dynamics and Statistics}, A. Mees ed.,
  Birkhauser, Boston (2000) 31--64.


\bibitem{Smith02}
  L.A.~Smith:
  What might we learn from climate forecasts?,
  \emph{Proc. Natl. Acad. Sci.} \textbf{99} (2002), 2487--2492.

\bibitem{Boy}
  J.P.~Boyd:
  \textit{Chebyshev \& Fourier Spectral Methods},
  Lecture Notes in Engineering \textbf{49}, Springer-Verlag, Berlin, 1989.



\bibitem{NAKA}
  N. Nakamura:
  Momentum flux, flow symmetry, and the nonlinear barotropic governor,
  \textit{J. Atmos. Sci.} \textbf{50} (1993), 2159--2179.




\bibitem{Stone}
  P.~Stone:
  Baroclinic adjustment,
  \emph{J. Atmos. Sci.} \textbf{35} (1978), 561--571.



\bibitem{GO}
  D.~Gottlieb, S.A.~Orszag:
  \textit{Numerical Analysis of Spectral Methods: Theory and Applications},
  CBMS-NSF Regional Conference Series in Applied Mathematics \textbf{26},
  SIAM Publications, Philadelphia, 1977.


\bibitem{fftw}
  M.~Frigo, S.G.~Johnson:
  The Design and Implementation of FFTW3,
  Proceedings of the IEEE \textbf{93(2)}, 216--231 (2005).
  Invited paper, Special Issue on Program Generation, Optimization, and
  Platform Adaptation.


\bibitem{SWD}
  L.F.~Shampine, H.A.~Watts, S.~Davenport:
  Solving Non-stiff Ordinary Differential Equations - The State of the Art,
  \textit{SIAM Review} \textbf{18} (1976), 376--411.


\bibitem{BGGS}
  G.~Benettin, L.~Galgani, A.~Giorgilli, J.-M.~Strelcyn:
  Lyapunov characteristic exponents for smooth dynamical systems
  and for Hamiltonian systems; a Method for computing all of them,
  Part 2: numerical applications,
  \textit{Meccanica} \textbf{15} (1980), 21--30.


\bibitem{Si1}
  C.~Sim\'o:
  On the Analytical and Numerical Approximation of Invariant Manifolds,
  \textit{Les M\'ethodes Modernes de la Mec\'anique C\'eleste
    (Course given at Goutelas, France, 1989)},
  D.~Benest and C.~Froeschl\'e eds., Editions Fronti\`eres, Paris (1990),
  285--329.
  Available at \texttt{www.maia.ub.es/dsg/2004/index.html}.



\end{thebibliography}
\end{document}